\documentclass[useAMS,letterpaper,fleqn,usenatbib]{mn2e}
\usepackage[T1]{fontenc}
\usepackage{ae,aecompl}
\usepackage{ulem}
\usepackage{graphicx}
\usepackage{amsmath}
\usepackage{amssymb}
\usepackage{xspace}
\usepackage{xcolor}
\usepackage{aas_macros}
\usepackage{hyperref}

\newcommand{\hii}{H$\,${\sc ii}\xspace}

\newcommand*\diff{\mathop{}\!\mathrm{d}}

\newcommand{\blue}[1]{\textcolor{black}{#1}}

\newcommand{\chk}{$\checkmark$\xspace}
\newcommand{\SFE}{$\epsilon$\xspace}

\newcommand{\Mcl}{$M_{\rm{cl}}$\xspace}

\newcommand{\Msun}{$M_{\odot}$\xspace}

\title[Winds and radiation in unison]{Winds and 
radiation 
in unison: a new semi-analytic feedback model for cloud dissolution}

\author[D. Rahner et al.]{Daniel 
Rahner$^{1}$\thanks{daniel.rahner@uni-heidelberg.de},  Eric W.\ 
Pellegrini$^{1}$\thanks{eric.pellegrini@uni-heidelberg.de}, Simon C. O. Glover$^{1}$, Ralf S. Klessen$^{1,2}$ \\
$^{1}$ Universit{\"a}t Heidelberg, Zentrum f{\"u}r Astronomie, Institut f{\"u}r Theoretische Astrophysik, \\
Albert-Ueberle-Stra{\ss}e 2, 69120 Heidelberg, Germany\\
$^{2}$ Universit{\"a}t Heidelberg, Interdisziplin{\"a}res Zentrum f{\"u}r Wissenschaftliches Rechnen,\\
Im Neuenheimer Feld 205, 69120 Heidelberg, Germany}

\date{Accepted 2017 June 15. Received 2017 June 8; in original form 2017 April 13}

\pubyear{2017}
\begin{document}

\defcitealias{Draine2011}{D11}
\defcitealias{Weaver1977}{W77}
\defcitealias{Silich2013}{ST13}
\defcitealias{Castor1975}{C75}
\defcitealias{Kim2016}{K16}
\defcitealias{Murray2010}{M10}
\defcitealias{Pellegrini2007}{P07}

\label{firstpage}
\pagerange{\pageref{firstpage}--\pageref{lastpage}}
\maketitle

\begin{abstract}
	
Star clusters interact with the interstellar medium (ISM) in various ways, 
most importantly in the destruction of molecular star-forming clouds, resulting 
in inefficient star formation on galactic scales. On cloud scales, ionizing 
radiation creates \hii regions, while stellar winds and supernovae drive the 
ISM into thin shells. These shells are accelerated by the combined effect of 
winds, radiation pressure and supernova explosions, and slowed down by gravity. 
Since radiative and mechanical feedback is highly interconnected, they must be 
taken into account in a self-consistent and combined manner, including the 
coupling of radiation and matter.
We present a new semi-analytic one-dimensional feedback 
model for isolated massive clouds ($\geq 10^5\,M_{\odot}$) to calculate shell dynamics and shell structure simultaneously. It allows us to 
scan a large range of physical parameters (gas density, star formation efficiency, metallicity) and to estimate escape fractions of 
ionizing radiation $f_{\rm{esc,i}}$, the minimum star formation efficiency 
$\epsilon_{\rm{min}}$ required to drive an outflow, and recollapse time scales 
for clouds that are not destroyed by feedback. 
Our results show that there is no simple answer to the question 
of what dominates cloud dynamics, and that each feedback process significantly 
influences the efficiency of the others. We find that variations in natal cloud 
density can very easily explain differences between dense-bound and 
diffuse-open star clusters. 
We also predict, as a consequence of feedback, a $4-6$ Myr age difference for massive clusters with multiple 
generations.

\end{abstract}


\section{Introduction}

The formation of stars from the cold, dense interstellar medium (ISM) marks the onset of the conversion of
nuclear binding energy into radiative and mechanical energy. Injected back 
into the immediate surroundings of the stars, this energy drives a rapid chemical 
and dynamic evolution of the very molecular cloud from which the stars formed. 
This chain of events, where the creation of stars leads to energy injection by 
stars which disrupt the clouds, is known as stellar feedback. In the case of massive 
stellar clusters  ($M_{*} > 10^3 M_{\odot}$), the energetic processes are dominated by 
three main forms of feedback: ultraviolet radiation, colliding stellar winds, 
and supernovae (SNe). Each of these processes provides a source of energy and momentum that acts in opposition to gravity \blue{(for a review about stellar feedback, see \citealt{Krumholz2014})}. 

Around young massive clusters, confined interacting winds produce hot ($T \sim 10^6 - 10^8$~K) bubbles \blue{(\citealt[][hereafter W77]{Weaver1977}; \citealt{Dunne2003})}. 
These adiabatically expand, compressing the gas ahead of them into a thin dense shell. The bubbles 
are characterized by a rarefied, collisionially ionized gas. While this gas remains hot, its high thermal 
pressure drives the expansion of the surrounding shell \citepalias{Weaver1977}. Once the gas cools, however, the winds from
the central cluster push the remainder of the gas from the bubble into the shell. Thereafter, the wind
momentum is deposited directly into the shell in the form of ram pressure.
Supernovae exploding within the bubble add their energy to the existing thermal and mechanical 
energy of the gas in the bubble.

The optical depth of the gas inside a wind bubble is very low, and so radiation from the central stellar 
cluster easily reaches the dense shell surrounding the bubble \blue{\citep{Townsley2003,Gupta2016}}. Ultraviolet photons with energies 
$E > 13.6$~eV photoionize hydrogen in this shell, resulting in one of two outcomes: either the entire shell
becomes ionized, or only the inner layers become so, with the outer layers of the shell remaining neutral \blue{\citep[e.\,g.][]{Martinez-Gonzalez2014}}.

Photons that are absorbed in the shell not only heat it and potentially change its chemical state, but also
deposit momentum \blue{\citep{Lebedew1901}}. Essentially, the radiation exerts a pressure force on the gas and dust that acts radially
outwards from the central stellar cluster. If this radiation pressure is sufficiently large, then it can become
dynamically significant and can play a major role in driving the evolution of the shell \blue{\citep{Mathews1967, Draine2011,Kim2016}}. One of the key
factors that determines whether or not radiation pressure becomes significant is the efficiency with which
radiation couples with the shell \blue{\citep{Krumholz2009a}}.  
For ionizing radiation, this is determined by the amount of neutral and molecular material \blue{as well as dust} absorbing the radiation. 
When the column density of the gas is high enough to absorb all the ionizing photons (i.e.\ when the layer is optically thick to ionizing
radiation), the system is ``radiation bounded", coupling is efficient and momentum is transferred effectively. 
However, the shells surrounding many observed star-forming regions are optically thin to ionizing radiation, suggesting that coupling is
not always effective \blue{\citep{Pellegrini2012, Seon2009}}.
For non-ionizing radiation \blue{($E <  13.6$~eV)}, the optical depth is again the main factor determining whether or not coupling
is efficient, but in this case the dominant source of opacity is provided by dust \blue{unless the radiation field is weak \citep{Krumholz2008}}.

Previous simplified models of the growth of shells and bubbles around young massive clusters have typically assumed that
the dynamics of the shell are dominated by the effect of winds (e.\,g. \citetalias{Weaver1977}; \blue{\citealt{Chevalier1985}}; \citealt{MacLow1988,Koo1992}; \blue{\citealt{Canto2000}}; \citealt[][hereafter ST13]{Silich2013})
or radiation pressure \citep[e.\,g.][]{Krumholz2009a,Murray2010,Kim2016}. However, as we will see later, in the general case both must be included in order for the model to be self-consistent and hence both processes are important. In addition, in the
treatments that do account for radiation pressure, the shell is often assumed to be completely opaque to radiation \blue{\citep{Krumholz2009a, Murray2010}}, whereas in
reality the escape fraction can often be significant (see Section~\ref{sec: escape fraction}).

In this paper, we present a new model for the growth of shells around clusters that properly accounts for both winds and
radiation, and that carefully treats the structure of the shell and its influence on the fraction of the radiation that is absorbed.
In Section~\ref{sec:model}, we present our model for the structure and dynamics of the shell, and in Section~\ref{sec:Timeline},
we discuss the evolution of an exemplary cloud and compare to analytic solutions. In Section~\ref{sec:radiation_coupling},
we examine how well coupled radiation is to the shell and use those results in Section~\ref{sec:leading} to explore the conditions in which each of the different feedback processes (winds, SNe and radiation) dominates,
examining this both as a function of time during the expansion, and in an integrated form over the entire lifetime of the cloud. Our model
also allows us to make predictions for the evolution of the escape fraction of ionizing radiation during the growth
of the shell, which we present in Section~\ref{sec: escape fraction}. In Section~\ref{sec:Feedback Fails}, we discuss what we can
learn from our model about the star formation efficiency $\epsilon$ of the cloud, and how this varies as a function of the mass, mean
density and metallicity of the cloud. We conclude in Section~\ref{sec:conc} with a summary of the key results of our study.

\section{Model} \label{sec:model}

For our model we consider a spherical cloud  with a constant density 
$\rho_{\rm{cl}}$.  We assume the ISM of the cloud has a standard chemical composition of 1\,He atom per 10\,H atoms; thus the mean mass per nucleus $\mu_{\rm{n}} = (14/11)\,m_{\rm H}$  and the mean mass per particle $\mu_{\rm{p}} = (14/23) \,m_{\rm H}$, where $m_{\rm H}$ is the proton mass. The cloud's radius is given by
\begin{equation}
R_{\rm{cl}} = 19.7\,\rm{pc}\times \left(\frac{M_{\rm{cl}}/10^5M_{\odot}}{n_{\rm{cl}}/100\,\rm{cm}^{-3}}\right)^{1/3},
\end{equation}
where $M_{\rm{cl}}$ is the cloud mass, and $n_{\rm{cl}} = \rho_{\rm{cl}}/\mu_{\rm{n}}$ is the number density of atoms/ions in the cloud.
At $t = 0$ a star cluster of mass $M_*$ forms at the cloud's center. It injects feedback into the surrounding ISM in the form of 
stellar winds, radiation and eventually supernova explosions.
As outlined in the Introduction, the combined effects of radiation and winds from a massive cluster will create an expanding bubble of tenuous and hot ionized gas which is surrounded by a much denser and colder shell of swept-up cloud material.
In order to calculate the resulting expansion speed, or -- if gravity starts to dominate at some stage --
to compute the corresponding contraction velocity, we need to have a detailed understanding of the strength of the different forces acting on the shell. 
For this we need to take into account the aging population of the star cluster, the morphological and kinematical structure of the bubble and the shell, and their chemical composition. 
In this Section, we first outline our physical model for the shell dynamics, then discuss the structure of the dense shell, and finally introduce our scheme to couple both together.

\subsection{Shell Dynamics} \label{sec:ShellDynamics}
\begin{figure*}
\centering
\includegraphics[width=1.0\textwidth]{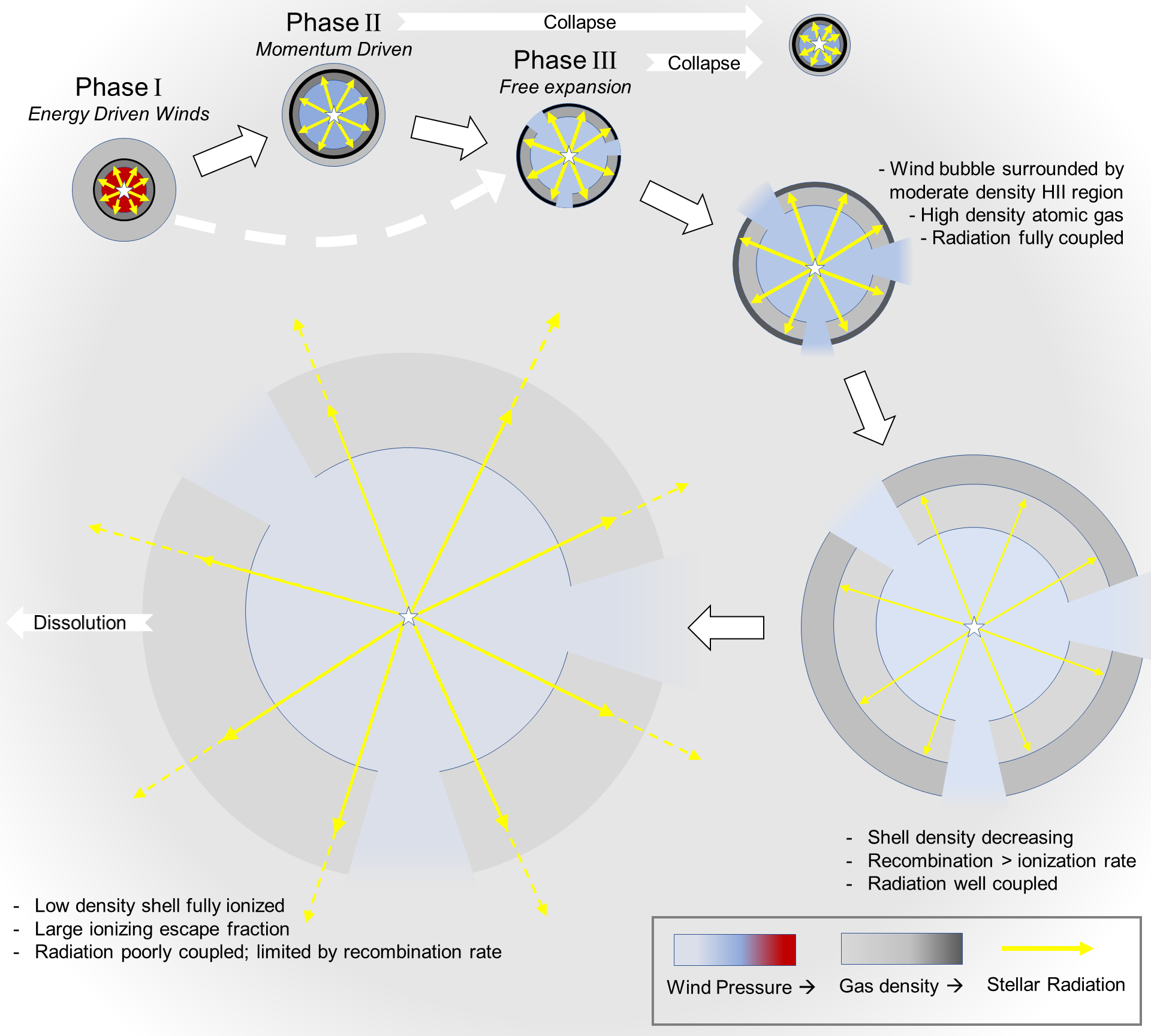}
\caption{Overview of the shell evolution from the initial adiabatic phase to recollapse or dissolution.}
\label{fig:ModelCartoon}
\end{figure*}

We model three phases of expansion of the natal cloud around the cluster. Early 
expansion is adiabatic and dominated by wind energy which sweeps the cloud 
interior into a thin shell (Phase\,I). This phase last so long as the energy is 
confined and radiative losses are small. After that, shell acceleration is 
determined by momentum input by winds, radiation and eventually by SN 
explosions opposing gravity (Phase\,II \& III). In Phase\,II the expanding 
shell continues to sweep-up material. Once the whole cloud has been swept up,   
the shell can freely expand into the ambient ISM (Phase\,III). These phases are 
outlined in Figure~\ref{fig:ModelCartoon}, and are now discussed in more detail. Since we only model isolated clouds we do not take into account any effects of an external galactic potential like shearing, which would introduce differential rotation and tidal torques, or the coupling to the larger-scale turbulent flows in the ISM. 

\subsubsection{Phase I: Energy-dominated winds}
Initially, radiation with $E>13.6$\,eV creates a large ionized region around the cluster (the so-called Str{\"o}mgren sphere). At the same time, winds from the star cluster expand freely into the ISM. Due to its very short duration, however, this initial phase can be neglected \citep{Lamers1999}. Soon, several distinct zones form around the cluster \citepalias{Weaver1977}: An inner free wind zone is surrounded by a hot shocked wind region. Together they make up the wind bubble \blue{(red region in Figure~\ref{fig:ModelCartoon})} which works against a dense shell consisting of swept-up material. Since the density in the shell is higher than in the cloud, the recombination rate increases and the ionization front travels inwards until it lies inside the shell. The shocked wind material reaches temperatures of $10^6 - 10^8$\,K causing a fast, adiabatic expansion.
During this phase we can ignore the effect of gravity and 
 radiation pressure as they are second order effects. If the shell runs into 
 ISM of a constant density, the equation of motion in the 
 thin shell limit
 according to \citet{Bisnovatyi-Kogan1995} is
\begin{equation} \label{BK95}
\frac{\diff ^2}{\diff t^2}\left(R^3\dot{R}\right) + (3\gamma -2) 
\frac{\dot{R}}{R}\frac{\diff }{\diff t} \left(R^3\dot{R}\right) = 
\frac{9(\gamma -1)L_{\rm{w}}}{4\pi\rho_{\rm{cl}}}\frac{1}{R}.
\end{equation}
Here, $R$ is the (inner) radius of the shell and 
$\gamma$ is the adiabatic index, with $\gamma = 5/3$ for an ideal gas. If the 
mechanical luminosity of the winds $L_{\rm{w}}$ is a constant, eq. (\ref{BK95}) 
can be solved analytically, yielding $R \propto t^{3/5}$ 
(\citealt{Avedisova1972}, \citealt{Castor1975}, hereafter C75, and \citetalias{Weaver1977}). 
However, stellar evolution models \citep[e.g.][]{Leitherer2014} show that 
$L_{\rm{w}}$ is time dependent, especially in the Wolf-Rayet and pre-SN phases 
and we will thus use eq. (\ref{BK95}) instead of the analytic solution for 
constant $L_{\rm{w}}$. From \citet{Bisnovatyi-Kogan1995}, during the adiabatic 
phase of the shell expansion the pressure of the hot bubble is
\begin{equation} \label{Pb1}
P_{\rm{b}} = 7\rho_{\rm{cl}}^{1/3} 
\left[\frac{3\left(\gamma-1\right)L_{\rm{w}}}{28\left(9\gamma-4\right)\pi R^2} 
\right]^{2/3} . 
\end{equation}

Evaporative flows from the shell gradually increase the density in the shocked wind region, leading to strong radiative cooling.
When radiative losses become comparable to the wind energy input, the bubble loses the driving pressure of the hot gas and the adiabatic phase ends.
The cooling time $t_{\rm cool}$ of a hot 
wind bubble is given by 
\citet{MacLow1988} as
\begin{equation}
t_{\rm{cool}} = 16\,\textup{Myr} \times  (Z/Z_{\odot})^{-35/22} 
n_{\rm{cl}}^{-8/11} L_{38}^{3/11},
\end{equation}
 where $Z$ is the metallicity, $n_{\rm{cl}}$ is given
 in cm$^{-3}$ and $L_{38}=L_{\rm{w}}/(10^{38}$\,erg\,s$^{-1}$). 
 
Alternatively, as the shell expands, inhomogeneities or asymmetries in the 
cloud may provide low density pathways along which the hot gas can escape \blue{\citep{Rogers2013}}. If 
this occurs, instead of expanding and doing work, the hot gas will escape into 
the low density/pressure ambient ISM. However, here we argue that in a  rather high 
density environment as investigated in this paper, and given the resulting 
rapid expansion in the adiabatic phase it is reasonable to assume the bubble 
does not ``burst" until the expansion is of the order of the initial cloud 
radius.
At this time, $t_{\rm{sweep}}$, the entire cloud has been 
swept up in the shell. Further expansion begins to stretch the shell \blue{without significantly adding to its mass. The shell's} average density begins to decrease, possibly becoming unstable and leading to 
the formation of channels.  Modeling the formation of low density channels goes beyond the scope 
of a 1D model. For simplicity, we assume that before $t_{\rm{sweep}}$, the  
formation of any leaks gets hampered. After $t_{\rm sweep}$ we assume the 
remaining shell structure is coherent, but does not effectively confine the 
winds. The time, when Phase I transitions to the next phase is thus given by $t_{\rm{tran}} = \min(t_{\rm{cool}}, t_{\rm{sweep}})$.

\subsubsection{Phase II: Momentum-dominated sweeping} \label{sec:phase2}

Once the hot X-ray emitting gas in the bubble cools, causing its thermal pressure to drop dramatically, the reverse shock quickly moves towards the shell as the shocked wind region is pushed into the shell \citepalias{Silich2013}. This evacuates almost all of the remaining gas from the bubble \blue{(now represented by the blue region in Figure~\ref{fig:ModelCartoon})}, and therefore during Phase~II and Phase~III, it is a good approximation to treat the bubble as if it were completely empty. This allows us to assume that the wind thereafter imparts its momentum directly on the shell and that no absorption of radiation occurs before the radiation reaches the shell. 
\blue{In reality, the transition between energy-driving in Phase~I and momentum-driving in Phase~II will be gradual and even at $t>t_{\rm{tran}}$ some thermal pressure from the shocked wind material will be present. However, the remaining hot gas is dynamically weak (\citealt{Gupta2016}, Rahner et al., in prep.) and we will ignore it here.}

Following the evacuation of the bubble, the further expansion of the shell is 
driven by a combination of radiation pressure and ram pressure from winds and 
-- at later times -- SNe, all of which act to oppose gravity. If the hot gas 
cooled before the cloud was swept up, the shell continues to expand into high 
density ISM so that the mass of the shell grows as 
$M_{\rm{sh}}=(4\pi/3)R^3\rho_{\rm{cl}}$ (as in Phase~I). During this phase, the 
shell's equation of motion is

\begin{equation} \label{MomODE}
\frac{\textup{d}}{\textup{d} t}\left(M_{\rm{sh}} \dot{R}\right) = F_{\rm{ram}} + F_{\rm{rad}} - F_{\rm{grav}},
\end{equation}
where $F_{\rm{ram}}$, $F_{\rm{rad}}$, and $F_{\rm{grav}}$ are the forces 
corresponding to ram pressure from stellar winds and type II SNe, radiation 
pressure, and gravity, respectively. Since we assume that the 
bubble is efficiently evacuated by feedback from the cluster, its density is 
too low to exert any significant amount of thermal pressure on the shell. Also, \citet{Dale2012} have shown that massive clouds are largely unaffected by thermal pressure from ionizing radiation. 
In our model we hence assume that thermal pressure from the bubble is negligible for the dynamics of the shell (thermal pressure does however influence the shell structure, as described in Section \ref{sec:Shell Structure Modell}). We note that this argument does not apply for 
low-mass systems where thermal pressure from H$\,${\sc ii} regions plays a significant role in driving outflows \citep[e.g.][]{Walch2012, Dale2012}.
	
The star clusters investigated in this work are large enough that as soon as the first SNe occur, treating them as a continuum process rather than distinct explosions is a good approximation. The ram pressure force term is then
\begin{eqnarray} \nonumber
F_{\rm{ram}} &=& F_{\rm{wind}} + F_{\rm{SN}} \\ \label{ram-term}
 		&=& \dot{M}_{\rm{w}} v_{\rm{w}} + \dot{M}_{\rm{SN}}v_{\rm{SN}}.
\end{eqnarray} 
Here $\dot{M}_{\rm{w}}$ and $\dot{M}_{\rm{SN}}$ are the mass loss rates due to stellar winds and SNe and $v_{\rm{w}}$ and $v_{\rm{SN}}$ are the terminal velocities of the winds and SN ejecta.
The ram pressure at the edge of the bubble is then
\begin{equation} \label{Pb2}
  P_{\rm{b}}=\frac{F_{\rm{ram}}}{4\pi R^2}.
\end{equation}
The full amount of the ram pressure is always transmitted to the shell. However, the shell does not absorb all photons emitted by the cluster. Consequently, it will feel only a fraction $f_{\rm abs}$ of the maximum radiation pressure that the photons from the stellar cluster can potentially exert (c.f.\ Section~\ref{sec:Shell Structure Modell}). Additionally, radiation absorbed by dust grains is re-emitted isotropically in the infrared (IR) which leads to an enhancement of radiation pressure. The total force due to radiation pressure is thus given by a direct and an indirect term,
\begin{eqnarray} \nonumber
F_{\rm{rad}} &=& F_{\rm{direct}} + F_{\rm{indirect}}\\ \label{rad-term}
        &\approx& f_{\rm{abs}}\frac{L_{\rm{bol}}}{c} \left(1+\tau_{\rm{IR}}\right),
\end{eqnarray}
where $L_{\rm{bol}}$ is the bolometric luminosity of the star cluster and $c$ is the speed of light.
The quantity $f_{\rm{abs}}(1+\tau_{\rm{IR}})$ is sometimes referred to as the trapping factor \citep[e.g.][]{Krumholz2009a}. 
The optical depth of the shell in the IR is given by 
\begin{equation}
\tau_{\rm{IR}} = \kappa_{\rm{IR}} \int\limits_R^{R_{\rm{out}}}\mu_{\rm{n}} n_{\rm{sh}} \diff r,
\end{equation} 
where $\kappa_{\rm{IR}}$ is the Rosseland mean dust opacity, $n_{\rm{sh}}$ is the number density of atoms/ions in the shell, and $R_{\rm{out}}$ is the shell's outer radius. For simplicity we do not relate $\kappa_{\rm{IR}}$ to the dust temperature but use a constant $\kappa_{\rm{IR}} = 4$\,cm$^2$\,g$^{-1}$ as would be appropriate for M17. For more details about the M17 model, see \citet[hereafter P07]{Pellegrini2007}. 

In our treatment of gravity we consider both gravity between the cluster and the shell and the self-gravity of the shell. Thus,
\begin{equation} \label{grav-term}
F_{\rm{grav}} = \frac{GM_{\rm{sh}}}{R^2}\left( M_*+\frac{M_{\rm{sh}}}{2}\right),
\end{equation}
where $G$ is the gravitational constant. We do not, however, consider any gravitational collapse by the parts of the cloud that have not yet been incorporated into the shell \blue{as we assume the cloud is in virial equilibrium}. 

\subsubsection{Phase III: Free expansion into low-density ISM or recollapse} \label{sec:phase3}
If feedback is strong enough, the shell eventually overtakes the initial cloud radius $R_{\rm{cl}}$. The shell then expands into the low-density ambient ISM. It is assumed to become leaky at $t_{\rm{sweep}}$ so that any shocked, hot wind material cools after $t_{\rm{sweep}}$ at the latest. Thus, if $t_{\rm{sweep}} < t_{\rm{cool}}$, Phase~III follows directly after Phase~I \blue{(indicated by the dashed white arrow in Figure~\ref{fig:ModelCartoon})}.

 Here, we take the ambient ISM to have a mass density $\rho_{\rm{ISM}}=1.67 \times 10^{-25}$\,g\,cm$^{-3}$, corresponding to a number density of $\sim 0.1$\, cm$^{-3}$. The equation of motion is still given by eq.~(\ref{MomODE}) but the mass of the shell is now
\begin{equation} \label{free_exp_mass}
M_{\rm{sh}}=M_{\rm{cl}}+ \frac{4\pi}{3}\left(R^3-R_{\rm{cl}}^3 \right) \rho_{\rm{ISM}} .
\end{equation}
We also ran tests with $\rho_{\rm{ISM}}=1.67 \times 10^{-24}$\,g\,cm$^{-3}$ and found that this leads to somewhat slower expansions but overall the effect is small.

There are two options now. If feedback is strong enough the shell will expand to very large radii. As the shell expands, it thins out, its density drops and it \blue{eventually} becomes indistinguishable from the diffuse ambient ISM. Even before this, we can no longer represent the shell using the thin shell limit, and so eq.~(\ref{MomODE}) does not adequately describe its dynamics any longer. To account for this, we stop the integration if the density of the densest part of the shell drops below 1\,cm$^{-3}$ for an extended period of time (more than 1\,Myr) as we consider the shell dissolved. If we would immediately stop, we might miss the reformation of a shell, e.g.\ during the Wolf-Rayet phase which drastically increases the wind ram pressure.
We call the time when the shell dissolves the dissolution time $t_{\rm{dis}}$.

If, on the other hand, gravity overcomes stellar feedback, the shell collapses back on itself. The equation of motion during the collapse is the same as before except that the mass of the shell is kept constant.
Collapse can happen either during Phase~II or Phase~III (but not during Phase~I as no gravity is included there) and we follow the collapse until the inner radius of the shell has shrunk to 1\,pc. We define the time when this happens as the collapse time $t_{\rm{collapse}}$.
We stop the integration at this point but already note that a collapse leads to more star formation (see Section~\ref{sec:Feedback Fails}) and thus possibly renewed expansion.

\subsection{Shell Structure} \label{sec:Shell Structure Modell}
In order to determine how well-coupled radiation is to the shell and its momentum deposition rate, we need to 
determine the fraction of absorbed radiation $f_{\rm{abs}}$. Numerical codes 
like {\sc cloudy} \citep{Ferland2013} provide powerful tools for calculating 
the chemistry, density, and temperature structure of shells. However, here we 
choose a simpler set of equations which sacrifice a detailed treatment of the 
chemical and thermal structure of the shell in exchange for a great increase in 
the speed with which one can calculate the volume of ionized gas. Our simple 
approach here also makes it easier to assess the relative importance of the 
different forms of feedback responsible for driving the dynamical evolution of the shell.

During Phase~I, dust inside the hot bubble is destroyed by sputtering and hydrogen is collisionally ionized, allowing
radiation to pass through unattenuated. During Phases II and III, the density inside the bubble quickly drops below 
1$\,$cm$^{-3}$ (see Section \ref{sec:ShellDynamics}), so that only little attenuation of radiation occurs. Thus, 
ionizing photons from the cluster can reach and ionize at least the inner edge of the 
shell (and potentially the whole of the shell, as we explain below).

Beyond the wind bubble, the momentum carried by radiation has a pronounced 
effect on the density structure of the ISM. Our model assumes the ionized and 
neutral/molecular phases of the shell are in the quasi-hydrostatic equilibrium 
described by the equation of state outlined in \citetalias{Pellegrini2007} 
(hereafter the P07-EOS). 
The work by \citetalias{Pellegrini2007} was the 
first to validate a hydrostatic equation of state by reproducing
an observed $\rm H^+/H/H_2$ star-forming ISM interface. The final pressure law 
defining a hydrostatic shell subject to external radiation states that the total pressure $P_{\rm{tot}}$ at a radius $r>R$, measured from the star cluster to a point in the shell, equals the sum of the pressure $P_0$ at the inner boundary of the shell and a term arising from radiative acceleration $a_{\rm{rad}}$ from photons deposited in the shell:
\begin{eqnarray} \nonumber
P_0(R) + \int\limits_{R}^{r} a_{\rm rad}\rho_{\rm{sh}}\,\diff r' &=& P_{\rm tot}(r)\\ \label{P07-EOS}
&=& P_{\rm therm} + P_{\rm turb} + P_{\rm mag}.
\end{eqnarray}
Here, $\rho_{\rm{sh}}$ is the density of the shell and $P_{\rm therm}$, $P_{\rm turb}$, and $P_{\rm mag}$ are the thermal pressure, the turbulent pressure\footnote{Note that 
this assumes that the turbulence is dominated by motions on scales that are 
small compared to the shell thickness.}, and the magnetic pressure in the shell, respectively.

It is important to understand that a hydrostatic shell is not at constant 
pressure when exposed to a radiation field. By definition, the condition of hydrostatic equilibrium
implies that there is no differential acceleration within the shell. In a 
hydrostatic shell, at any interior point the net external force (excluding 
gravity) acting on a layer with thickness $\diff r$ is proportional to the amount of stellar 
radiation absorbed. Since absorption by each previous layer reduces the transmitted 
flux of ionizing and non-ionizing UV flux, if we want the amount of radiation per unit mass
absorbed in each layer (and hence the amount of momentum deposited per unit mass) to 
remain constant, then the optical depth $\tau$ of each layer must progressively increase. In ionized
gas, this means increasing the gas density of the layer. However, if we increase the density of
the layer, we also increase its mass, and hence require an even higher momentum deposition
rate in order to keep it accelerating at the same rate as the previous layer. This implies that
the density of the layer must increase even more, in order to provide the necessary increase 
in $\tau$. 
In shells with an outward density gradient due to radiation pressure, a 
monotonically increasing total pressure is required to produce uniform 
acceleration.

The terms in the P07-EOS have been validated against the density, 
chemical and velocity structures of observed multi-phase shells. A very strong 
magnetic field could provide additional 
pressure support even in the ionized gas, lowering the gas densities and 
recombination rate. Following P07, we can estimate the potential importance of the 
magnetic field by examining the peak magnetic field
\begin{equation}
B = \sqrt{8\pi P_0+\frac{2Q_{\rm{i}}h\bar{\nu}}{R_{\rm{i}}^2 c}},
\end{equation}
where $Q_{\rm{i}}$ is the rate at which ionizing photons are emitted by the central source, $h\bar{\nu}$ is the average energy of a stellar photon, and $R_{\rm{i}}$ is the radius of the ionization front.

We have computed the peak magnetic pressure predicted by this equation
for the clusters and gas densities modeled here and find that magnetic  pressure is only marginally 
significant in the ionized gas while $R_{\rm{i}} \leq R_{\rm cl}$. At larger radii 
and/or late times when the winds are momentum-driven, magnetic pressure is 
much smaller than the radiation pressure, and decreases in significance as the
shell evolves. The magnetic field may still provide a dominant source of pressure 
in the atomic gas, but the momentum deposited there is proportional to the dust 
column only (cf. eq.~(\ref{dtau_nonion})), and is therefore not affected by the structure of the atomic gas layer. 
Thus, in our calculations we ignore the effect of magnetic fields. 

We also neglect the effects of turbulence, which is unlikely to be important in the ionized gas, 
unless the turbulence velocity dispersion is large ($\sigma_{\rm rms} \gg  10 \: {\rm km \, s^{-1}}$ in the ionized shell). \blue{However, in star forming regions like Orion, the turbulent velocities in the HII-region are clearly subsonic \citep{Arthur2016} and turbulence is thus of limited importance for determining the structure of the ionized shell.
In the atomic gas turbulence may play an important role in structuring the material but since, as mentioned above, there the absorbed fraction of radiation depends only on the column density, turbulence does not play a significant role in the overall dynamics of the shell.}

Detailed studies of observations find that the inner edge of the shell and the wind bubble are in pressure equilibrium
\citepalias[see e.g.][]{Pellegrini2007}.
In this case, $P_0(R) = P_{\rm{b}}$. Neglecting magnetic and turbulent pressure, the number density of the atomic nuclei $n_{\rm{sh}}$ at the inner radius of the shell $R$ must then satisfy
\begin{equation} \label{nsh0}
  n_{\rm{sh}}(R) = \frac{\mu_{\rm{p}}}{\mu_{\rm{n}}kT_{\rm{i}}}P_{\rm{b}}, 
\end{equation} 
where $k$ is the Boltzmann constant and $T_{\rm{i}}$ is the temperature of the 
inner (ionized) region of the shell. The pressure of the bubble $P_{\rm b}$ is 
given by eq.~(\ref{Pb1}) during Phase I and by eq.~(\ref{Pb2}) during Phases II 
and III. Note that pressure equilibrium  implies that the shell is expanding at 
the same rate as the bubble. 
 
For simplicity, we also assume that the ionized gas is at a constant 
temperature of $T_{\rm{i}}=10^4$\,K. Under these 
assumptions, the condition of hydrostatic equilibrium, eq.~(\ref{P07-EOS}), dictates that the 
gradient in the total pressure be offset by the
external forces, in this case the force from radiation pressure, leading to 
\begin{equation} \label{P07-simple-gradient}
a_{\rm rad} \rho_{\rm{sh}} =  \frac{\diff}{\diff r}\left( 
\frac{\mu_{\rm{n}}}{\mu_{\rm{p}}}n_{\rm{sh}}kT_{\rm{i}}\right).
\end{equation}

The radiative transfer of eq.~(\ref{P07-simple-gradient}), can be reduced to two energy bands:
Ionizing radiation (photons with energies above 13.6\,eV) which is absorbed by hydrogen and dust, and non-ionizing radiation which is absorbed by dust only.\footnote{Photons in the energy range 11.2--13.6~eV can also be absorbed in the Lyman-Werner bands of H$_{2}$, but this is significant in comparison to dust absorption only when the radiation field is relatively soft \citep{Krumholz2008}.} 
Recombination is assumed to occur only via case~B recombination with a 
recombination coefficient $\alpha_{\rm{B}} = 2.59 \times 
10^{-13}\,$cm$^3$\,s$^{-1}$ \citep{AGN3}. 
These simplifying assumptions, and a conversion from acceleration times density to force per 
volume, give rise to the following set of coupled differential equations for the number density of the shell $n_{\rm{sh}}(r)$, the attenuation function for ionizing radiation $\phi(r)$ and the optical depth $\tau_{\rm{d}}(r)$ of dust in the shell, which have been applied to dusty \hii regions by \citet{Draine2011} and to 
shells by \citet{Martinez-Gonzalez2014}:
\begin{eqnarray} \label{dn}
\frac{\diff}{\diff r}\left(\frac{\mu_{\rm{n}}}{\mu_{\rm{p}}}n_{\rm{sh}}k T_{\rm{i}}\right) &=& \frac{1}{4\pi r^2 c}\frac{\diff}{\diff r}\left( L_{\rm{n}} e^{-\tau_{\rm{d}}} + L_{\rm{i}} \phi\right), \\ \label{dphi}
\frac{\diff \phi}{\diff r} &=& -\frac{4\pi r^2}{Q_{\rm{i}}}\alpha_{\rm{B}} n_{\rm{sh}}^2  - n_{\rm{sh}}\sigma_{\rm{d}} \phi, \\ \label{dtau}
\frac{\diff \tau_{\rm{d}}}{\diff r} &=& n_{\rm{sh}}\sigma_{\rm{d}}.
\end{eqnarray}
Here, $L_{\rm{n}}$ and $L_{\rm{i}}$ are the luminosities of non-ionizing and ionizing radiation. We assume the dust cross section $\sigma_{\rm{d}}$ scales linearly with metallicity, $\sigma_{\rm{d}} = \sigma_0 Z/Z_{\odot}$ where $\sigma_0 = 1.5 \times 10^{-21}\textup{cm}^{2}$ \citep{Draine2011} and neglect any formation or destruction of dust in the shell. \blue{During Phase~I, with temperatures of the shocked wind material in excess of $10^6$\,K, neglecting dust sublimation is certainly not correct. However, we treat this early phase as being dominated by ram-pressure anyway and ignore radiation pressure on dust altogether. At later times, gas temperatures in the shell reach at most $10^4$\,K, at which point the dust-to-gas ratio is not so different from the general ISM \citep{AGN3}. Destruction of dust is only important close to the illuminated face of the shell and even if dust destruction is taken into account, the majority of ionizing photons will continue to be absorbed by dust \citep{Arthur2004}. The formation of dust is never significant at the densities considered in this paper.}

\blue{Equations (\ref{dn}), (\ref{dphi}), and (\ref{dtau})} hold at all radii $r < R_{\rm i}$ within the shell. The radius of the ionization front corresponds to the transition between the ionized and non-ionized parts of the shell and hence marks the point at which the ionizing photon flux drops to zero, i.e.\ $\phi(R_{\rm{i}}) = 0$. Beyond the ionization front we assume the 
gas is purely atomic with a temperature of $T_{\rm{a}}=100\,$K. At radii $r > R_{\rm i}$, we then have
\begin{eqnarray} \label{dn_nonion}
\frac{\diff}{\diff r}\left(n_{\rm{sh}} k T_{\rm{a}} \right) &=& \frac{1}{4\pi r^2 c}\frac{\diff}{\diff r}\left( L_{\rm{n}} e^{-\tau_{\rm{d}}} \right), \\ \label{dtau_nonion}
\frac{\diff \tau_{\rm{d}}}{\diff r} &=& n_{\rm{sh}}\sigma_{\rm{d}}.
\end{eqnarray}
Note that the condition of pressure equilibrium between the ionized and the non-ionized gas leads to a discontinuous increase in $n_{\rm sh}$ by a factor $\mu_{\rm{n}}T_{\rm{i}}/(\mu_{\rm{p}}T_{\rm{a}})$ at $R_{\rm i}$.

\begin{figure}
\centering
\includegraphics{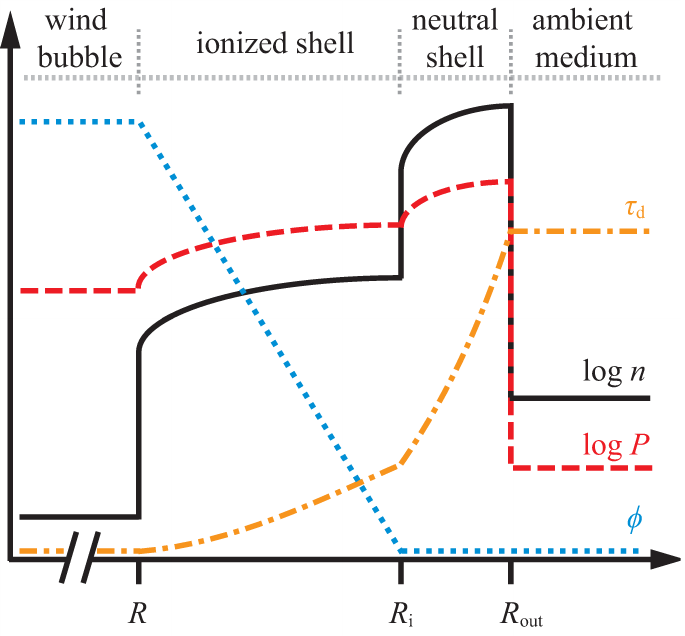}
\caption{Sketch of number density $n$, dust optical depth $\tau_{\rm{d}}$, and attenuation of ionizing radiation $\phi$ as a function of radius. The red dashed line shows the pressures of the wind bubble $P_{\rm{b}}$, the thermal gas pressure $P_{\rm{therm}}$ of the shell and lastly of the ambient medium. At very early and late times when the column density of the shell and/or the pressure from winds is low, the shell may be fully ionized (not shown). \blue{See also \citet{Martinez-Gonzalez2014}.}}
\label{fig:Shell Structure}
\end{figure}

Since the density inside the bubble is assumed to be very low, any absorption inside the bubble is negligible and the boundary conditions used for solving eqs.~(\ref{dn}), (\ref{dphi}), and (\ref{dtau}) are given by eq.~(\ref{nsh0}), $\phi(R) = 1$, and $\tau_{\rm{d}}(R) = 0$. We stop the integration at a radius $R_{\rm{out}}$, once we have accounted for all of the shell's mass, i.e.\
 \begin{equation} 
4\pi\mu_{\rm{n}} \int\limits_{R}^{R_{\rm{out}}} n_{\rm{sh}}(r) r^2 \diff r = M_{\rm{sh}}.
\end{equation} 
Figure \ref{fig:Shell Structure} shows a sketch of the density, pressure and attenuation of radiation across the shell as obtained from eqs.\ (\ref{dn}) to (\ref{dtau_nonion}).

We can now calculate the fraction of absorbed ionizing and non-ionizing radiation:
\begin{eqnarray} \label{fi}
f_{\rm{abs,i}} &=& 1-\phi(R_{\rm{out}}), \\ \label{fn}
f_{\rm{abs,n}} &=& 1-\exp\left[-\tau_{\rm{d}}\left(R_{\rm{out}}\right)\right].
\end{eqnarray}
Finally, the total absorption fraction $f_{\rm{abs}}$ is defined as a luminosity weighted average of $f_{\rm{abs,i}}$ and $f_{\rm{abs,n}}$,
\begin{equation} \label{fabs}
f_{\rm{abs}} = \frac{f_{\rm{abs,i}}L_{\rm{i}}+f_{\rm{abs,n}}L_{\rm{n}}}{L_{\rm{bol}}},
\end{equation}
where $L_{\rm{bol}} = L_{\rm{i}}+L_{\rm{n}}$. 

\blue{By ignoring absorption of Lyman-Werner radiation on H$_2$ we underestimate $f_{\rm{abs}}$. We recalculated some of our shell structure models with {\sc cloudy} to explore the effect chemistry has on opacity and find that a significant amount of H$_2$ only forms when the shell is dense and quite optically thick, i.\,e. if $f_{\rm{abs}} \sim 1$. In lower density, expanded shells the interstellar radiation field suppresses the formation of H$_2$, and a more detailed chemical model does not lead to substantially different escape fractions.}

\blue{A larger caveat is that we fix the dust cross section $\sigma_0$ (for a fixed metallicity). In reality, $\sigma_0$ is a function of the effective stellar temperature and decreases as the massive stars die \citep{Draine2011}. Again using {\sc cloudy}, we find that at later times ($t\gtrsim 3\,$Myr) in our simplified treatment we are overestimating $f_{\rm{abs}}$ by $\sim 25\,$\%. But since, as we will show, at late times radiation pressure is rarely the dominating source of feedback, this does not strongly affect the dynamics of shells. In a future iteration of our method we plan to self-consistently calculate $\sigma_0$ from the time-variable stellar spectrum.}

\subsection{Coupling Structure and Dynamics}

There have been many attempts to model the dynamic evolution of feedback-driven shells (see Table~\ref{tab:ModelSummary}). In the wind energy-driven model by \citetalias{Castor1975} and \citetalias{Weaver1977} mentioned above,
\begin{equation} \label{Weaver_propto}
R \propto t^{3/5}
\end{equation}
if cooling is neglected and the ISM is assumed to be infinite and homogeneous. \citetalias{Silich2013}, expanded that model and included momentum feedback from winds after the wind energy has been radiated away. Still for an infinite, homogeneous ISM, they show that
\begin{equation} \label{Silich_propto}
 R \propto \left( At^2 + Bt + C\right)^{1/4} \quad \textrm{if}\ t>t_{\rm{cool}},
 \end{equation} 
 where $A$, $B$ and $C$ depend on wind parameters, the cloud density and the 
 cooling time. Both these models neglect the influence of gravity, radiation 
 pressure and SNe on the dynamics. \citet{Kim2016} study the combined effect of 
 radiation pressure and gravity but neglect winds, similar to 
 \citet{Murray2010} who include energy from hot winds in one of their models 
 but always neglect wind momentum. We also note \citet{Krumholz2009a} who 
 calculated the dynamics under the influence of radiative momentum deposition, 
 albeit under the assumption of full absorption and while neglecting gravity.

At one point or another all of these models fall short of a full, self-consistent treatment of feedback. The expansion rate of the shell depends on how well-coupled it is to radiation, 
which in turn depends on the shell structure. However, as we have seen, the shell 
structure itself depends on the expansion rate of the shell. To complicate 
things even further winds, radiation and SNe output depend on cluster mass and 
age. It is therefore necessary to simultaneously solve for the expansion rate 
and structure of the shell while accounting for an evolving stellar population.

\begin{table*}
	\caption{Summary of 1D shell dynamical models. \blue{Included and neglected physical processes are marked with \chk and -, respectively.} \label{tab:ModelSummary}}
\begin{tabular}{lc|c|cccccccc}
\hline
Model & Mass  & Gravity & Wind  & Wind  & Radiation & Radiation& Shell & Stellar & SNe & Turbulence \\
& reservoir & & (${\dot{E}}$) & (${\dot{p}}$) &  ($\dot{E}$) & ($\dot{p}$)  & structure & evolution & & \\
\hline
		This Work & Finite & \chk& \chk& \chk& \chk& \chk& \chk& \chk&  \chk & - \vspace{0.05in}\\
		$^1$\citetalias{Kim2016} & Finite & \chk & - & - & \chk & \chk  & - & - & - & - \vspace{0.05in}\\
		$^2$\citetalias{Silich2013}  & Infinite   & - & \chk & \chk & - & - & - & - & - & - \vspace{0.05in}\\
		$^3$\citetalias{Murray2010} & Finite   & \chk & (\chk) & - & \chk & \chk & - & - & - & \chk \vspace{0.05in}\\
 		$^4$\citetalias{Weaver1977}   & Infinite    & - & \chk & - & - & - & - & - & - & -\\
\hline	
\end{tabular}

{\footnotesize Notes. $^1$\citet{Kim2016}; $^2$\citet{Silich2013}; $^3$\citet{Murray2010}; $^4$\citet{Weaver1977}. $\dot{E}$ and $\dot{p}$ stand for energy- and momentum-driven, respectively.}

\end{table*}

Expanding shells in the ISM are not truly hydrostatic -- in the sense that parts of the shell do not move radially with respect to each other -- as they tend to become thicker over time. If the ``thickening velocity" $v_{\rm{t}} \equiv \diff (R_{\rm{out}}-R)/\diff t$ is lower than the shell's sound speed $c_{\rm{s}}$, the pressure distribution within the shell can readjust itself on a timescale short compared to that on which the shell thickness changes, and the shell therefore rapidly settles into a quasi-hydrostatic equilibrium. In such a case, the assumption of local hydrostatic equilibrium is a good approximation.

We find that in our models $v_{\rm{t}}$ is subsonic except for short times when we switch from the adiabatic phase to Phase~II or III and around the occurrence of the first SNe. Over the whole simulated time span, the short periods when the quasi-hydrostatic assumption breaks down are expected to be negligible for the dynamics. Additionally, observations suggests that hydrostatic models as adopted here provide reasonable approximations for expanding gas shells (e.\,g. \citealt{Pellegrini2007}).

In order to self-consistently model the dynamics of feedback-driven shells we thus take the following approach:
\begin{enumerate}
\item[\textbf{1)}]A star cluster with mass $M_*$, \blue{following a Kroupa initial mass function \citep{Kroupa2001}}, forms at $t=0$ in the center of a gas cloud. All stars in the cluster are assumed to be coeval. We do not consider any ongoing star formation. The cloud has mass $M_{\rm{cl}}$ and constant density $n_{\rm{cl}}$.
\item[\textbf{2)}] We take the relevant parameters for stellar feedback $L_{\rm{w}}$, $L_{\rm{i}}$, $L_{\rm{n}}$, $Q_{\rm{i}}$, $\dot{M}_{\rm{w}}$, $\dot{M}_{\rm{SN}}$ and $v_{\rm{\rm{w}}}$ from the population synthesis code Starburst99 \citep{Leitherer1999, Leitherer2014}, v7.0.1, using the Geneva evolution tracks \citep{Ekstrom2012, Georgy2012, Georgy2013} for non-rotating stars (fiducial model) and rotating stars (see Appendix~\ref{sec: stellRot}). The terminal velocity of the SN ejecta $v_{\rm{SN}}$ is set to a constant $10^4$\,km\,s$^{-1}$. These feedback parameters as well $\tau_{\rm{IR}}$ and $f_{\rm{abs}}$ (which are 0 at $t=0$ as no shell yet exists) are used to calculate the shell dynamics via the expansion equations (\ref{BK95}) and (\ref{MomODE}).
\item[\textbf{3)}] After a certain time step $\Delta t$ the feedback parameters are updated and the shell structure is modeled via eq. (\ref{dn}) -- (\ref{dtau_nonion}). From the shell structure we get $\tau_{\rm{IR}}$ and $f_{\rm{abs}}$. The time step is adaptive: It is small ($\sim 0.02\,$Myr) during the early phase and around the time of the first SNe (at $t\sim 3$\,Myr), when $f_{\rm{abs}}$ is strongly time-dependent. 
\end{enumerate}
Steps 2 and 3 are repeated until the end of the simulation is reached at a time $t_{\rm{end}}$. The code \blue{\textsc{warpfield} (\textbf{W}inds \textbf{A}nd \textbf{R}adiation \textbf{P}ressure: \textbf{F}eedback \textbf{I}nduced \textbf{E}xpansion, col\textbf{L}apse and \textbf{D}issolution) developed for} this work is publicly available for download under \href{https://bitbucket.org/drahner/warpfield}{https://bitbucket.org/drahner/warpfield}\,.

\subsection{Investigated parameter space}

We explore the evolution of shells in clouds with masses in the range $10^5M_{\odot}\leq M_{\rm{cl}} \leq 10^8M_{\odot}$, \blue{i.\,e. giant molecular clouds (GMCs) and giant molecular associations. For simplicity, we will refer to them as clouds, independent of their mass.} 
The masses are equally spaced in log-space with $\Delta \log (M_{\rm{cl}})$ = 0.25. We investigate star formation efficiencies 
\begin{equation}
\epsilon \equiv \frac{M_*}{M_{\rm{cl}}+M_*}
\end{equation} 
varying from 0.01 to 0.25 with $\Delta \epsilon = 0.01$. The investigated parameter space thus includes a small region where the star clusters are not massive enough to fully sample the IMF ($M_* \lesssim 10^4M_{\odot}$), namely clouds with $M_{\rm{cl}}<10^6M_{\odot}$ and with very low star formation efficiencies. In the stochastic regime the assumption of continuous SN explosions after $t\sim 3$\,Myr and the values for $L_{\rm{i}}$ and $L_{\rm{w}}$ obtained from scaling down a fully sampled star cluster are not valid any more and hence we do not include this regime in our analysis.
Also, shells around low mass \blue{GMCs} ($M_{\rm{cl}} \sim 10^5M_{\odot}$) with very high star formation efficiencies ($\epsilon \gtrsim 0.2$) are not in quasi-hydrostatic equilibrium as $v_{\rm{t}}>c_{\rm{s}}$ after the stellar winds of the most massive stars disappear and the pressure at shell's inner edge drops significantly, thus leading to a rapid increase in the shell's thickness. However, these are shells which are close to dissolution and for which radiation pressure is already negligible. Hence, the absolute error we make when calculating the amount of momentum deposited by radiation into such shells is small. 

We examine two different natal cloud densities, $n_{\rm{cl}}=100$\,cm$^{-3}$ and $n_{\rm{cl}}=1000$\,cm$^{-3}$, \blue{corresponding to diffuse and translucent molecular clouds, respectively \citep{Snow2006}. In later sections we will refer to these as low- and high-density runs. Some GMCs contain clumps and cores in excess of $n = 10^4$\,cm$^{-3}$ but on average their density is $\sim 100-1000$\,cm$^{-3}$. We do not yet include a density profile for our clouds but plan to do so in the future.} We also model two different metallicities, $Z=Z_{\odot}$ and $Z = 0.15Z_{\odot}$. Note that $Z$ refers to both the metallicity of the cloud, affecting the amount of dust and the time-scale for radiative cooling, and to the metallicity of the cluster, affecting the energy and momentum output by stellar winds and to a lesser extent by radiation. We call these the solar $Z$ and low $Z$ runs, respectively.

\begin{table}
	\caption{Investigated parameter space \label{tab:ParameterTable}}
\begin{tabular}{l|lll}

\hline
		cloud number density & $n_{\rm{cl}}$ & 100, 1000 & cm$^{-3}$\\
		metallicity & $Z$ & 0.15, 1 & $Z_{\odot}$\\
		cloud mass & $M_{\rm{cl}}$ & $10^5 - 10^8$ & $M_{\odot}$\\
		star formation efficiency & $\epsilon$ & $0.01-0.25$ & \ldots \\
\hline	
\end{tabular}
\end{table}

Table \ref{tab:ParameterTable} lists the parameter space described above. The expansion of the shell is modelled until either it dissolves into the ambient ISM, or it recollapses, or 7 free-fall times $\tau_{\rm{ff}}$ have passed; thus, $t_{\rm{end}} = \min(t_{\rm{dis}}, t_{\rm{collapse}}, 7\tau_{\rm{ff}})$. The free-fall time is defined as
\begin{equation}
\tau_{\rm{ff}} \equiv \sqrt{\frac{3\pi}{32G\rho_{\rm{cl}}}},
\end{equation}
and so $7 \tau_{\rm{ff}}$ corresponds to $t \sim10\,$Myr and $t \sim 32\,$Myr for the high density and low density runs, respectively. Note that in the low density case, this time is close to the time at which the last of the SNe associated with the cluster would have exploded, which marks the point at which the effects of stellar feedback drop to a very low value.

\section{A Feedback-Driven Dynamic Timeline} \label{sec:Timeline}

\begin{figure}
\begin{center}
\includegraphics[width=0.48\textwidth]{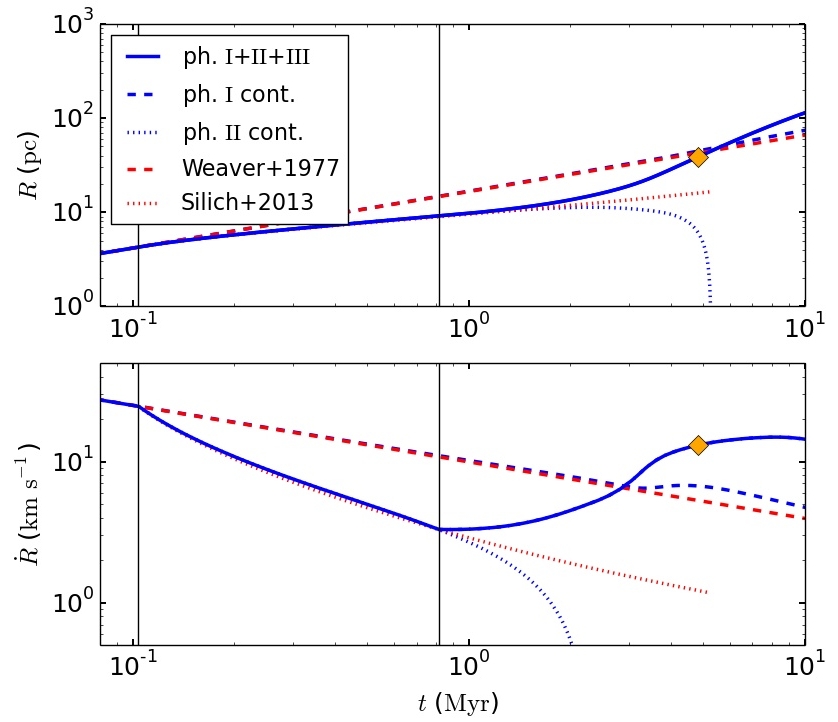}
\caption{\textit{Top}: Evolution of the inner radius of the shell as a function of time for a model with $M_{\rm{cl}}=10^5M_{\odot}$,  $\epsilon = 0.1$, $Z=Z_{\odot}$ and $n_{\rm{cl}}=1000$\,cm$^{-3}$. \textit{Bottom:} Expansion velocity of the shell. The vertical black lines mark the transition between the expansion phases (marked by the Roman numerals I, II and III) at $t_{\rm{cool}}$ and $t_{\rm{sweep}}$. The yellow diamond indicates where the shell becomes fully ionized and ionizing radiation starts to leak out. The blue dashed and dotted lines show a continuation of Phase~I (assuming no cooling and an infinite mass reservoir) and Phase~II (assuming an infinite mass reservoir only), respectively. The actual evolution of the shell is shown by the solid blue line. The red dashed and dotted lines show analytic solutions for comparison, eq. (21) in \citetalias{Weaver1977} and eq. (13) in \citetalias{Silich2013}.}
\label{fig:summary_expansion}
\end{center}
\end{figure}

We will now attempt to summarize the contribution of each feedback mechanism and its variation with time. Our aim is to highlight the different physical regimes where simple scaling relations fall short. There is no simple answer to the question of which feedback mechanism is dominant. Instead this complex problem must be addressed by quantifying how their relative contributions vary with time in an effort to combat gravity.

We start by showing the expansion of a shell that is driven by feedback from a cluster in a dense molecular cloud with cloud mass $M_{\rm{cl}}=10^5M_{\odot}$ and star formation efficiency $\epsilon = 0.1$ (see Figure~\ref{fig:summary_expansion}). An overview of the shell dynamics for a large number of other models can be found in Appendix \ref{sec:appendix_allmodels}. In this example, both the cloud and the cluster have solar metallicity. Rapid expansion in the adiabatic phase (Phase~I) is followed by strong deceleration after $t_{\rm{cool}}\sim 0.1\,$Myr as the thermal pressure from the shocked wind bubble vanishes and the shell accumulates more and more mass (Phase~II). At $t_{\rm{sweep}}\sim 0.8\,$Myr the whole cloud has been swept up by the shell. Expanding into low density ISM (Phase~III), the shell now accelerates again. 

We have also simulated how the cloud would evolve if Phase~I (Phase~II) would continue indefinitely as one would expect for an infinite ISM reservoir without cooling (after cooling). This allows us to compare our results with analytic solutions for the equation of motion, i.e.\ eq.~(21) in \citetalias{Weaver1977} and eq.~(13) in \citetalias{Silich2013}. For the particular cloud shown here, \citetalias{Weaver1977} and \citetalias{Silich2013} provide good approximations for the shell expansion in Phase~I and II (some small deviations towards faster expansion in our model in Phase~II are due to radiation pressure). However, for a model with the same cluster mass but a larger cloud size, \citetalias{Silich2013} would seriously overestimate the shell's velocity and radius at late times (due to their neglect of gravity). Even though we follow \citetalias{Weaver1977} in neglecting gravity in Phase~I, we do always take into account stellar evolution. This is why at late times our continued Phase~I model differs from eq.~(\ref{Weaver_propto}). 

An important consequence of including gravity is that for all models investigated here, shells expanding into a infinite ISM reservoir will always recollapse. Sweeping up more and more mass, the shell eventually becomes too massive for gravity to be balanced by the outward forces. If the shell approaches this point asymptotically, it can keep roughly that size until the massive stars have died and feedback decreases. Usually, however, the shell passes the point of force balance with a positive velocity. As soon as this happens, the shell starts to lose momentum and eventually recollapses. This is shown by the blue dotted line in Figure~\ref{fig:summary_expansion}. If gravity is included and the mass reservoir is infinite, the shell reaches a turning point at $t\sim 2$\,Myr as the expansion velocity becomes negative and the radius of the shell starts to significantly deviate from the \citetalias{Silich2013} model. 

In some models ionizing radiation can completely overpower the shell. This is the moment when ionizing radiation starts to leak out (see yellow diamond in Figure~\ref{fig:summary_expansion}). Coupling of radiation and the escape fraction of ionizing radiation will be discussed in the following sections.

\section{Radiation coupling} \label{sec:radiation_coupling}
For young star clusters, momentum carried by radiation exceeds momentum carried by winds by a factor of a few for solar metallicity and by a few decades at $0.15 \, Z_{\odot}$ \citep{Leitherer2014}. However, this does not mean that radiation always dominates over winds as a source of feedback. Rather, it is the coupling between radiation and the ISM, quantified by $f_{\rm{abs}}$ in our model,
that ultimately determines which of these feedback sources dominates. Any attempt to determine how radiation pressure and ram-pressure compare to each other must therefore begin by quantifying $f_{\rm{abs}}$.

Ionizing and non-ionizing radiation behave differently. While $f_{\rm{abs,n}}$ is only influenced by the column density of the shell, $f_{\rm{abs,i}}$ depends also on the volume density (since the recombination rate is proportional to $n_{\rm{sh}}^2$) and on the rate of ionizing photons $Q_{\rm{i}}$ emitted by the cluster, cf. eq.~(\ref{dphi}) and (\ref{dtau}). Thus, $f_{\rm{abs,n}}$ is solely set by how far out the shell has expanded and how much mass it has swept up in the process, whereas $f_{\rm{abs,i}}$ is also dependent on the cluster's current output in terms of ram pressure from winds and radiation pressure, which set $n_{\rm{sh}}(r)$ via eqs.~(\ref{nsh0}) and (\ref{dn}), and its current emission rate of ionizing photons. Since the shell expansion is a result of the history of feedback, we might say that $f_{\rm{abs,n}}$ only cares about the past while $f_{\rm{abs,i}}$ is determined by both the past and the present.

\begin{figure}
\begin{center}
\includegraphics[width=0.48\textwidth]{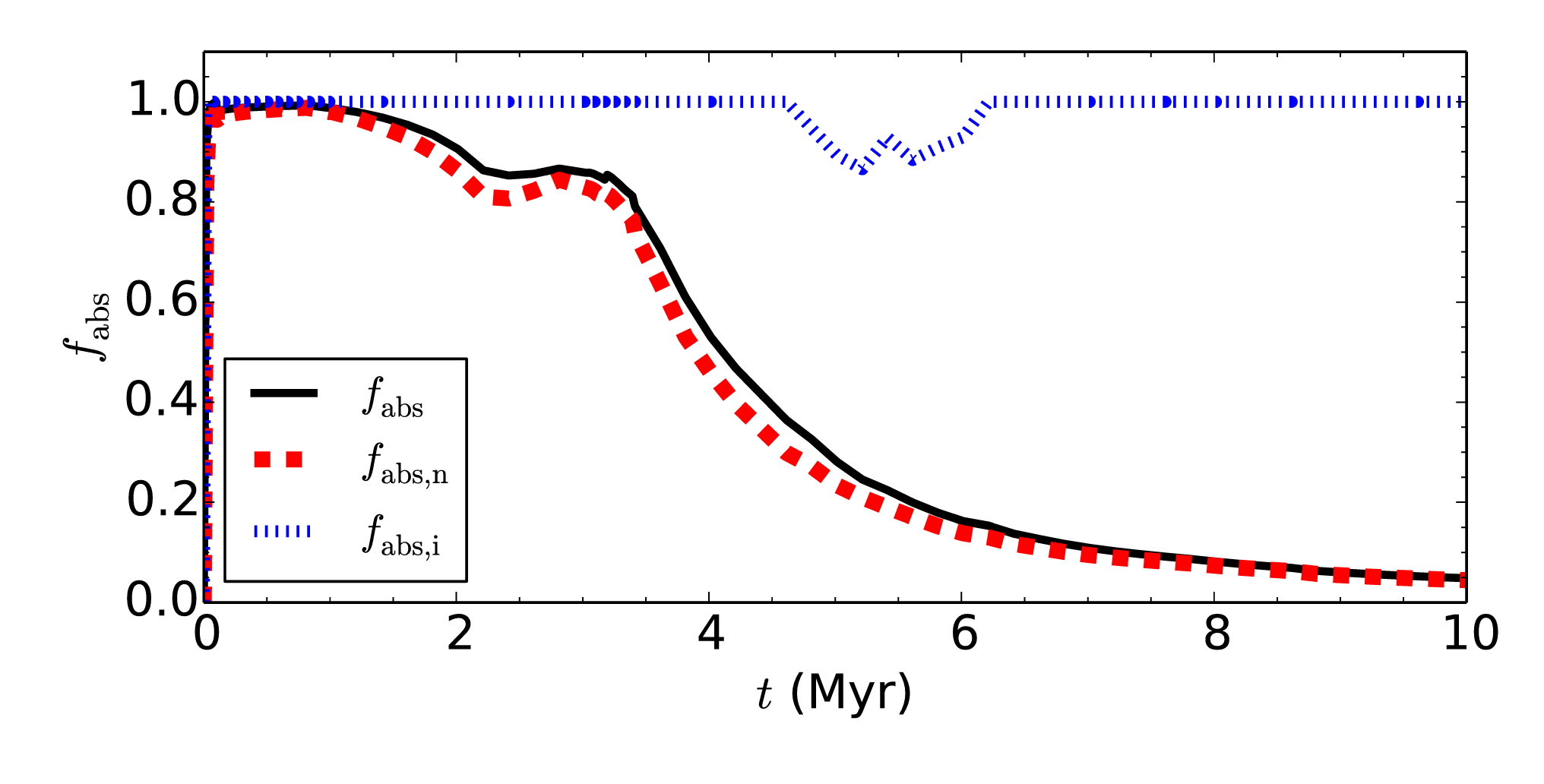}
\caption{Absorbed fraction of non-ionizing radiation $f_{\rm{abs,n}}$, ionizing radiation $f_{\rm{abs,i}}$, and the luminosity weighted mean $f_{\rm{abs}}$. The model is the same as in Figure \ref{fig:summary_expansion}.}
\label{fig:summary_fabs}
\end{center}
\end{figure}

After a dense shell has formed, radiation is initially well-coupled (see Figure~\ref{fig:summary_fabs}). However, after the shell enters the free expansion phase (Phase~III), the expansion velocity increases while at the same time the mass growth nearly stalls. The gas in the shell thus stretches over an ever-increasing surface area, reducing the shell's surface density and leading to a decrease of $f_{\rm{abs}}$. At the same time, ram pressure drops as $R^{-2}$, the volume density decreases and the shell becomes thicker. As a result, $f_{\rm{abs}}$ decreases even further. In the particular example shown in Figure~\ref{fig:summary_fabs}, $f_{\rm{abs}}$ starts to differ significantly from unity at $t\sim 1$\,Myr, just after the cloud has been swept up. The bump at $t \sim 3\,$Myr is caused by the increase in ram pressure during the Wolf-Rayet and pre-SN phases. At $t\sim 5$\,Myr, ionizing radiation decouples from the shell. At that point the whole shell is ionized. However, the time period during which ionizing radiation can pass through the shell is short: At late times the output of ionizing radiation is greatly reduced due to the death of the very massive stars. Ionizing radiation is then again fully absorbed by the ISM. At $t\sim 8$\,Myr, $f_{\rm{abs}}$ drops below 0.1. At this point, less than 10\,\% of the radiation, which has already been diminished due to the aging of the cluster, is transmitted to the shell, greatly reducing the efficiency of radiation pressure as a source of feedback. 
\newline

As explained above, the gas density of the shell which determines radiation 
momentum-coupling depends on many quantities in a non-linear way. To reduce the 
result into a digestible statement, it is useful to examine a fit to the 
absorption fraction as a function of the most important model parameters. 
Between 1 and 10\,Myr and for fully sampled IMFs ($M_* \gtrsim 10^4M_{\odot}$), 
$f_{\rm{abs}}$ is well fitted by
\begin{equation} \label{fabs_fit}
   f_{\rm{abs}} =
   \begin{cases}
     0 & \hspace{.5in} \text{if } \tilde{f} \leq 0, \\
     \tilde{f}  & \hspace{.5in} \text{if } 0 < \tilde{f} < 1, \\
     1 & \hspace{.5in} \text{if } \tilde{f} \geq 1,
   \end{cases}
\end{equation}
with
\begin{eqnarray}
\tilde{f} & = & \left(a\log\epsilon + b\log\left(\frac{M_{\rm{cl}}}{M_{\odot}}\right) + c\right)\frac{t}{\rm{Myr}} \nonumber \\ \label{fabs_fit_equation}
 & + & d\log\left(\frac{M_{\rm{cl}}}{M_{\odot}}\right) + e.
\end{eqnarray}
The fit parameters $a, b, c, d,$ and $e$ are provided in Table \ref{tab:fitpar} for the combinations of density and metallicity examined in this study. We also list the reduced chi squared statistic in each case, to indicate the goodness of fit.

\begin{table*}
\caption{Fit parameters for $f_{\rm{abs}}$ for the investigated parameter space (see eq.~(\ref{fabs_fit_equation})). The reduced chi squared statistic $\chi^{2}_{\nu}$
has been calculated assuming a variance of 0.01. For further details, see Section~\ref{sec: escape fraction}. \label{tab:fitpar}}
\begin{tabular}{lc|ccccc|c}
\hline
$n$ (cm$^{-3}$) & $Z$ ($Z_{\odot}$) & $a$ & $b$ & $c$ & $d$ & $e$ & $\chi_{\nu}^2$ \\
\hline
1000 & 1    & -0.323 &  0.129 & -1.119 & -0.143 & 1.975 & 1.07 \\
1000 & 0.15 & -0.118 &  0.085 & -0.695 & 0.102 & 0.140 & 2.01 \\
100  & 1    & -0.109 &  0.063 & -0.579 & 0.084 & 0.363 & 1.18 \\
100  & 0.15 & -0.020 & 0.037 & -0.312 & 0.097 & -0.034 & 3.18 \\
\hline
\end{tabular}
\end{table*}

From the signs of the fit parameters $a$ (negative) and $b$ (positive) we can already draw two conclusions:
\begin{enumerate}
\item[\textbf{a)}] Keeping the cloud mass constant, an increase in star formation efficiency leads to a faster decoupling with time.
\item[\textbf{b)}] Keeping the star formation efficiency constant, an increase in cloud mass leads to a slower decoupling with time.
\end{enumerate}

To understand these trends, imagine a cloud with a given mass and density. If the cloud has a high star formation efficiency, two effects play a role: First, as a more massive cluster outputs more energy and momentum in winds and SNe, the ram pressure at the inner edge of the shell rises. The shell thus becomes denser and ionizing radiation is more coupled. However, there is a second, competing effect. Stronger feedback (both ram and radiation pressure) leads to a faster expansion of the shell. The column density thus drops faster (as soon as the cloud has been swept up), leading to weaker coupling of radiation. The negative sign of $a$ shows that on average the second effect dominates.

Now consider a fixed cluster mass but a variable cloud mass. The higher the cloud mass, the higher the column density radiation has to pass through. Also, gravity becomes more important as the cloud mass is scaled up, slowing the expansion of the shell down and increasing the coupling of radiation. If instead of a fixed cluster mass, $\epsilon$ is kept constant, the same arguments applies, albeit in a somewhat weakened form as the cluster mass and its feedback also increase as we increase the cloud mass. In summary, radiation coupling is stronger in massive clouds, explaining the positive sign of $b$.

\section{Which type of feedback dominates?} \label{sec:leading}

Now that we have quantified radiation coupling we can start answering the question ``Which type of feedback dominates?" When asking this, it is crucial to distinguish between the instantaneous and the cumulative effect of feedback. The current density/chemical structure of the ISM is a bellwether of instantaneous feedback while cumulative feedback is traced by shell dynamics. 

\begin{figure}
    \begin{center}
        \includegraphics[width=0.49\textwidth]{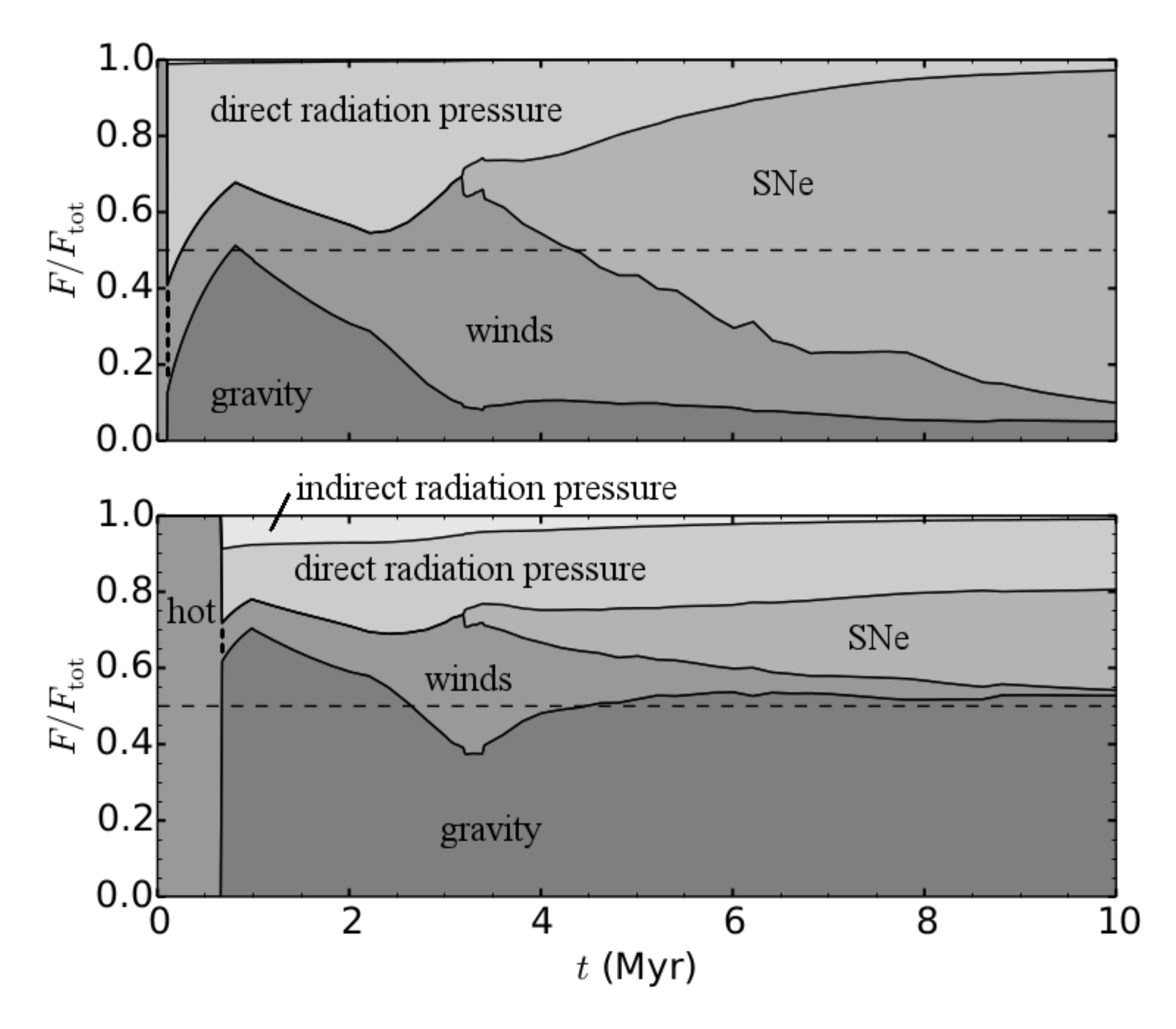} 
    \end{center}
\caption{Comparison of relative forces from direct and
indirect radiation pressure, winds, SNe, and gravity. If the
contribution from gravity is above the 50\,\% margin (dashed
horizontal line) the shell loses momentum. \textit{Top}: $M_{\rm{cl}} = 10^5M_{\odot}$, $\epsilon = 0.1$, $Z = Z_{\odot}$ and $n_{\rm{cl}} = 1000\,$cm$^{-��3}$ (same parameters as in Figure \ref{fig:summary_expansion}). The contribution from indirect radiation pressure fraction is so small it is barely visible ($< 1$\,\%). \textit{Bottom}: Same $n_{\rm{cl}}$ and $Z$ as in the top panel but with a higher cloud mass and star formation efficiency ($M_{\rm{cl}} = 3 \times 10^7M_{\odot}$, $\epsilon = 0.25$). For more information see Section~\ref{sec:leading}.}
\label{fig:force_comparison}
\end{figure}

Instantaneous feedback, as measured by its exerted force, is highly time-dependent. It is therefore necessary to specify what evolutionary stage one is interested in. To demonstrate this, we show in Figure \ref{fig:force_comparison} for two examples the relative contributions from the various forces influencing the shell. These are the forces associated to winds and SNe, $F_{\rm{wind}}$ and $F_{\rm{SN}}$, direct and indirect radiation pressure, $F_{\rm{direct}}$ and $F_{\rm{indirect}}$, as well as gravity $F_{\rm{grav}}$ (cf.\ Section~\ref{sec:phase2}). To allow easy comparison between the various terms, the forces are normalized to their sum, $F_{\rm{tot}} = F_{\rm{wind}} + F_{\rm{SN}} + F_{\rm{direct}} + F_{\rm{indirect}} + F_{\rm{grav}}$. The feedback term that dominates at a given time $t$ can be read off from the vertical width in Figure~\ref{fig:force_comparison}. Note that here for the sake of comparison gravity receives a positive sign. Therefore, if $F_{\rm{grav}}/F_{\rm{tot}} < 0.5$ the shell gains momentum, otherwise it loses momentum. 
During the adiabatic phase, the force associated to thermal pressure from shocked winds $F_{\rm{hot}}$ is the only force we consider in our model.

\begin{figure}
    \begin{center}
        \includegraphics[width=0.48\textwidth]{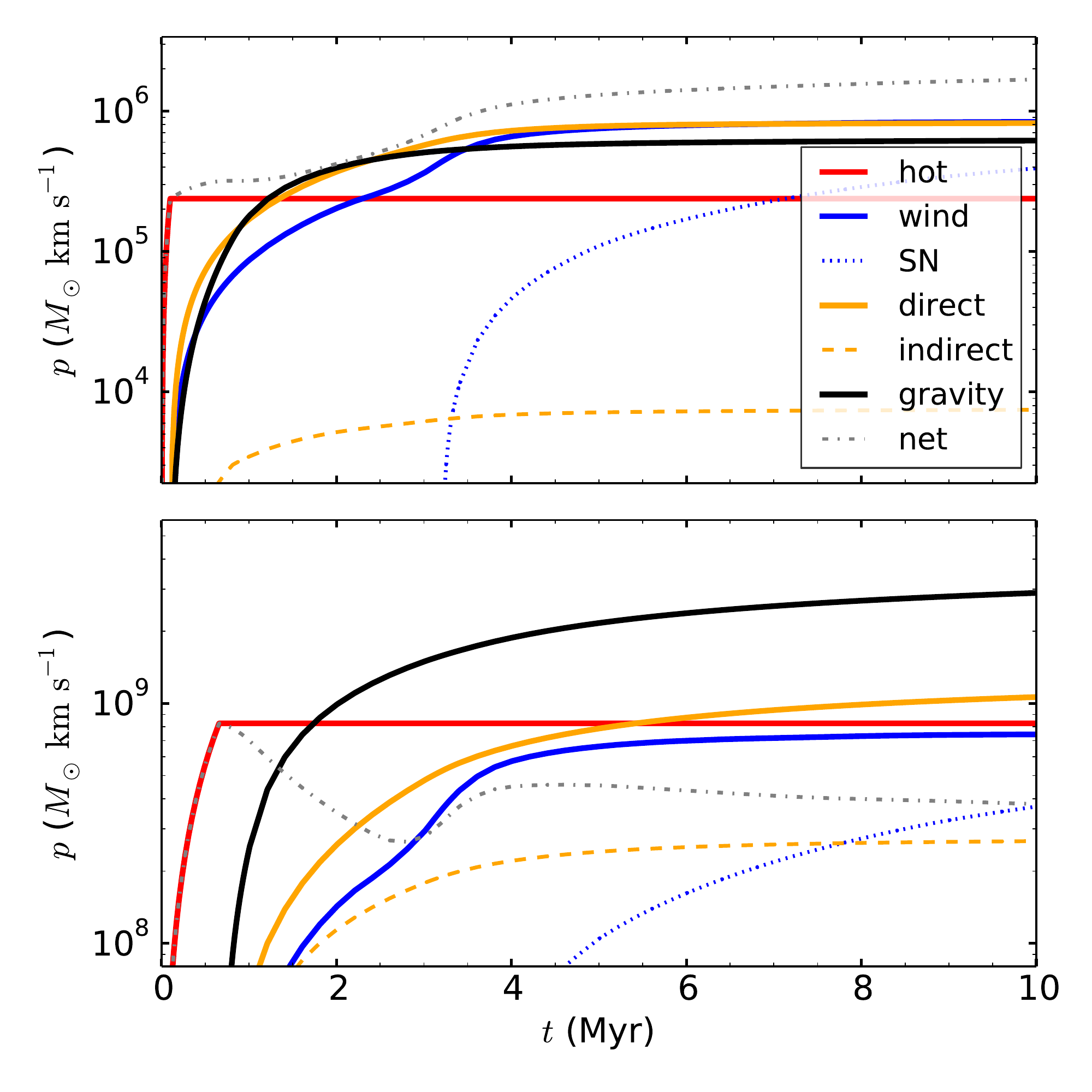} 
    \end{center}
\caption{Comparison of momentum $p$ deposited by the various feedback terms. The red line labeled ``hot" corresponds to feedback from hot shocked wind material during the adiabatic phase, the other terms are as in eq. (\ref{MomODE}), i.e.\ ram pressure in blue, radiation pressure in yellow, and gravity, which has a negative contribution, in black. The parameters of the clouds examined in the two panels are the same as in Figure~\ref{fig:force_comparison}.}
\label{fig:mom_comparison}
\end{figure}

Before we discuss the importance of the different feedback terms, it is also instructive to consider the integrated forces. The momentum $p$ injected by the various feedback terms (or removed in case of gravity) up to a time $t$ can be calculated via
\begin{equation}
p_i(t) = \int\limits_0^t F_i \diff t',
\end{equation}
where the index $i$ stands for the particular feedback term (wind, SN, etc.). The net momentum of the shell is
$p_{\rm{net}} = p_{\rm{hot}} + p_{\rm{wind}} + p_{\rm{SN}} + p_{\rm{direct}} + p_{\rm{indirect}} - p_{\rm{grav}}$. 
The evolution of $p$ is shown in Figure~\ref{fig:mom_comparison} for the same models as in Figure~\ref{fig:force_comparison}.

During Phase~I, gas pressure from hot winds is the only source driving the shell (cf.\ Figure~\ref{fig:force_comparison}) but as soon as the shell enters Phase~II this force is shut off so that $p_{\rm{hot}}$ remains constant. After the adiabatic phase, direct radiation pressure becomes the main driving force until at $t\sim 2-3$\,Myr first momentum from winds and then from SNe starts to dominate the feedback budget. At the end of the simulation, the cumulative contribution from direct radiation pressure equals that from wind ram pressure in the case of the low-mass cloud (Figure~\ref{fig:mom_comparison}, top panel) and exceeds the contribution from wind ram pressure by a factor of 1.5 in the case of the high-mass cloud with higher star formation efficiency (Figure~\ref{fig:mom_comparison}, bottom panel). In the low-mass cloud case shown, the absorption fraction drops rapidly after 3\,Myr (cf.\ Figure~\ref{fig:summary_expansion}) making radiation pressure a very ineffective feedback process at late times. This coincides with the death of massive stars marking a reduction in wind feedback and an increase in ram pressure from SNe. This additional pressure is not sufficient to raise the shell density, leading to a weak coupling between radiation and the swept-up ISM. 

Although SNe become the main driving force at late times, the momentum injected by them over the whole simulation time is lower than that injected by winds or direct radiation pressure, albeit still of the same order of magnitude. In massive clouds, the relative importance of SNe is lower than in less massive clouds, as the exerted force associated with direct radiation pressure remained comparable with the force from SN feedback for a long time span.

Whereas feedback parameters like luminosity scale linearly with a cluster's mass for a fully sampled IMF, the gravitational force increases quadratically. With increasing cloud mass, $F_{\rm{grav}}$ thus undergoes a super-linear increase, in contrast to the radiation pressure and ram pressure output of a cluster.  This is the reason why in the massive cloud case shown, gravity dominates for most of the time after the end of Phase~I and the cloud loses momentum. However, the shell still expands with a positive velocity caused by the initial velocity kick from the adiabatic phase (and a smaller kick during the Wolf-Rayet phase). Due to the slow expansion, radiation remains well-coupled. Thus, feedback from radiation pressure continues to exceed wind ram pressure feedback.

In all but the most massive clouds ($M_{\rm{cl}} \gtrsim 10^7M_{\odot}$) which produce very massive and dense shells, the contribution from indirect radiation pressure is small. During the expansion phase, even for a shell that forms in a $10^8M_{\odot}$ cloud, $\tau_{\rm{IR}}$ never exceeds $0.8$, supporting findings by \citet{Skinner2015,Martinez-Gonzalez2014}; \blue{Rei{\ss}l et al., in prep}. Only at late times during recollapse can $\tau_{\rm{IR}}$ exceed unity, but indirect radiation is still not strong enough to stall the collapse. However, for certain cloud-cluster combinations it can provide just enough momentum to keep the expansion of the shell going until the entire cloud has been swept up and the shell accelerates again. In such a case, indirect radiation pressure can make the difference between continued expansion and recollapse. 

\begin{figure}
\begin{minipage}{0.5\textwidth}
    \begin{center}
        \includegraphics[width=\textwidth]{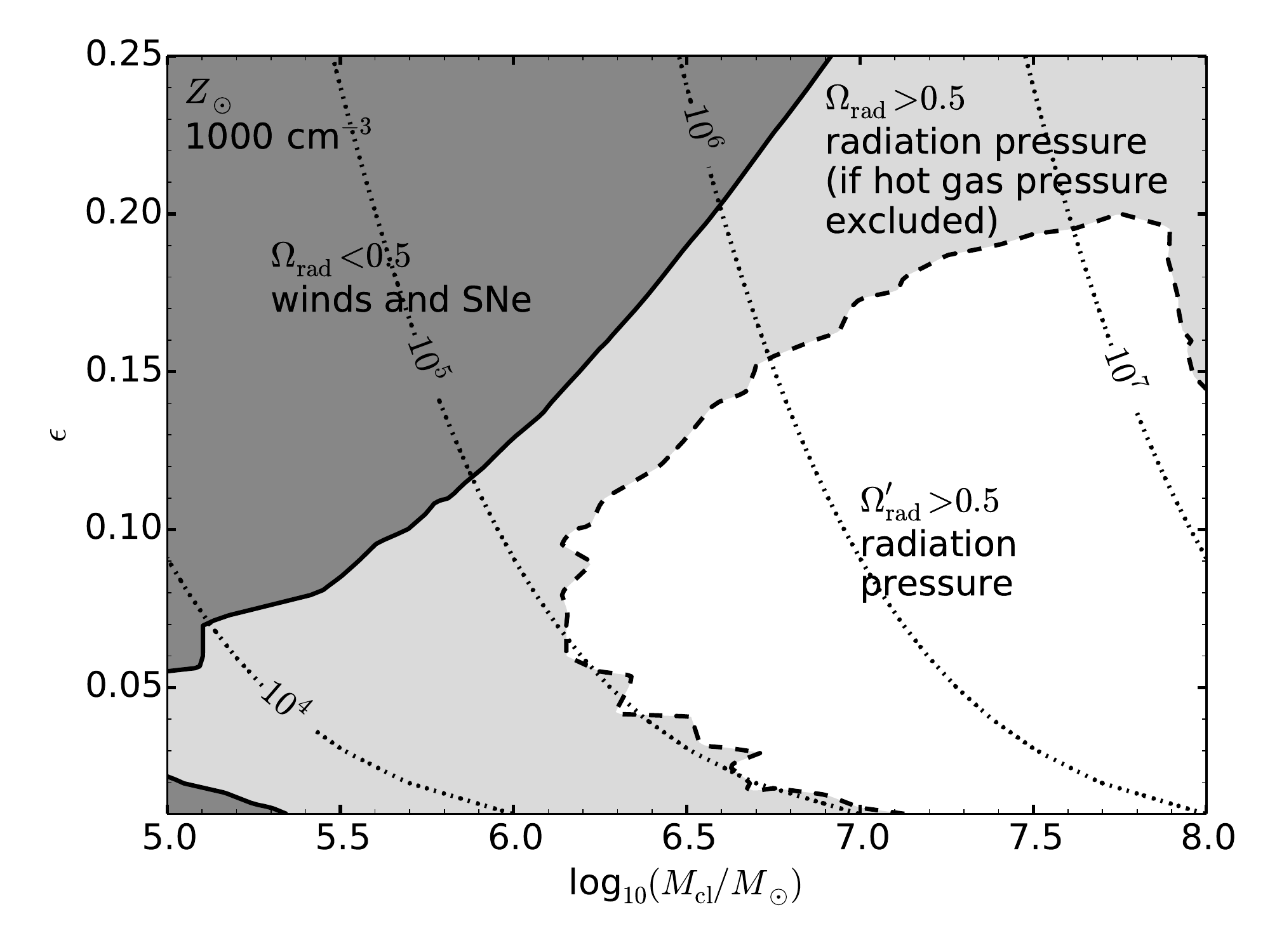} 
    \end{center}
\end{minipage}
\begin{minipage}{0.5\textwidth}
    \begin{center}
        \includegraphics[width=\textwidth]{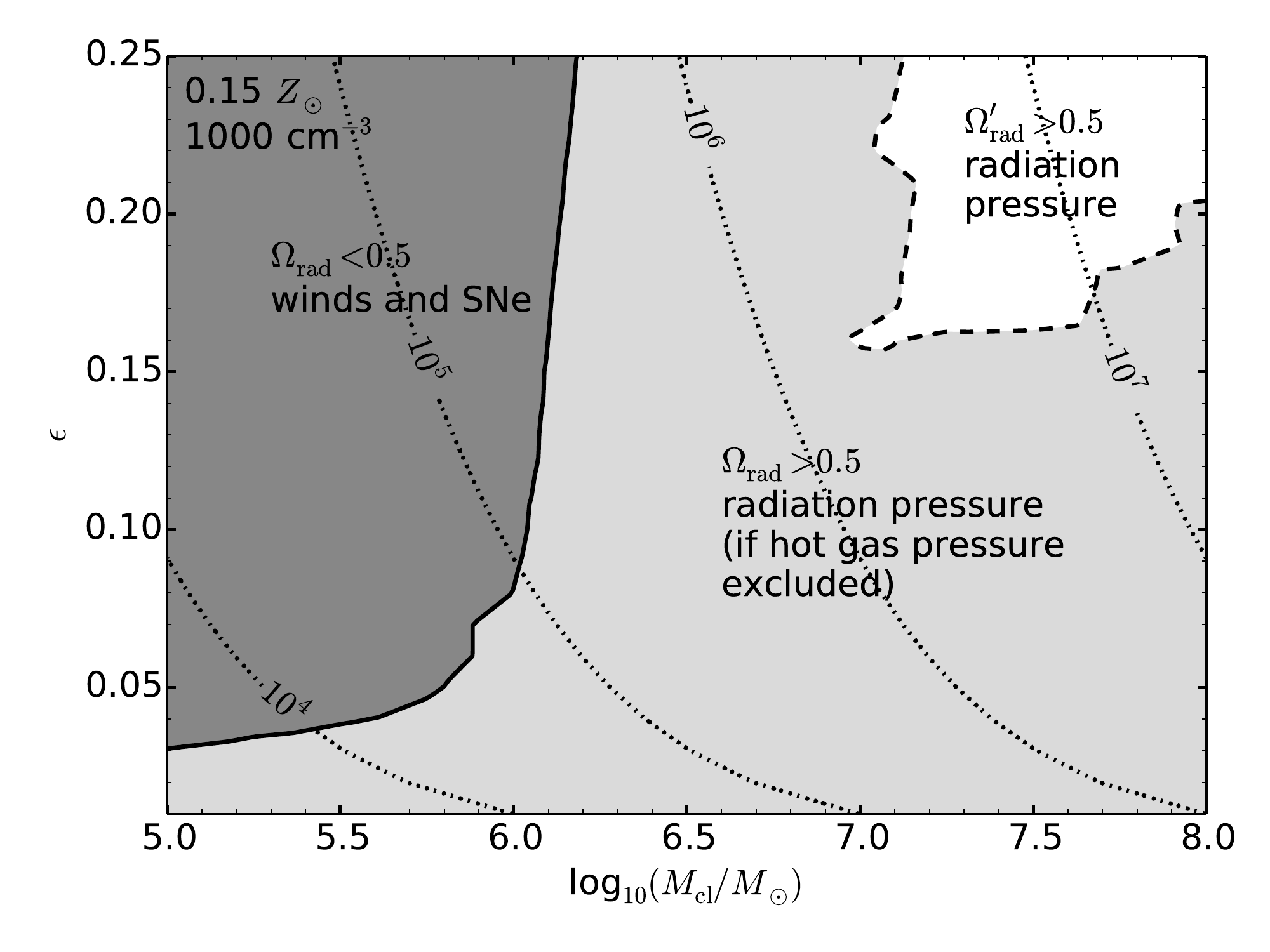}
    \end{center}
\end{minipage}
\caption{Regimes in which momentum, integrated over the whole simulation time $t_{\rm{end}}$, has mainly been injected by radiation or winds/SNe for the high-density runs with solar metallicity (\textit{top}) and low metallicity (\textit{bottom}). In white areas, the total momentum injected by radiation pressure exceeds the total momentum injected by ram pressure from winds/SNe and hot, shocked wind material ($\Omega_{\rm{rad}}'>0.5$). In light gray areas, momentum from radiation pressure exceeds momentum from ram pressure but not momentum from ram pressure and hot gas combined ($\Omega_{\rm{rad}}>0.5$). In dark gray areas, ram pressure dominates over radiation pressure ($\Omega_{\rm{rad}}<0.5$). Black dotted curves indicate lines of constant cluster mass from $10^4M_{\odot}$ to $10^7M_{\odot}$.}
\label{fig:Omega_rad}
\end{figure}

In order to determine whether the expansion of a shell up to a time $t$ was driven mainly by winds and SNe or by radiation pressure, it is instructive to compare $p_{\rm{ram}}$ and $p_{\rm{rad}}$ where, as before, $p_{\rm{rad}} = p_{\rm{direct}} + p_{\rm{indirect}}$ and $p_{\rm{ram}} = p_{\rm{wind}} + p_{\rm{SN}}$. We therefore introduce the relative radiation pressure strength parameter 
\begin{equation}
\Omega_{\rm{rad}}(t) \equiv \frac{p_{\rm{rad}}(t)}{p_{\rm{rad}}(t) + p_{\rm{ram}}(t) }.
\end{equation}
If $\Omega_{\rm{rad}}(t) > 0.5$, radiation pressure dominates over ram pressure from winds and SNe, in the sense that up to time $t$ more momentum has been injected by radiation pressure than by ram pressure. To include the contribution from winds during the adiabatic phase we also introduce the associated relative radiation pressure strength parameter
\begin{equation}
\Omega_{\rm{rad}}'(t) \equiv \frac{p_{\rm{rad}}(t)}{p_{\rm{rad}}(t) + p_{\rm{ram}}(t) + p_{\rm{hot}}(t) }.
\end{equation}
Following this definition, if $\Omega_{\rm{rad}}'(t) > 0.5$, radiation pressure has injected more momentum than ram pressure and hot gas pressure taken together. 
In Figure \ref{fig:Omega_rad} we show the regimes $\Omega_{\rm{rad}}'(t_{\rm{end}})>0.5$ (white area) which corresponds to the regime in which radiation pressure dominates over winds and SNe, $\Omega_{\rm{rad}}(t_{\rm{end}})>0.5$ (light gray area) which corresponds to the regime where radiation pressure only dominates if momentum injected during the adiabatic phase is not taken into account, and $\Omega_{\rm{rad}}(t_{\rm{end}})<0.5$ (dark gray area) which corresponds to the regime where winds and SNe dominate. 

Figure~\ref{fig:Omega_rad} shows that the dynamics of shells forming in high-mass natal clouds are dominated by radiation pressure while the dynamics of shells in low-mass clouds are dominated by winds (and to a lesser extent SNe). Also, ram pressure tends to dominate for high star formation efficiencies, as was expected from eq.~(\ref{fabs_fit}).

Interestingly, even in the low metallicity case, where momentum output from winds is roughly one order of magnitude lower than for solar metallicity, there is still a large regime where they dominate over radiation pressure (Figure \ref{fig:Omega_rad}, bottom panel). This has two reasons: First, the low amount of dust in metal-poor cloud leads to radiation being less coupled to the ISM. Second, the low ram pressure on the inner side of the shell \blue{causes the shell to be  extended and low density}; in such shells the recombination rate is small and ionizing radiation can easily escape without depositing its momentum. Thus, even though metallicity of a cluster does not strongly affect its radiative output, the entwinement between winds and radiation pressure still leads to a weakening of the efficiency with which radiation is deposited in the surrounding gas. A change in ram pressure output is always accompanied by a change in radiation coupling.

Our results show that for dense clouds there is a large parameter range in which radiation pressure dominates. This shed doubts on findings by \cite{Martinez-Gonzalez2014} who reported that radiation pressure is not the dominant feedback force for dense clouds. Their models, however, were not able to include radiation pressure in their shell expansion model. Instead they relied on an indirect diagnostic. 

For our low-density models, ram pressure dominates the whole parameter space. The main reason for this is not that these models were simulated up to later times when SN feedback increases but rather that the shells driven in low density environments have a lower density themselves and are thus less coupled to radiation. However, ram pressure only dominates by a factor of $1-4$ over radiation pressure, meaning that radiation is still not a negligible driving force.

\section{Escape fraction of ionizing radiation} \label{sec: escape fraction}

While $f_{\rm{abs}}$ determines how well-coupled the total radiation is to the shell, the 
escape fraction of ionizing radiation $f_{\rm{esc,i}}$ from the whole cloud is of 
particular interest for larger-scale simulations. For its calculation we have 
to take into account not only absorptions of ionizing photons by the shell but 
also -- at early times -- by the natal cloud. We can estimate the coupling of 
ionizing radiation at $t=0$ using a Str{\"o}mgren approximation \blue{\citep{Stromgren1939}}. For a classic 
Str{\"o}mgren sphere, the mass ionized in a constant density cloud $M_{\rm{Strom}} = (4\pi/3)R^3_{\rm{Strom}}\rho_{\rm{cl}}$, where $R_{\rm{Strom}}$ is the Str{\"o}mgren radius, can be formulated as  

\begin{equation}
M_{\rm{Strom}} = \frac{Q_{\rm{i}} \mu_{\rm{n}}}{\alpha_{\rm{B}} n_{\rm{cl}}}.
\end{equation}

We can calculate the star formation efficiency needed to ionize such a cloud 
($M_{\rm{cl}} = M_{\rm{Strom}}$), above which ionizing radiation is no longer fully 
coupled. Assuming an ionizing photon output that scales linearly with cluster mass
($Q_{\rm{i}} = 4 \times 10^{51}\,\rm{s}^{-1}\, \times M_{*} / 10^5 M_\odot$)
the star formation efficiency needed to fully ionize a constant density cloud 
and decouple radiation dynamically at early times is
\begin{equation}
\epsilon_{\rm{ionize}} = \left(\frac{\mu_{\rm{n}}}{n_{\rm{cl}}\alpha_{\rm{B}}}\frac{Q_{\rm{i}}}{M_*}+1\right)^{-1}.
\end{equation}
This corresponds to star 
formation efficiencies of 0.86 and 0.38 respectively for the 1000 and 100\,cm$^{-3}$ models examined here.

\begin{figure}
    \begin{center}
        \includegraphics[width=0.48\textwidth]{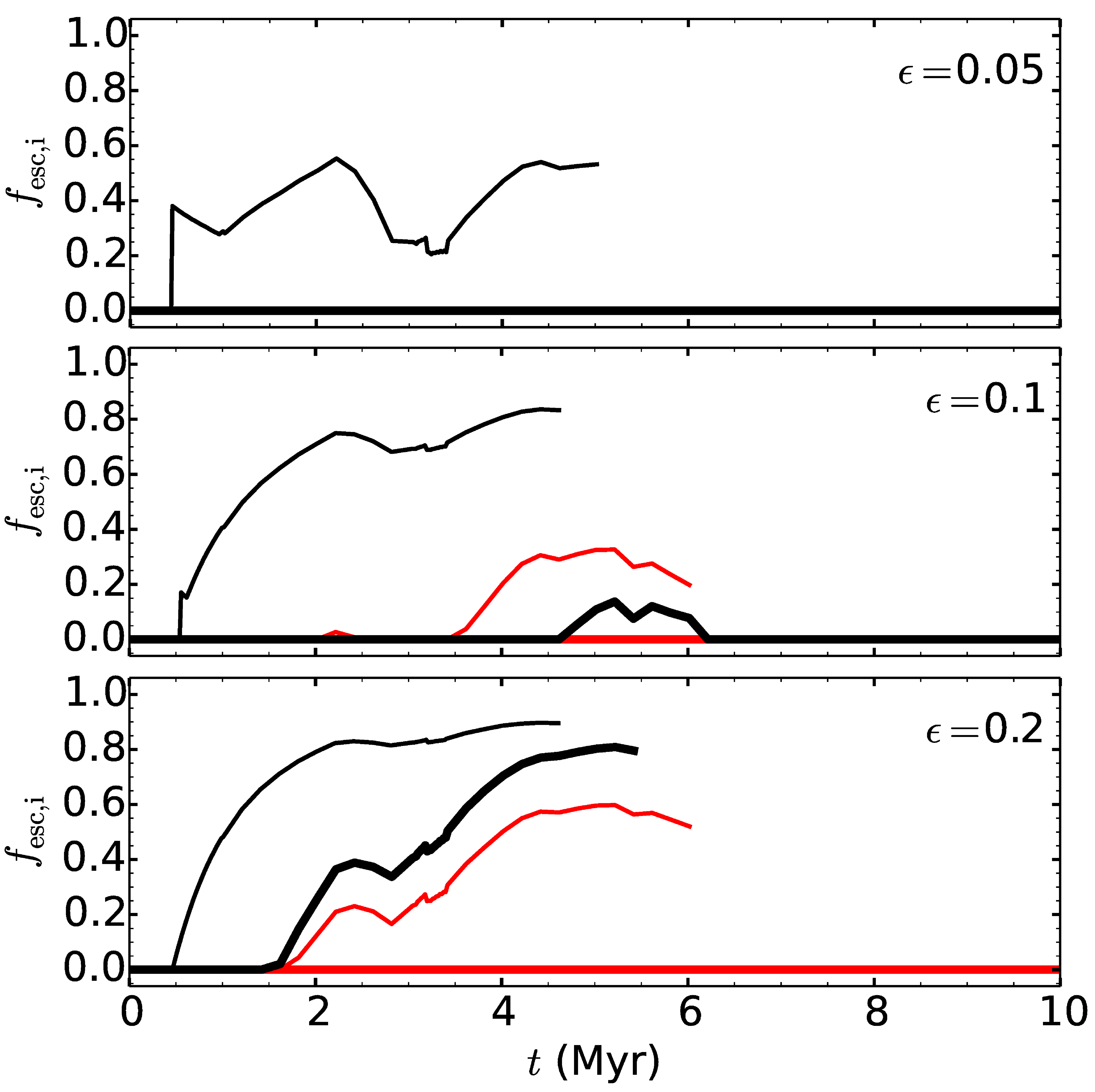} 
    \end{center}
\caption{Escape fractions for ionizing radiation $f_{\rm{esc,i}}$ for $\epsilon = 0.05$, 0.1, and 0.2 (top, middle and bottom panels, respectively) for $Z=Z_{\odot}$. The black lines are for a $10^5M_{\odot}$ cloud, and the red lines for a $10^6M_{\odot}$ cloud. Thick and thin lines correspond to cloud densities of $n_{\rm{cl}}=1000$\,cm$^{-3}$ and 100\,cm$^{-3}$, respectively. Lines that stop before 10\,Myr belong to shells that have dissolved into the ambient ISM before this time.}
\label{fig:fesc_highZ}
\end{figure}

\begin{figure}
    \begin{center}
        \includegraphics[width=0.48\textwidth]{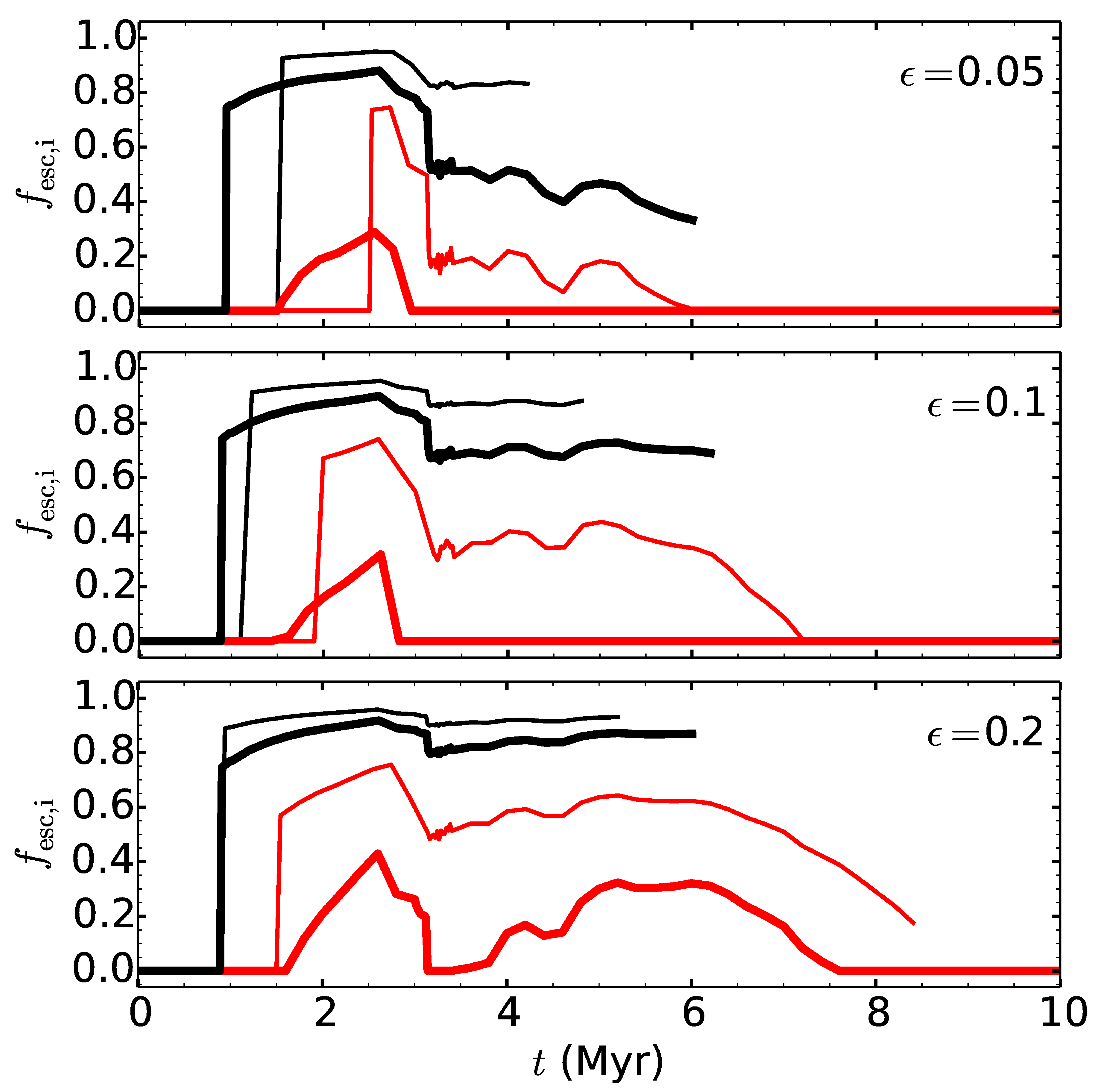} 
    \end{center}
\caption{Escape fractions for ionizing radiation $f_{\rm{esc,i}}$ for $\epsilon = 0.05$, 0.1, and 0.2 (top, middle and bottom panels, respectively) for $Z=0.15 \,Z_{\odot}$. The black lines are for a $10^5M_{\odot}$ cloud, and the red lines for a $10^6M_{\odot}$ cloud. Thick and thin lines correspond to cloud densities of $n_{\rm{cl}}=1000$\,cm$^{-3}$ and 100\,cm$^{-3}$, respectively. Lines that stop before 10\,Myr belong to shells that have dissolved into the ambient ISM before this time.}
\label{fig:fesc_lowZ}
\end{figure}

Initial expansion of the wind bubble increases the density of the shell and 
hence the global cloud recombination rate, which will not decrease until 
the expansion radius exceeds the initial cloud radius. Therefore, ionizing 
 radiation cannot escape in any of our models as long as the shell is still 
 confined by the cloud. Thus,
\begin{equation}
   f_{\rm{esc,i}} =
   \begin{cases}
     0 & \hspace{.5in} \text{if } t < t_{\rm{sweep}}, \\
     1 - f_{\rm{abs,i}}  & \hspace{.5in} \text{otherwise}. \\
   \end{cases}
\end{equation}
In Figures~\ref{fig:fesc_highZ} and \ref{fig:fesc_lowZ} we show how the escape fraction varies as a function of time for $10^5M_{\odot}$ and $10^6M_{\odot}$ clouds with a range of densities and metallicities. For clouds more massive than $10^7M_{\odot}$, $f_{\rm{esc,i}}$ remains 0 at all times. Note, however, that we do not take into account fragmentation of the shell. Hence, the escape fractions provided here purely consider radiation escaping through the isotropic shell ignoring any holes and clumps. Consequently, in most cases the escape fractions derived here will be lower limits on the true values.

For solar metallicity (Figure~\ref{fig:fesc_highZ}), $f_{\rm{esc,i}}$ reaches its highest values around 5\,Myr. We have tested how the escape fraction would evolve if we would continue the expansion of the ``shell" even after it has dissolved and found that $f_{\rm{esc,i}}$ always drops after $t\sim 5\,$Myr. At late times the strong reduction in $L_{\rm{i}}$ due to the death of the massive stars causes a decrease in $f_{\rm{esc,i}}$, even though the shell has a low column and volume density by then. Both the time span during which ionizing radiation can escape and the amount of escaping ionizing radiation depend on the cloud mass (more escape for low \Mcl) and cloud density (more escape for low $n_{\rm{cl}}$).  Additionally, the fact that the shell dissolves before 10\,Myr for some models does not mean that all ionizing radiation can escape. With a decrease in $L_{\rm{i}}$ at late times, even a diffuse medium of $\lesssim 1$\,cm$^{-3}$ can be enough to absorb a large part of the ionizing radiation.

Low metallicity models (Figure~\ref{fig:fesc_lowZ}) have higher integrated ionizing escape fractions than solar metallicity models and $f_{\rm{esc,i}}$ peaks earlier, at $\sim 2.5$\,Myr, as less radiation is absorbed by dust.
Also, even at low $\epsilon$, dense clouds become optically thin to ionizing radiation before the first SNe. Thus, the Wolf-Rayet phase and the first SNe lead to a significant reduction in $f_{\rm{esc,i}}$ between $\sim 3-4$\,Myr. Even though we neglect turbulence, which can open and close low density channels in the ISM through which radiation can escape, we show that some strong variability in $f_{\rm{esc,i}}$ is expected purely due to stellar evolution.

Our results are in good agreement with 3D MHD simulations by \cite{Howard2017} 
for a cloud with $M_{\rm{cl}}=10^6M_{\odot}$, $\epsilon = 0.1$ and 
$n_{\rm{cl}}=100$\,cm$^{-3}$ and solar metallicity even though they include 
turbulence but neglect stellar winds. Furthermore, our results are consistent 
with ionization parameter mappings of the Magellanic clouds\footnote{Typical 
sampled cloud masses associated with massive clusters in the LMC and SMC are 
$>10^4 M_{\odot}$ (the same as those 
shown in Figures~\ref{fig:fesc_highZ} and \ref{fig:fesc_lowZ}; see e.g. 
\citealt{Wong2011}) but the characteristic 
metallicity of the gas in these two galaxies is $0.5 \, Z_{\odot}$ (in between the metallicities we investigated) and $0.2 \, 
Z_{\odot}$ (slightly above our low-$Z$ model), respectively.} carried out by \cite{Pellegrini2012}, who find average ionizing 
escape fractions of 0.4. These escape fractions are dominated by  \hii regions 
with two types of geometries: blister type \hii and classical density-bounded 
nebulae. Our model is most applicable to the density-bounded regions, which are 
consistent with fully ionized shells.

\section{When Feedback Fails -- Recollapse and Sequential Star Formation} \label{sec:Feedback Fails}

It is not a given that stellar feedback is always able to overpower gravity and drive an outflow. If \SFE is lower than some minimum star formation efficiency $\epsilon_{\rm{min}}$ the shell eventually collapses back on itself, initiating more star formation. \blue{One possible example for this could be the core of 30 Doradus where two distinct stellar clusters of different age coexist \citep[e.\,g.][]{Sabbi2012}.}  The collapse time thus sets what we coin the cadence of star formation. Only when $\epsilon > \epsilon_{\rm{min}}$ can further star formation be shut off \blue{(neglecting triggered star formation in the shell)}. Since we cannot follow the expansion of each shell ad infinitum we regard shells as non-collapsing if they have either dissolved or have not collapsed by $t = t_{\rm{end}}$. We hence might miss a small number of shells that take longer than $7\tau_{\rm{ff}}$ to collapse.

\begin{figure}
\begin{minipage}{0.5\textwidth}
    \begin{center}
        \includegraphics[width=\textwidth]{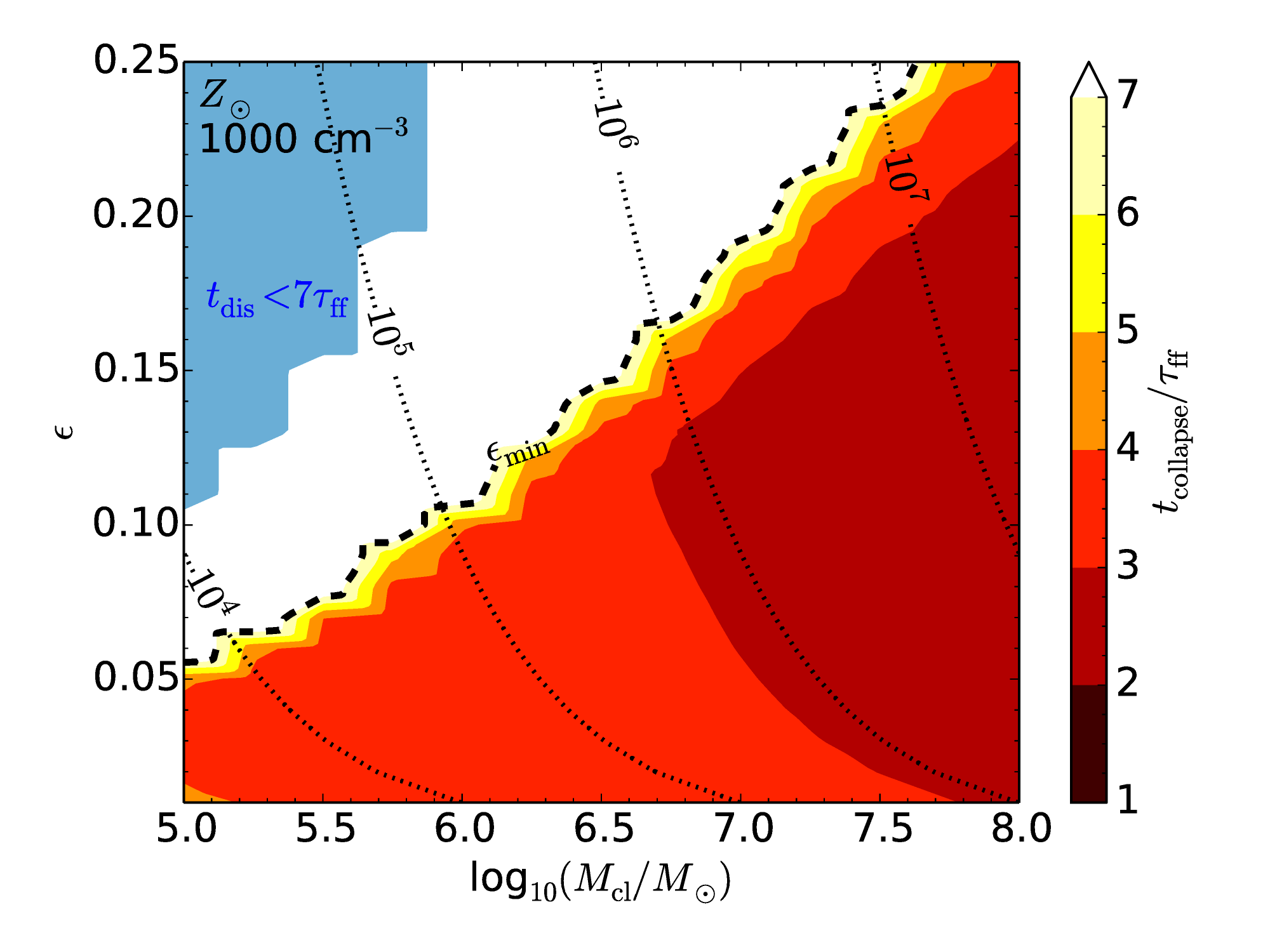} 
    \end{center}
\end{minipage}
\begin{minipage}{0.5\textwidth}
    \begin{center}
        \includegraphics[width=\textwidth]{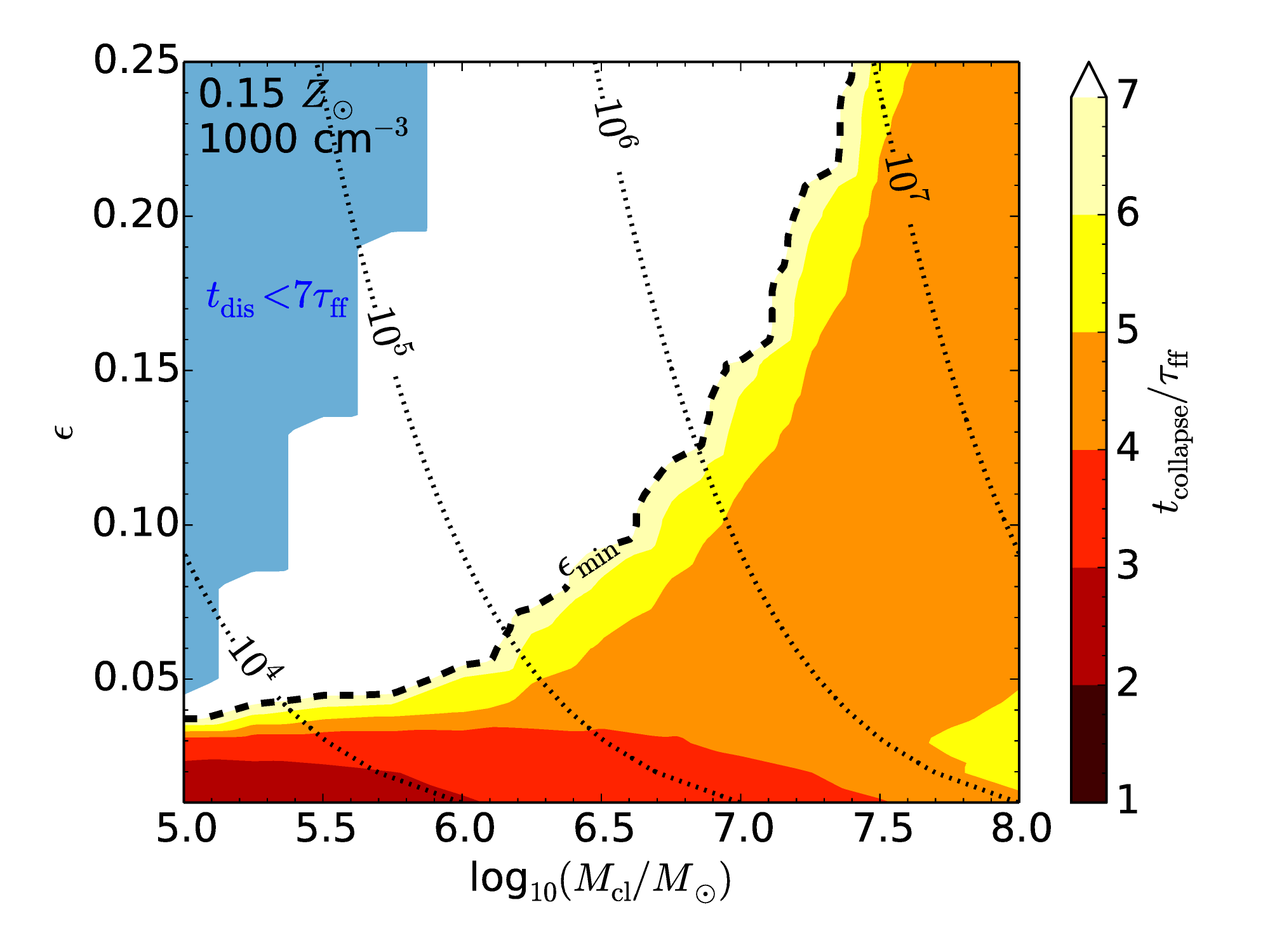}
    \end{center}
\end{minipage}
\caption{Collapse time $t_{\rm{collapse}}$ in multiples of $\tau_{\rm{ff}}$ (1.4\,Myr) as a function of cloud mass and star formation efficiency for high density runs with solar metallicity (\textit{top}) and low metallicity (\textit{bottom}). The black dashed line shows the minimum star formation efficiency $\epsilon_{\rm{min}}$ (see main text). Shells in the light blue regime have dissolved before $t=7\tau_{\rm{ff}}$ and are assumed to never recollapse. Black dotted curves indicate lines of constant cluster mass from $10^4M_{\odot}$ to $10^7M_{\odot}$.}
\label{fig:MapsZ1}
\end{figure}

Figure~\ref{fig:MapsZ1} shows the collapse time $t_{\rm{collapse}}$ for high-density runs. It is remarkable that a vast majority of models that collapse share a similar collapse time: $t_{\rm{collapse}} = 2 - 4\tau_{\rm{ff}}$ ($\sim 3-6$\,Myr) for solar metallicity and $t_{\rm{collapse}} = 4-5\tau_{\rm{ff}}$ ($\sim 6-7$\,Myr) in our low metallicity run. No shell in the investigated range collapsed in less than $2\tau_{\rm{ff}}$. 
Even though in our model all stars formed in an instantaneous star burst we can define a time averaged star formation rate $\langle \dot{M}_*\rangle \equiv M_*/t_{\rm{collapse}}$.
Following \citet{Krumholz2005} we then define the dimensionless star formation rate per free-fall time
\begin{equation}
\epsilon_{\rm{ff}} \equiv \frac{\langle\dot{M}_*\rangle}{M_*+M_{\rm{cl}}}\tau_{\rm{ff}},
\end{equation}
which can be rewritten as $\epsilon_{\rm{ff}} = \epsilon \tau_{\rm{ff}}/t_{\rm{collapse}}$.
Our re-collapsing models have $\epsilon_{\rm{ff}}$ of the order 0.01 and never exceed 0.07, in very good agreement with observations \citep[e.g.][]{Krumholz2007}.

The dashed contour line between re-collapsing and non-collapsing models shows the minimum star formation efficiency $\epsilon_{\rm{min}}$. It increases with increasing cloud mass as gravity prevents outflows in massive clouds.
We find that for solar metallicity, $\epsilon_{\rm{min}}$ scales linearly with $\log M_{\rm{cl}}$ while for the low $Z$, high density model $\log \epsilon_{\rm{min}}$ scales linearly with $\log M_{\rm{cl}}$. For all but the most massive clouds, $\epsilon_{\rm{min}}$ is lower for low metallicity.

The blue area in Figure~\ref{fig:MapsZ1} shows models in which the shells 
dissolve before $7\tau_{\rm{ff}}$ ($\sim 10$\,Myr). The earliest dissolutions 
take place after 4\,Myr. \blue{Using numerical simulations, this is also what \citet{Inutsuka2015} find for the destruction time of $\sim 10^5\,M_{\odot}$ clouds, albeit for lower star formation efficiencies.} 4\,Myr is clearly shorter than what observational studies usually estimate
for the lifetimes of molecular clouds \blue{after the onset of star formation}, i.e.\ $\sim 20$\,Myr (see 
\citealt{Dobbs2014} for an overview). We note this calls into question the 
existence of clouds with low masses and high star formation efficiencies.

\begin{figure}
    \begin{center}
        \includegraphics[width=0.48\textwidth]{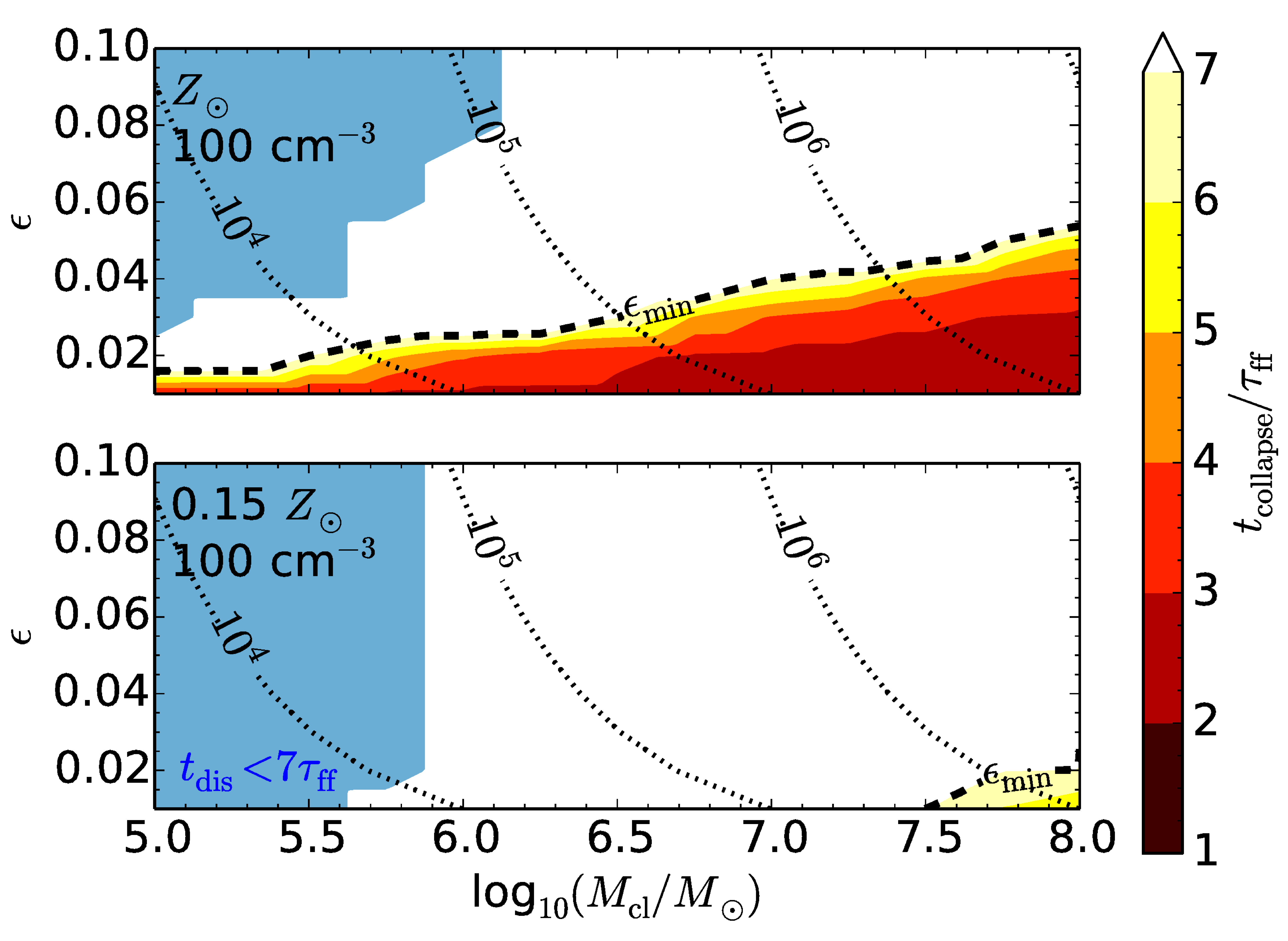} 
    \end{center}
\caption{Collapse time $t_{\rm{collapse}}$ in multiples of $\tau_{\rm{ff}}$ (4.6\,Myr) as a function of cloud mass and star formation efficiency for low density runs with solar metallicity (\textit{top}) and low metallicity (\textit{bottom}). The black dashed line shows the minimum star formation efficiency $\epsilon_{\rm{min}}$ (see main text). Shells in the light blue regime have dissolved before $t=7\tau_{\rm{ff}}$ and are assumed to never recollapse. Black dotted curves indicate lines of constant cluster mass from $10^4M_{\odot}$ to $10^7M_{\odot}$. Only star 
formation efficiencies up to $\epsilon = 0.1$ are shown.}
\label{fig:MapsZ2}
\end{figure}

In Figure~\ref{fig:MapsZ2}, we show $t_{\rm{collapse}}$ for our low density 
models. Recollapse is limited to the most massive clouds or small star 
formation efficiencies in the case of solar metallicity. At low metallicity, only 
shells that form in clouds with masses close to $10^8$\Msun and $\epsilon \leq 
0.02$ collapse. Re-collapsing low-density models have lower 
$\epsilon_{\rm{ff}}$ values than high density models but are still consistent 
with observations (e.g.\ \citealt{Murray2011}). 

The trend of increasing $\epsilon_{\rm{min}}$ for increasing cloud mass hints at star formation being more efficient for massive clouds. Observationally, this is hard to test. Some studies that found the opposite trend, i.e.\ lower $\epsilon$ with increasing cloud mass, were probably limited by sampling and selection effects \citep{Murray2011}.

\cite{Kim2016} present $\epsilon_{\rm{min}}$ for various cloud densities. As an example, for a $2\times10^6 M_{\odot}$ cloud with $n_{\rm{cl}}=1000\,$cm$^{-3}$ they find $\epsilon_{\rm{min}}$ anywhere between $0.2$ and $0.7$ depending on which of their definitions for $\epsilon_{\rm{min}}$ they use. Our results suggest a lower value of $\epsilon_{\rm{min}} = 0.12$ for such a cloud. This difference, however, is not surprising since \cite{Kim2016} ignore wind and SN feedback in their model.

Studies of the effect of gas expulsion on star cluster evolution show that a 
majority of stars remain bound only if $\epsilon\gtrsim 0.1-0.2$ 
\citep{Geyer2001, Baumgardt2007, Shukirgaliyev2017}. Since clouds with a low gas density or a low mass have a lower 
minimum star formation efficiency than this value, our model predicts that such 
clouds will form gravitationally unbound OB associations rather than gravitationally 
bound star clusters. Similarly, the lower values of $\epsilon_{\rm min}$ that we find in 
our lower metallicity models suggest that the formation of unbound associations rather
than bound clusters may be more common in these systems.

\section{Conclusions and Summary}\label{sec:conc}
We have developed a new model that simultaneously and self-consistently calculates the structure and the expansion of shells driven by feedback from stellar winds, supernovae and radiation pressure. 
The model has been put to use to investigate the conditions in which the various different sources of feedback dominate, the amount of radiation that can escape through the shell, and to derive minimum star formation efficiencies for a large parameter space of clouds and clusters. Our main results are summarized below.

\subsection{What is the dominant source of feedback?}
\begin{itemize}
\item Radiation pressure and ram pressure are interconnected. Any attempt to estimate the momentum that radiation injects into the ISM without accounting for ram pressure by winds and SNe will yield incorrect results. Changing the momentum imparted by winds always leads to a change in the efficiency of radiation pressure.
\item The evolution of a star-forming molecular cloud is strongly influence\blue{d} by the effects of stellar evolution. The Wolf-Rayet phase and SN explosions do not only increase the effect of ram pressure but also indirectly increase the effect of radiation pressure (see above). It is thus imperative to include proper stellar evolution when investigating feedback.
\item After the shocked wind material has cooled, radiation dominates the driving of the shell as long as the shell remains optically thick. \blue{This is usually the case when the star cluster is still young ($t \lesssim 2-3$\,Myr). In massive clouds, which tend to expand more slowly due to the quadratic dependence of the gravitational force on mass, radiation pressure remains dominant for an even longer time span.} Thus, in more massive clouds the \blue{time-integrated effect} of radiation pressure compared to ram pressure increases. Indirect radiation pressure is negligible for low mass clouds and is only of some importance during the early phases of massive cloud evolution or during recollapse.
\item Stellar winds are more important than radiation pressure in dense clouds only if the cloud mass is towards the lower end of the range studied here ($M \sim 10^{5}$--$10^{6} M_{\odot}$). They always dominate \blue{over radiation pressure} if the cloud density is low. At low metallicity, the momentum output by winds is decreased but radiation also couples more weakly with the shell, and so winds can still dominate over radiation.
\item SNe dominate at late times. However, in most cases, over the whole cloud lifetime SN feedback does not exceed either feedback from winds or from radiation pressure. \blue{Also, feedback from SNe is not always sufficient to destroy a molecular cloud.}
\end{itemize}

\subsection{How well-coupled is radiation to the shell?}
As we have demonstrated, classical Str{\"o}mgren calculations show a full ionization of a massive
molecular \blue{cloud} by a star cluster is practically impossible. Despite this, we find 
the escape of ionizing radiation from a spherically-symmetric
expanding cloud is significant, and a direct result of the shell structure 
responding to stellar feedback. This is an unavoidable consequence of the 
dynamic evolution caused by feedback driving an expansion and stretching the 
gas over a larger volume, decreasing its density. 

\begin{itemize}
\item Radiation decouples more rapidly from the ISM for higher star formation efficiency, lower metallicity, lower cloud density or lower cloud mass. This is true for both ionizing and non-ionizing radiation.
\item For our calculations of ionizing escape fractions $f_{\rm{esc,i}}$ we consider the radiation escaping through a shell but neglect any fragmentation of shell. Our escape fractions are thus independent of the solid angle on the sky and, in most cases, are lower limits to real total escape fractions.
\end{itemize}

\subsection{What is the minimum star formation efficiency required to prevent Recollapse?}
\begin{itemize}
\item We find minimum star formation efficiencies $\epsilon_{\rm{min}}$ of a few percent for low mass clouds, increasing to $\sim 25$\% or more for very massive clouds. Clouds with star formation efficiencies above these values are disrupted by the effects of stellar feedback and do not recollapse. 
\item The values we recover for $\epsilon_{\rm{min}}$ are considerably smaller \blue{than} those found by \cite{Kim2016}, likely because those authors do not account for the effects of stellar winds or SNe.
\item The cadence of star formation (i.e.\ the delay between episodes of star formation in clouds that recollapse) is $3$--$6$\,Myr ($2$--$4\, \tau_{\rm{ff}}$) for dense clouds with solar metallicity and is somewhat higher for lower metallicity clouds. Low-density clouds are much easier to disrupt by feedback (especially if they are metal-poor), thus suggesting that they earlier shut off further star formation and hence tend to have a lower star formation efficiency.
\item Our results suggest that dense, massive and/or metal-rich clouds are more likely to form gravitationally bound star clusters, while less dense, less massive and/or more metal-poor clouds are more likely to form unbound OB associations.
\end{itemize} \vspace{2mm}

We thank Sam Geen, Mordecai-Mark Mac Low, Avery Meiksin, Stefan Rei{\ss}l, Stefanie Walch, and Robin Williams for helpful discussions.
We acknowledge support from the Deutsche Forschungsgemeinschaft in the Collaborative Research Centre (SFB 881) ``The Milky Way System'' (subprojects B1, B2, and B8) and in the Priority Program SPP 1573 ``Physics of the Interstellar Medium'' (grant numbers KL 1358/18.1, KL 1358/19.2, and GL 668/2-1). RSK furthermore thanks the European Research Council for funding in the ERC Advanced Grant STARLIGHT (project number 339177).

\begin{footnotesize}
\bibliographystyle{mn2e}  
\bibliography{library_final}
\end{footnotesize}

\appendix
\section{}
\subsection{The Effect of Stellar Rotation} \label{sec: stellRot}

Models that include stellar rotation can better reproduce the observed main sequence width and stellar surface abundances and velocities than models of non-rotating stars and are therefore thought to provide a more realistic view \citep{Ekstrom2012}. Given that rotating stars produce more ionizing radiation at later times \citep{Levesque2012}, it is interesting to see how stellar rotation effects the escape fractions of ionizing radiation in our models.

\begin{figure}
    \begin{center}
        \includegraphics[width=0.48\textwidth]{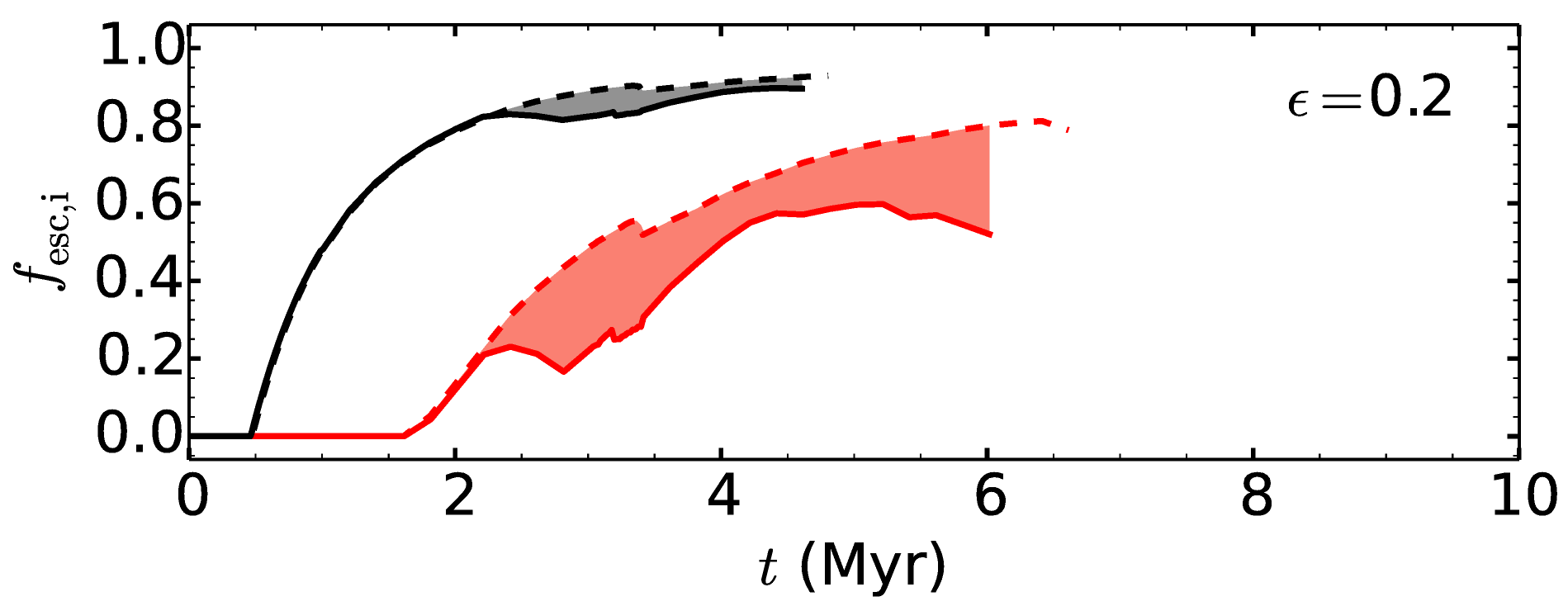} 
    \end{center}
\caption{Example of the dependence of $f_{\rm{esc,i}}$ on stellar rotation for $n_{\rm{cl}}=100$\,cm$^{-3}$ and solar metallicity. The dashed lines correspond to the model which includes stellar rotation. The solid lines correspond to the non-rotating model. We show results for clouds with masses $M_{\rm cl} = 10^5 \, M_{\odot}$ (black) and $M_{\rm cl} = 10^6 \, M_{\odot}$ (red), as in Figure~\ref{fig:fesc_highZ}. Since the assumed stellar rotation might be too high (see main text), realistic escape fractions are expected to lie in the gray and red shaded areas, respectively.}
\label{fig:fesc_rotation}
\end{figure}

We reran all models including stellar rotation and found that the effects on 
the dynamics of the shell are small. However, since most ionizing radiation 
gets emitted at late times when the density of the shell has already dropped, 
$f_{\rm{esc,i}}$ is larger at late times for rotating stars than for 
non-rotating stars (see Figure~\ref{fig:fesc_rotation}). On the other hand, at 
early times stellar rotation does not considerably decrease $f_{\rm{esc,i}}$. 
Taken together, the time-integrated escape fractions of ionizing radiation are 
higher if stellar rotation is included.

For our simulations we have used the rotating models by \cite{Ekstrom2012}, which assume a stellar rotation velocity of 40\,\% of the break-up velocity on the zero-age main sequence. However, as \cite{Martins2013} point out, this value might be too extreme. The results obtained from including such a high rotation velocity should thus be regarded as an upper limit for $f_{\rm{esc,i}}$ while non-rotating models provide a lower limit.

\subsection{Overview of Models} \label{sec:appendix_allmodels}

\begin{figure*}
	\includegraphics[width=0.33\textwidth, angle = 90]{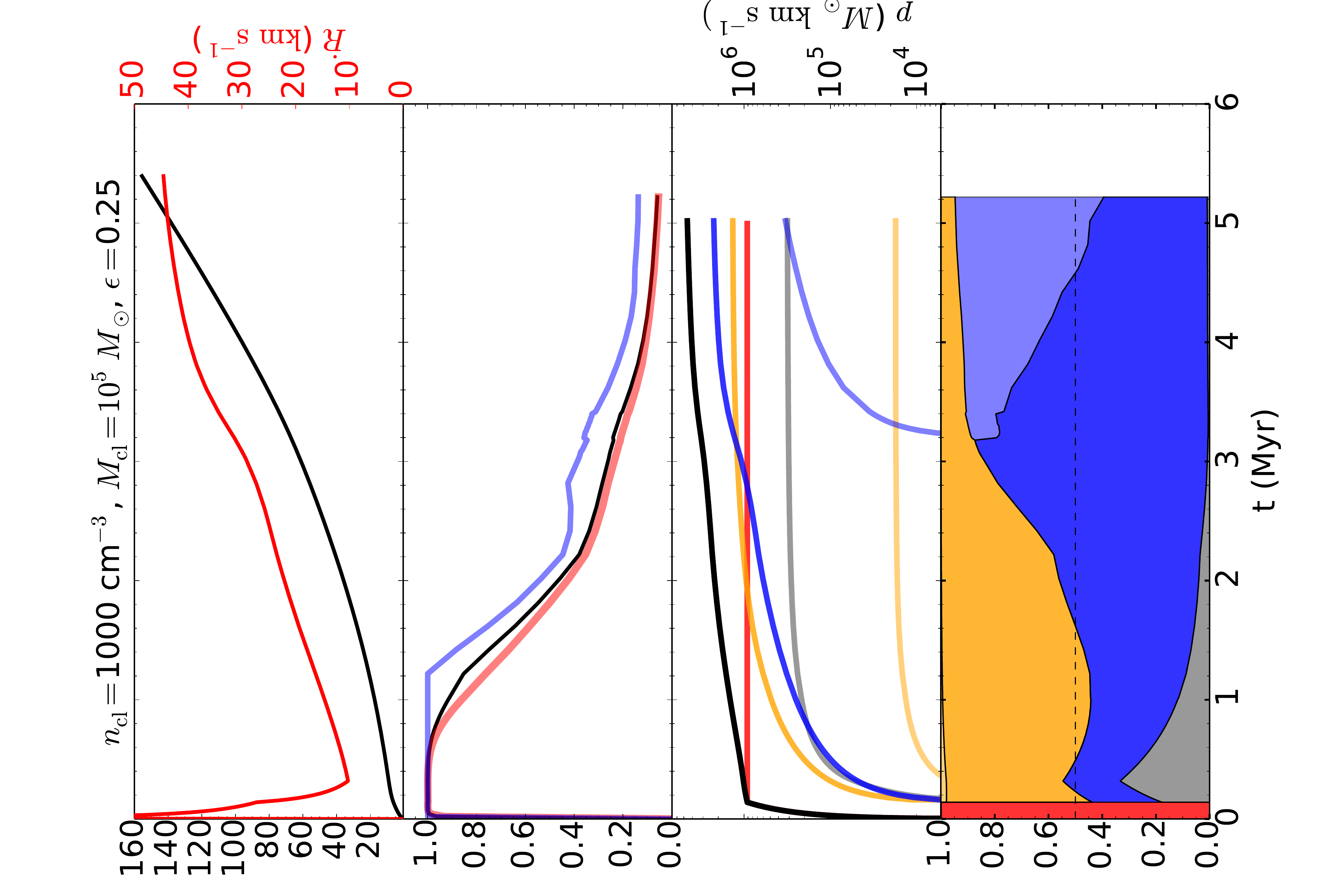}
	\hspace{-4mm}
	\includegraphics[width=0.33\textwidth, angle = 90]{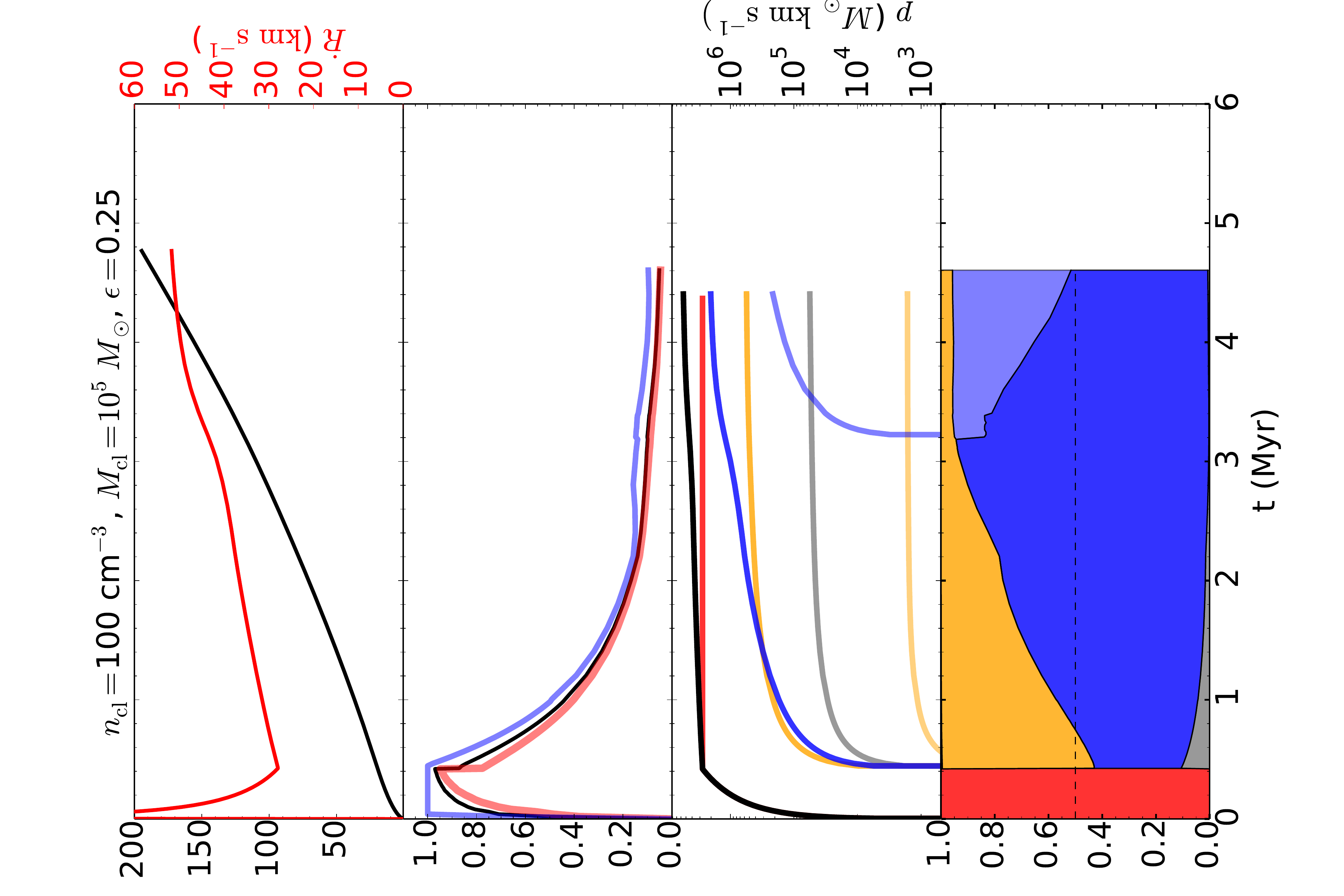}
	\vspace{-0.6mm}\\
	\includegraphics[width=0.33\textwidth, angle = 90]{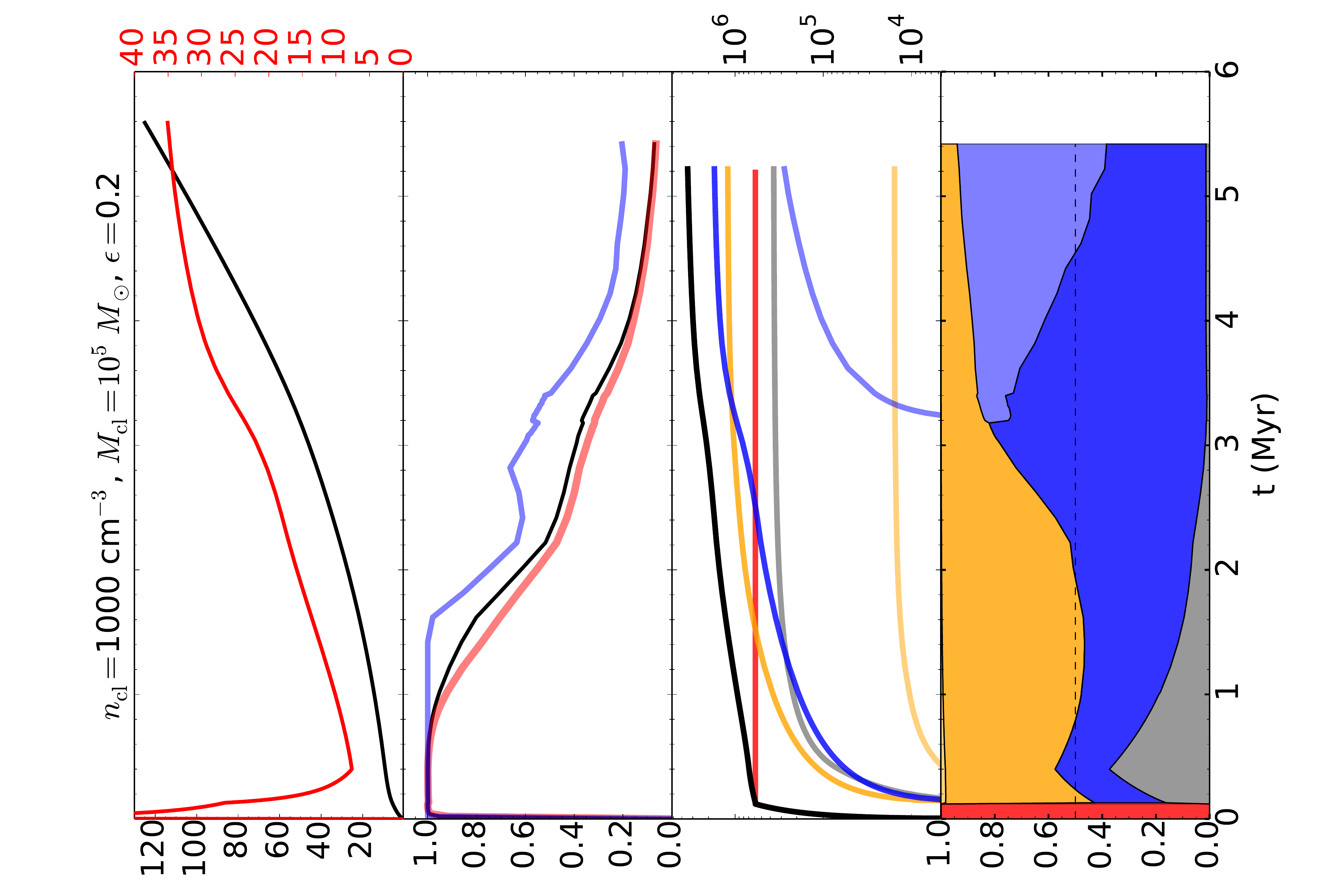}
	\hspace{-4mm}
	\includegraphics[width=0.33\textwidth, angle = 90]{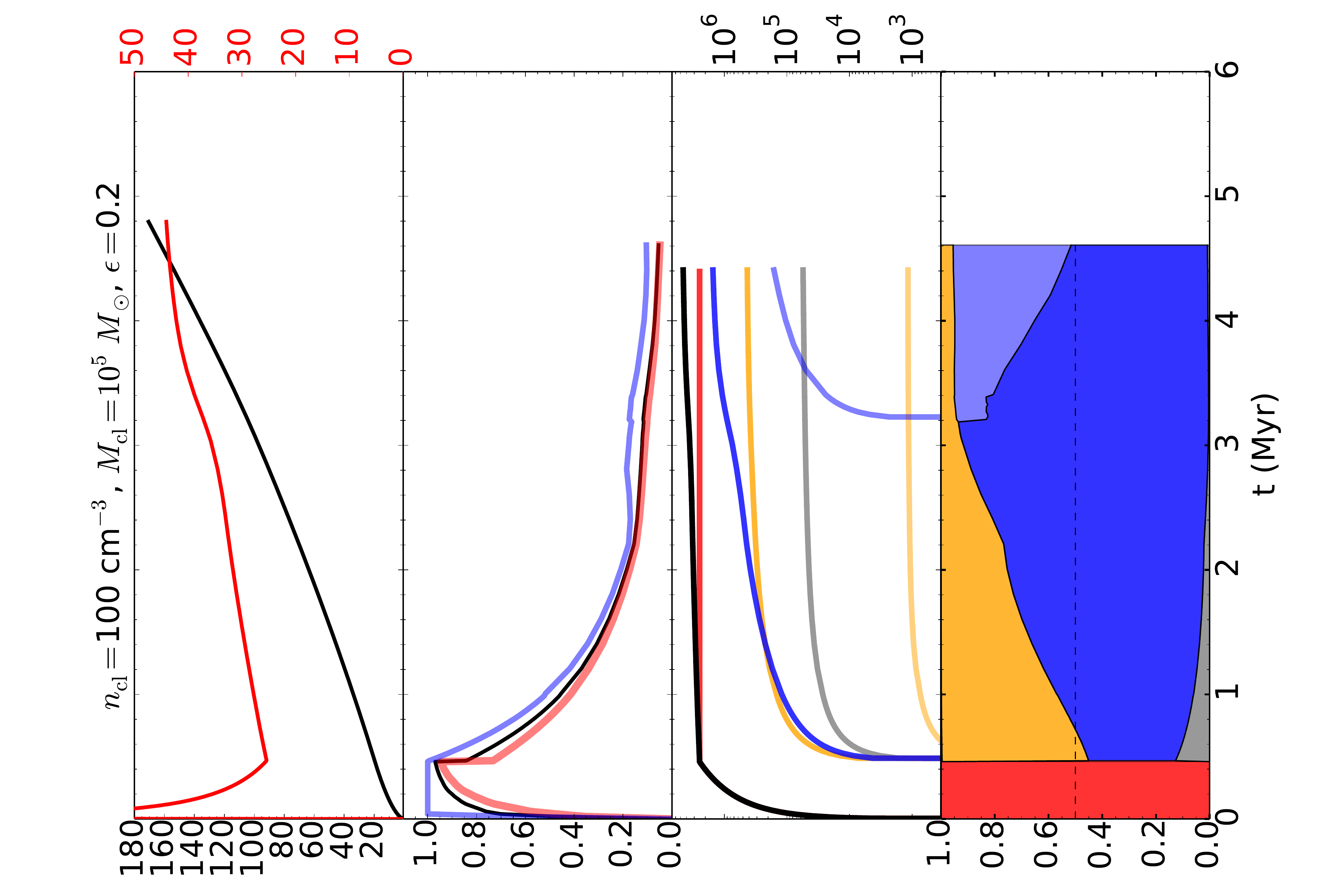}
	\vspace{-0.6mm}
	\includegraphics[width=0.33\textwidth, angle = 90]{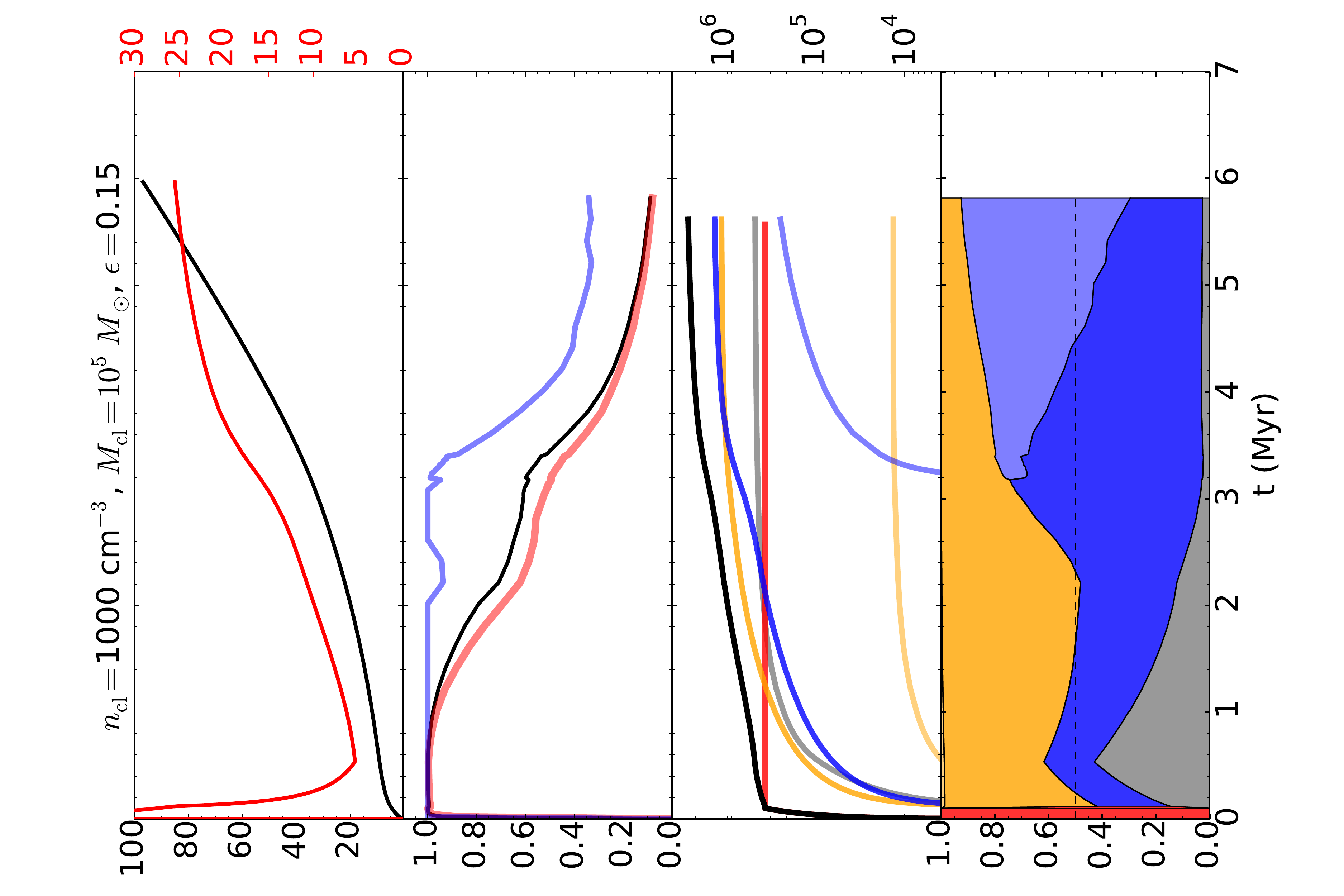}
	\hspace{-4mm}
	\includegraphics[width=0.33\textwidth, angle = 90]{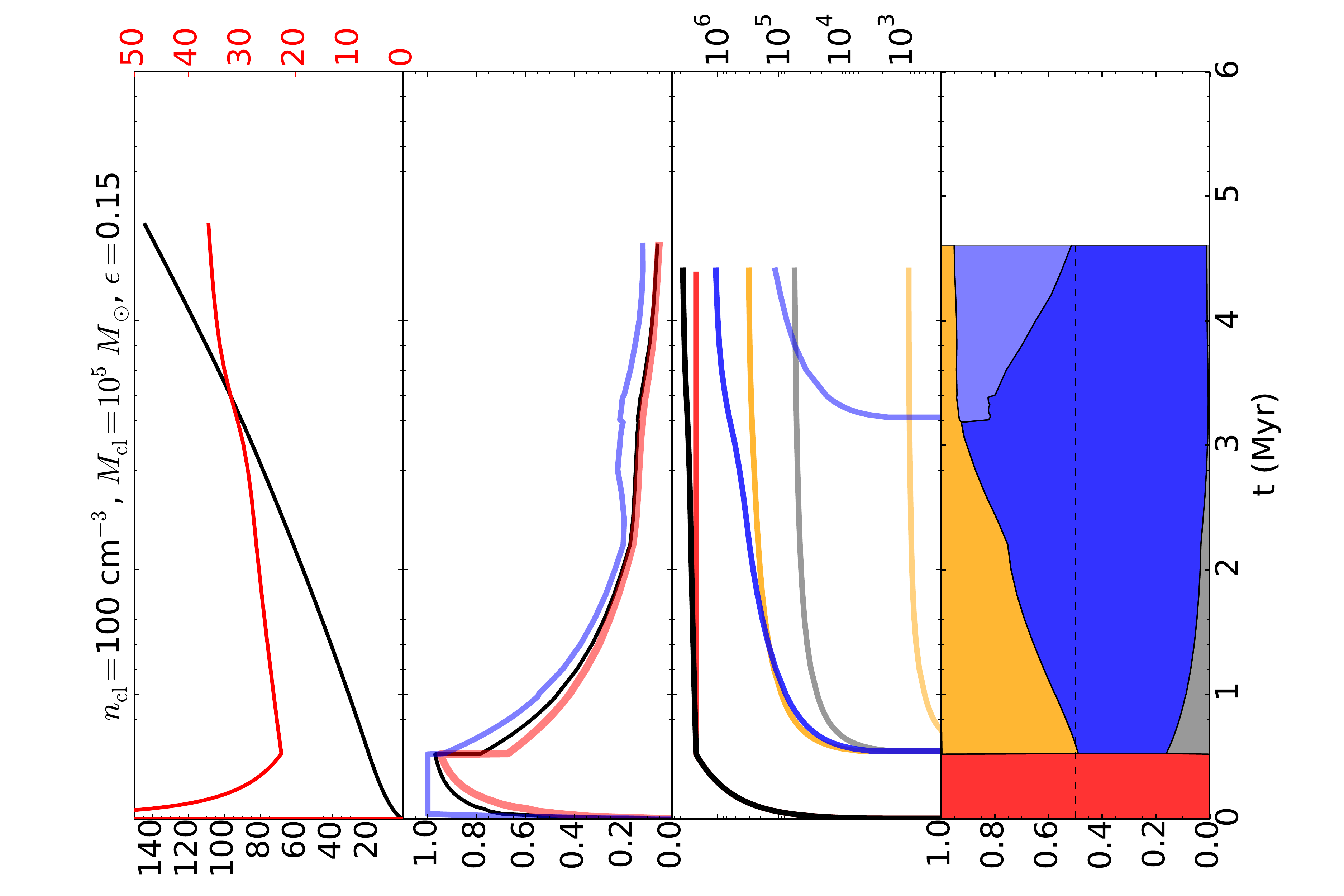}
	\vspace{-0.6mm}\\
	\includegraphics[width=0.33\textwidth, angle = 90]{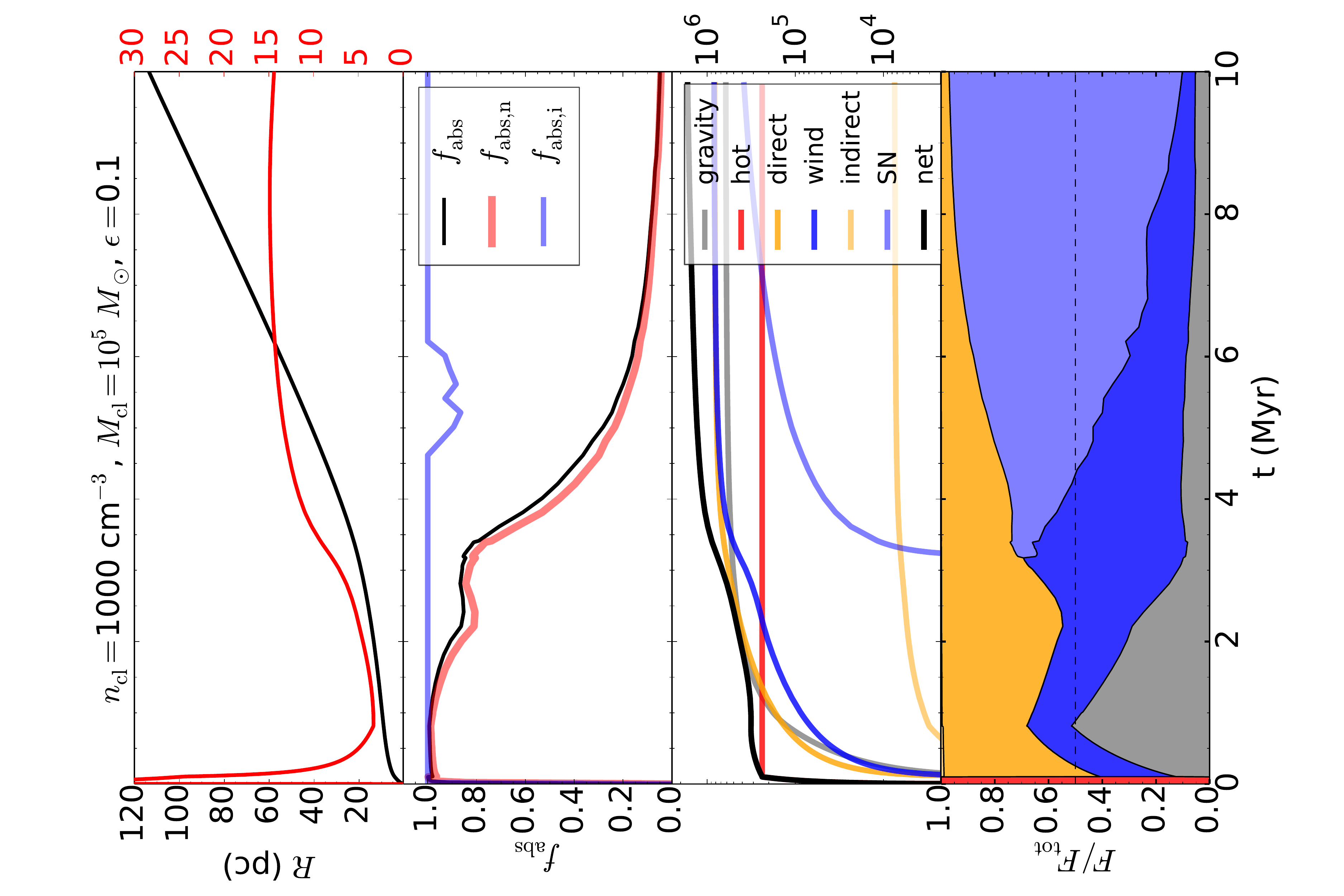}
	\hspace{-4mm}
	\includegraphics[width=0.33\textwidth, angle = 90]{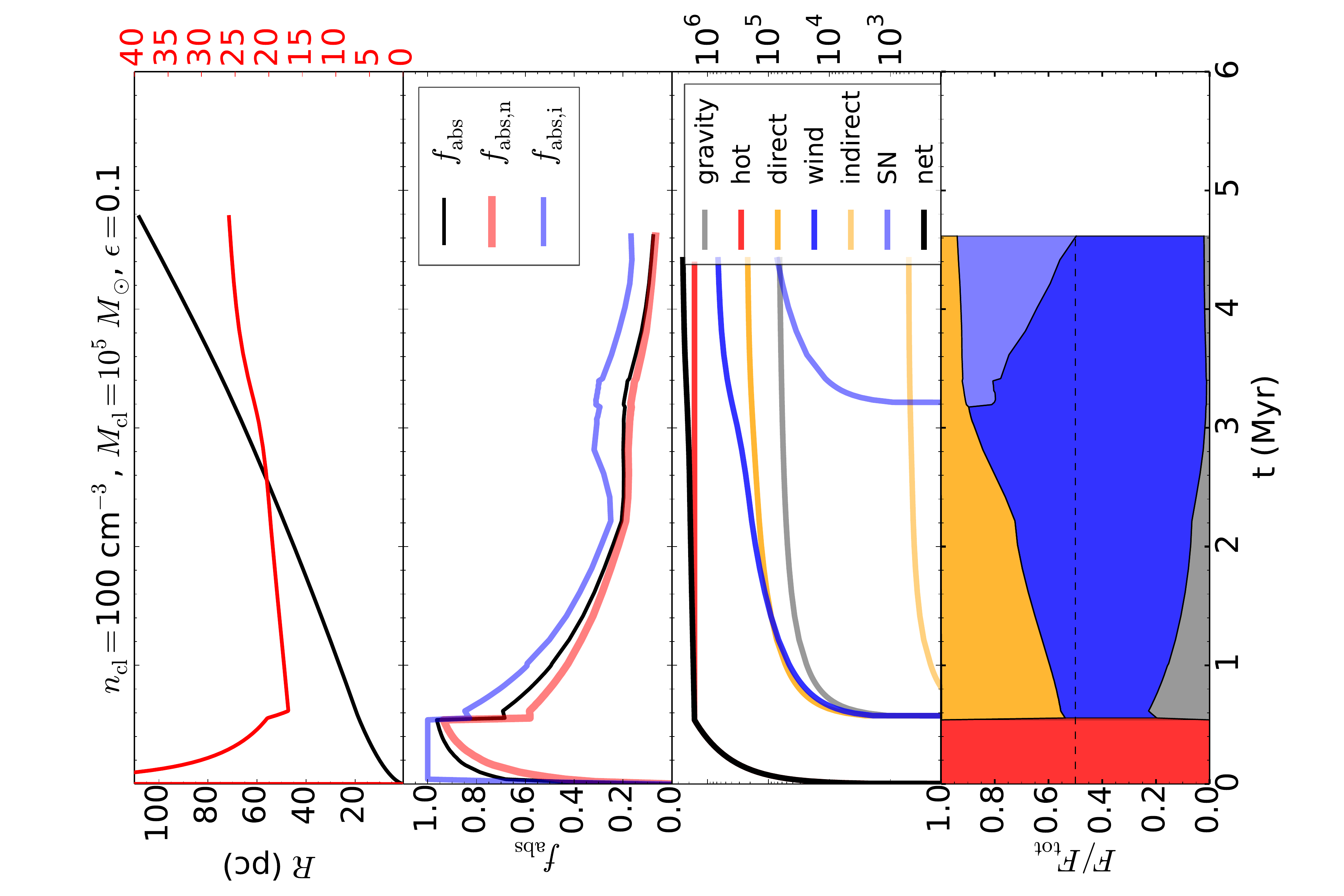}
	\vspace{-0.6mm}
	\caption{Models for clouds with $M_{\rm{cl}} = 10^5\,M_{\odot}$ and $\epsilon = 0.1, 0.15, 0.2,$ and 0.25.}
	\label{fig:Overview5}
\end{figure*}

\begin{figure*}
	\includegraphics[width=0.33\textwidth, angle = 90]{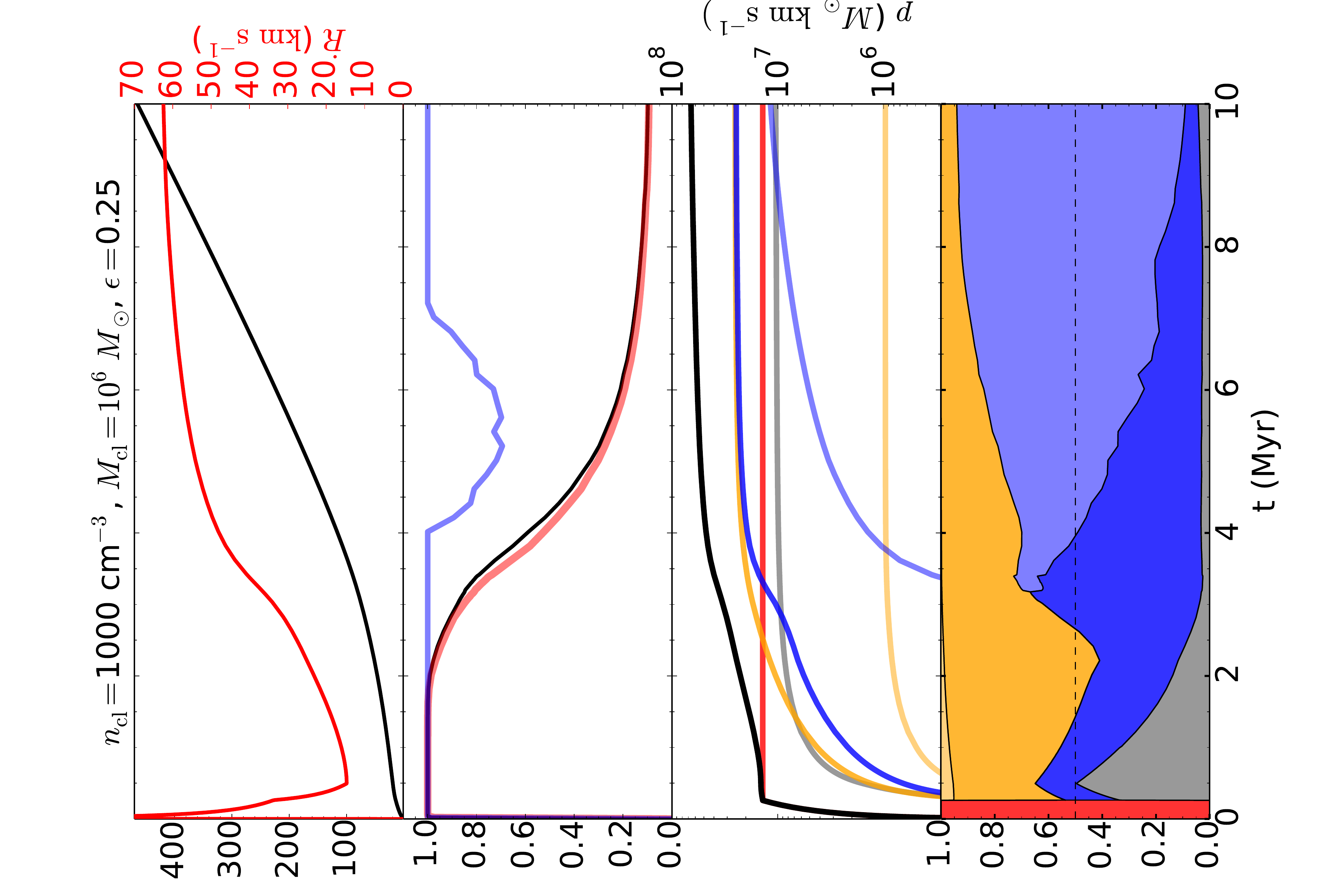}
	\hspace{-4mm}
	\includegraphics[width=0.33\textwidth, angle = 90]{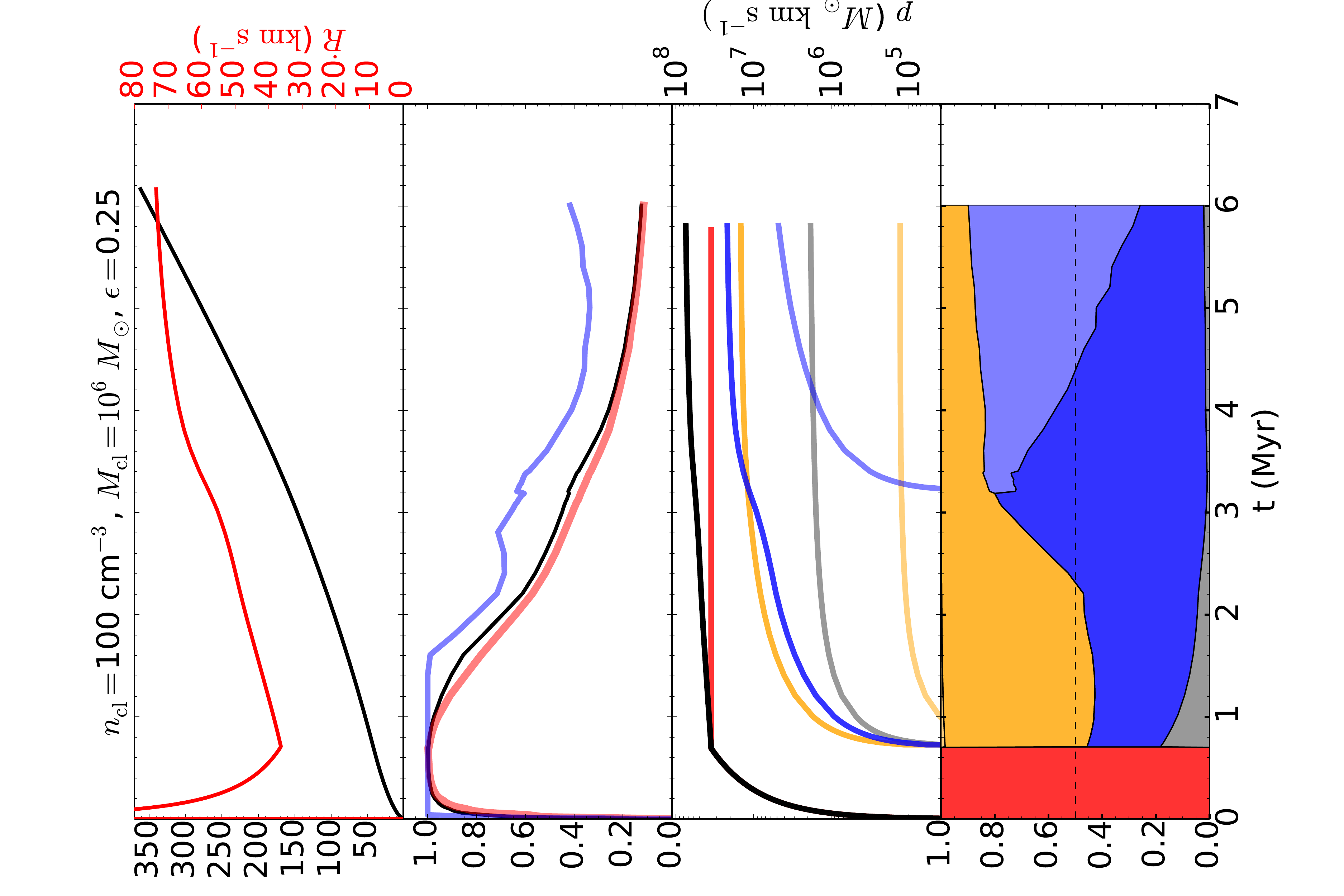}
	\vspace{-0.6mm}\\
	\includegraphics[width=0.33\textwidth, angle = 90]{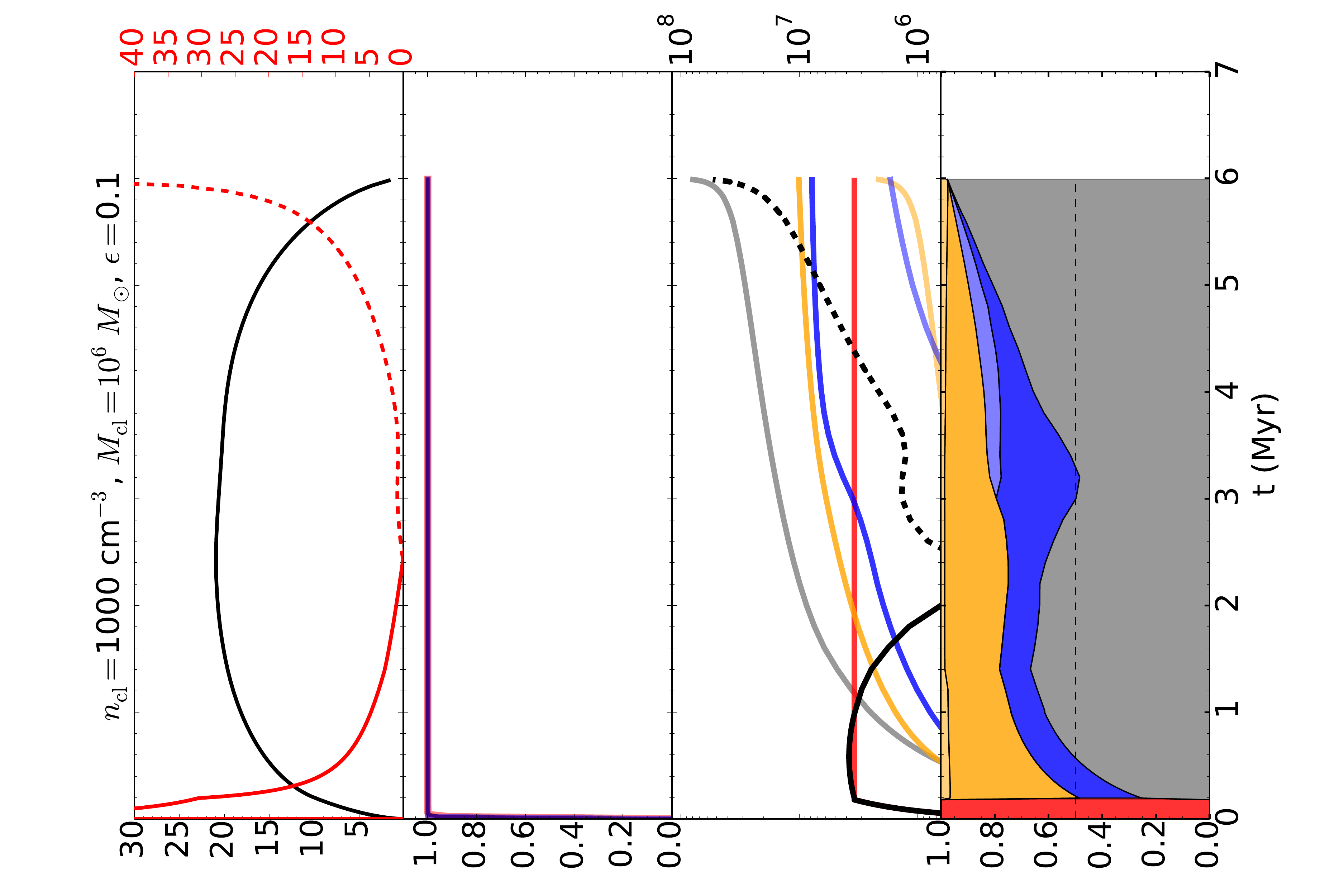}
	\hspace{-4mm}
	\includegraphics[width=0.33\textwidth, angle = 90]{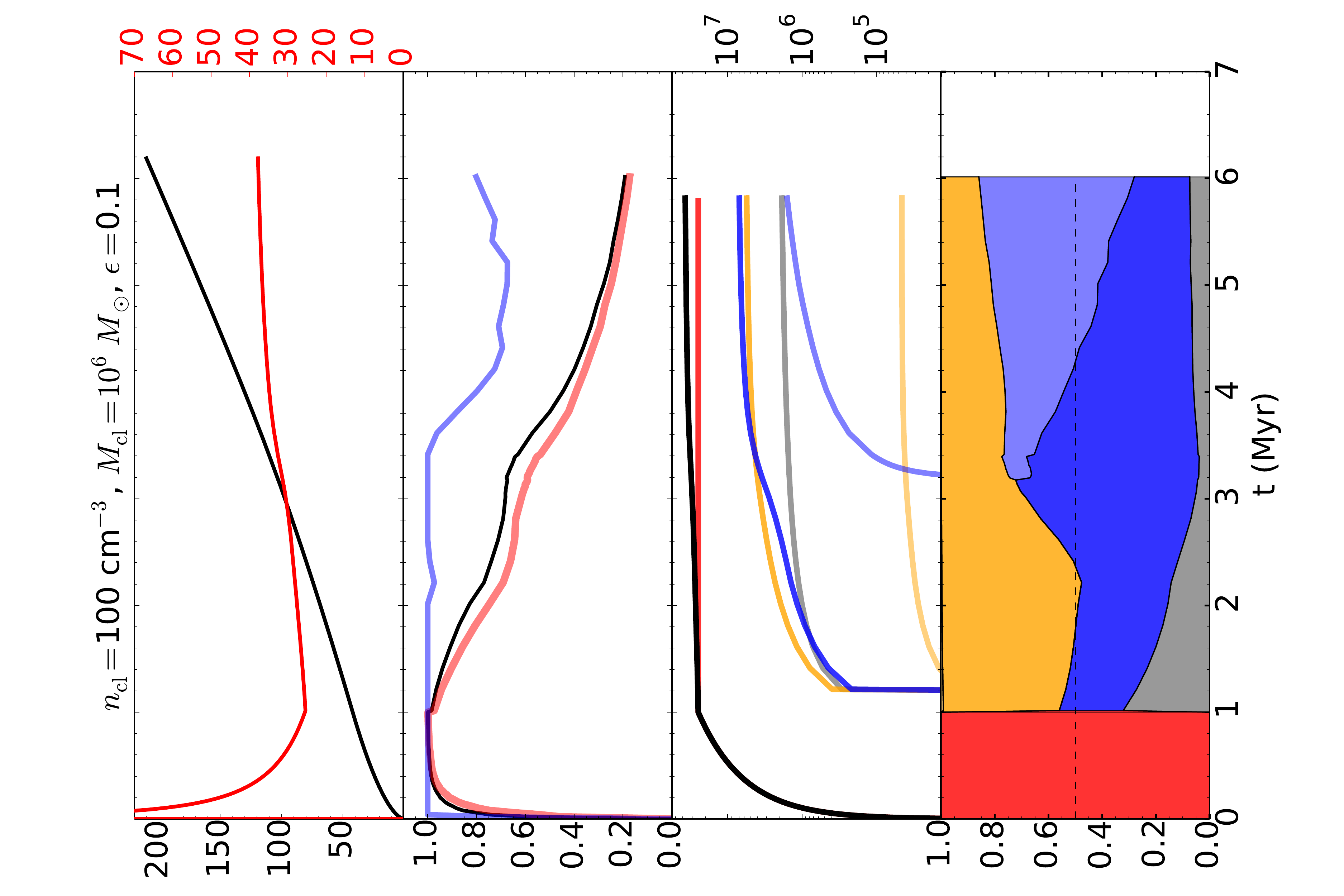}
	\vspace{-0.6mm}
	\includegraphics[width=0.33\textwidth, angle = 90]{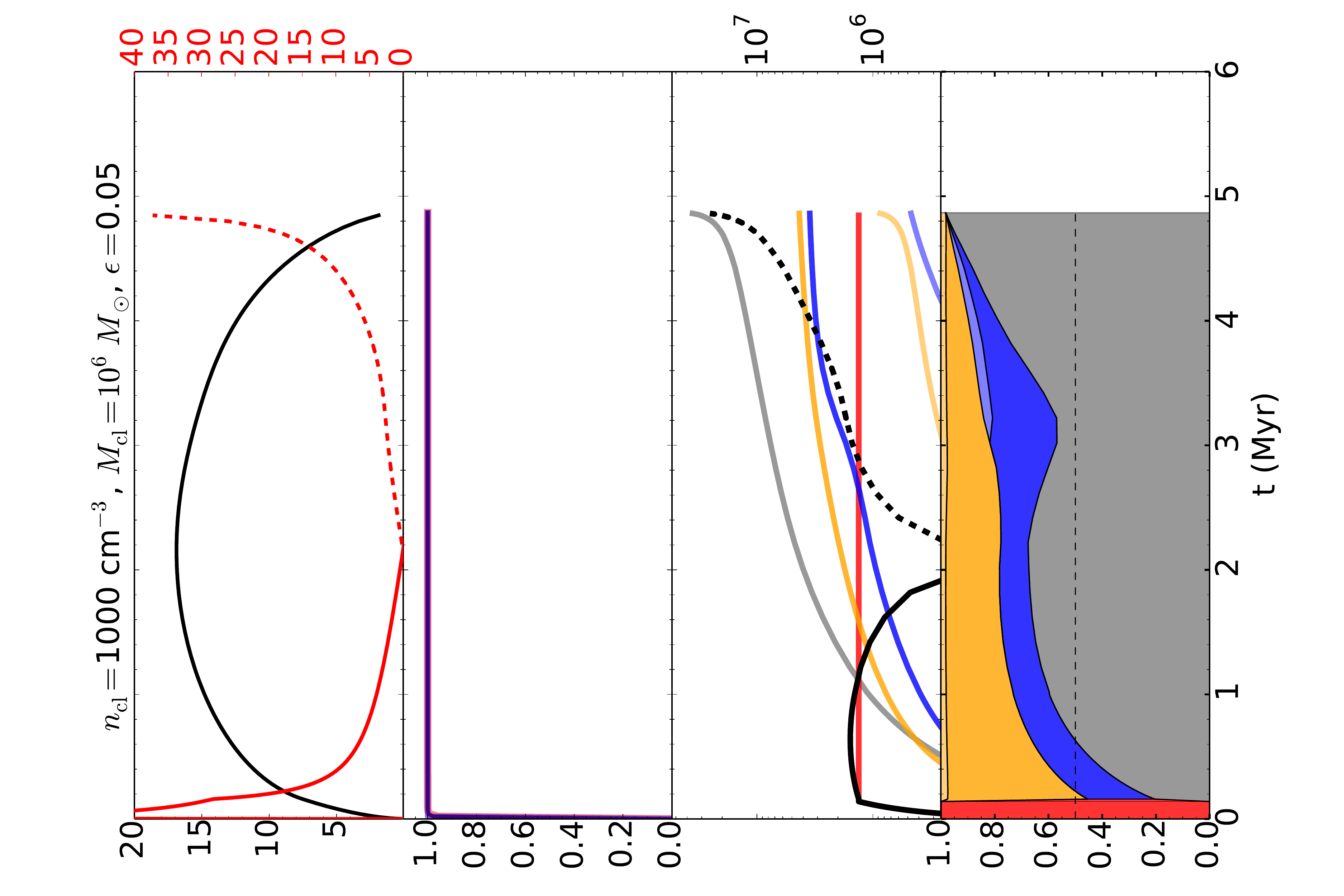}
	\hspace{-4mm}
	\includegraphics[width=0.33\textwidth, angle = 90]{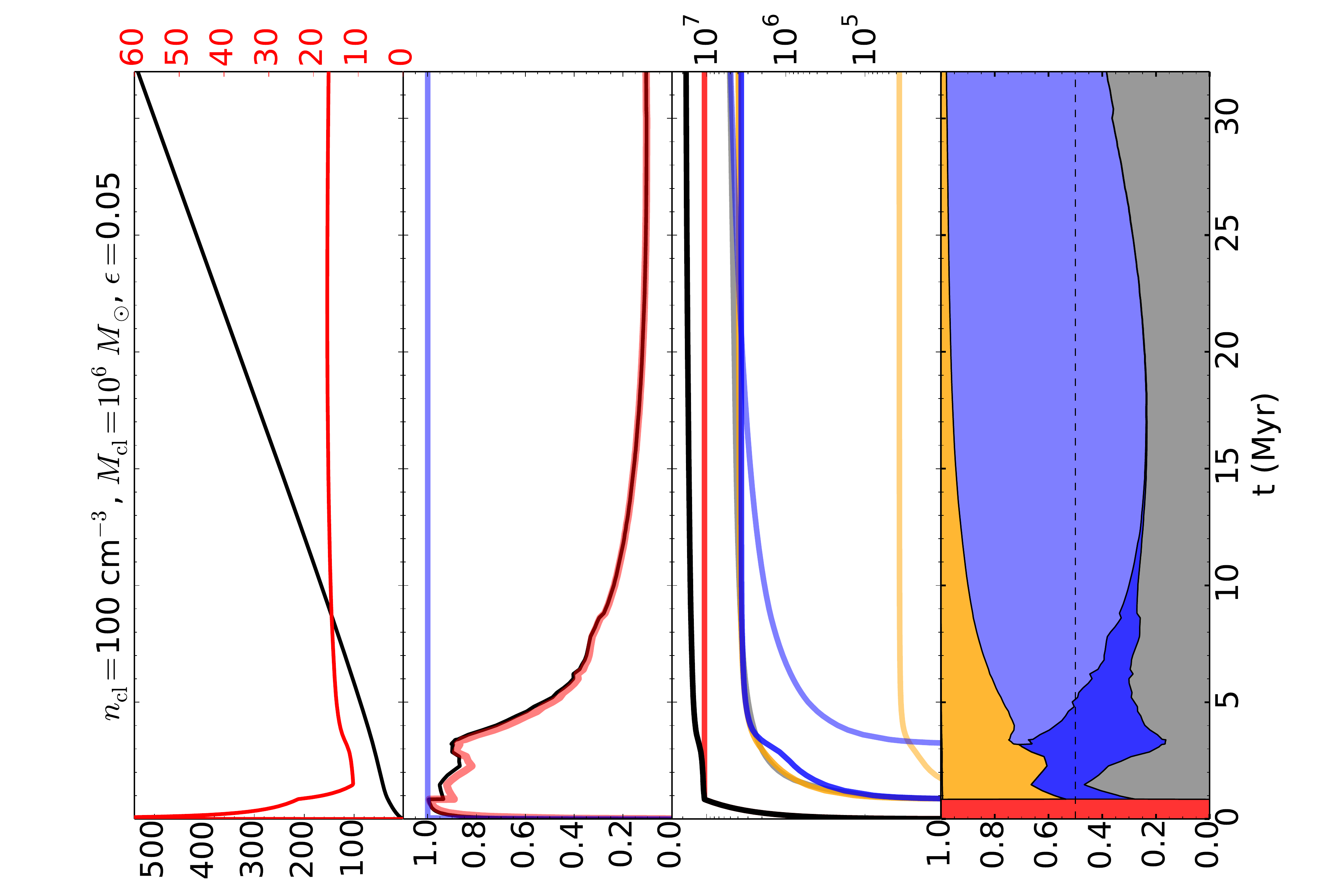}
	\vspace{-0.6mm}\\
	\includegraphics[width=0.33\textwidth, angle = 90]{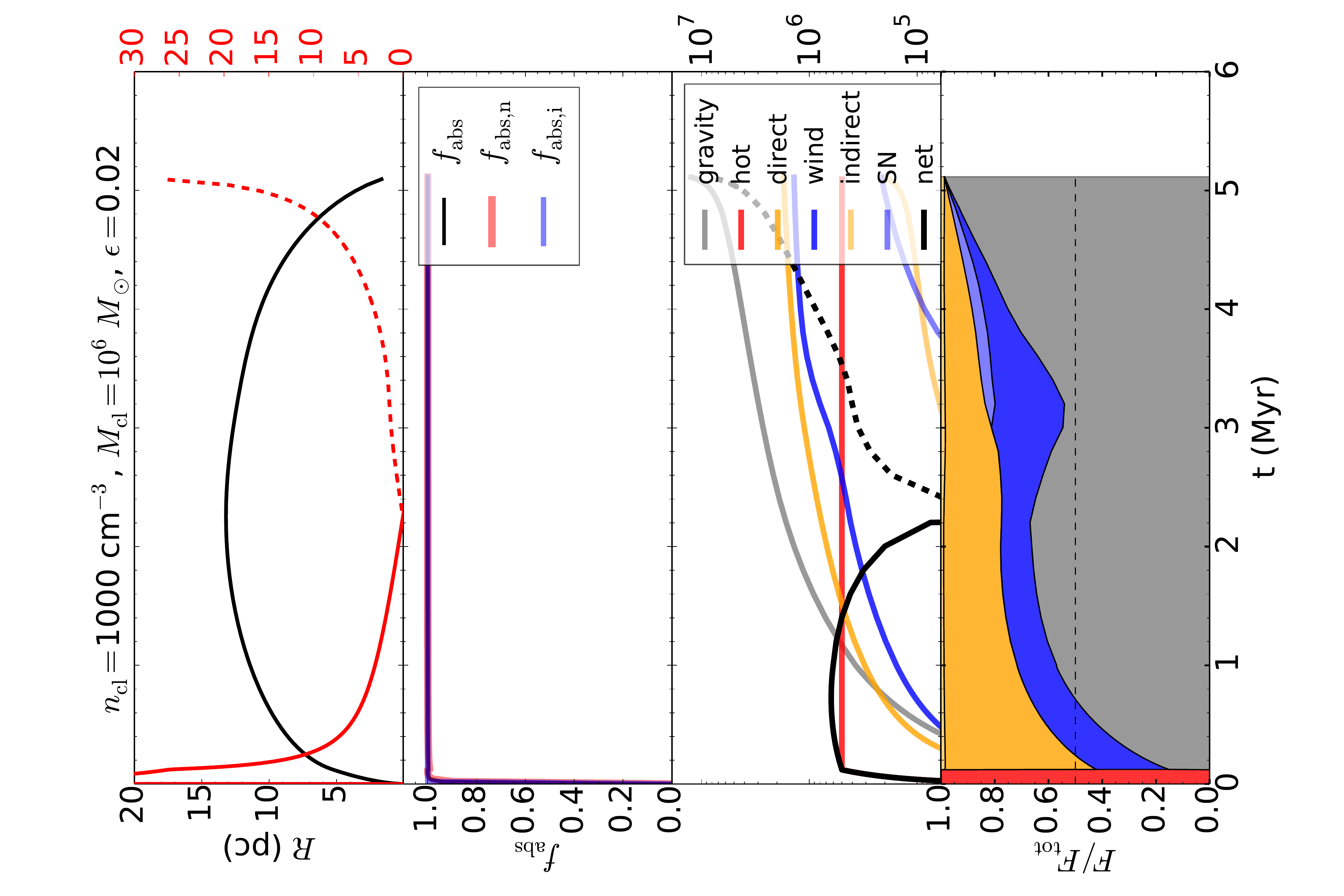}
	\hspace{-4mm}
	\includegraphics[width=0.33\textwidth, angle = 90]{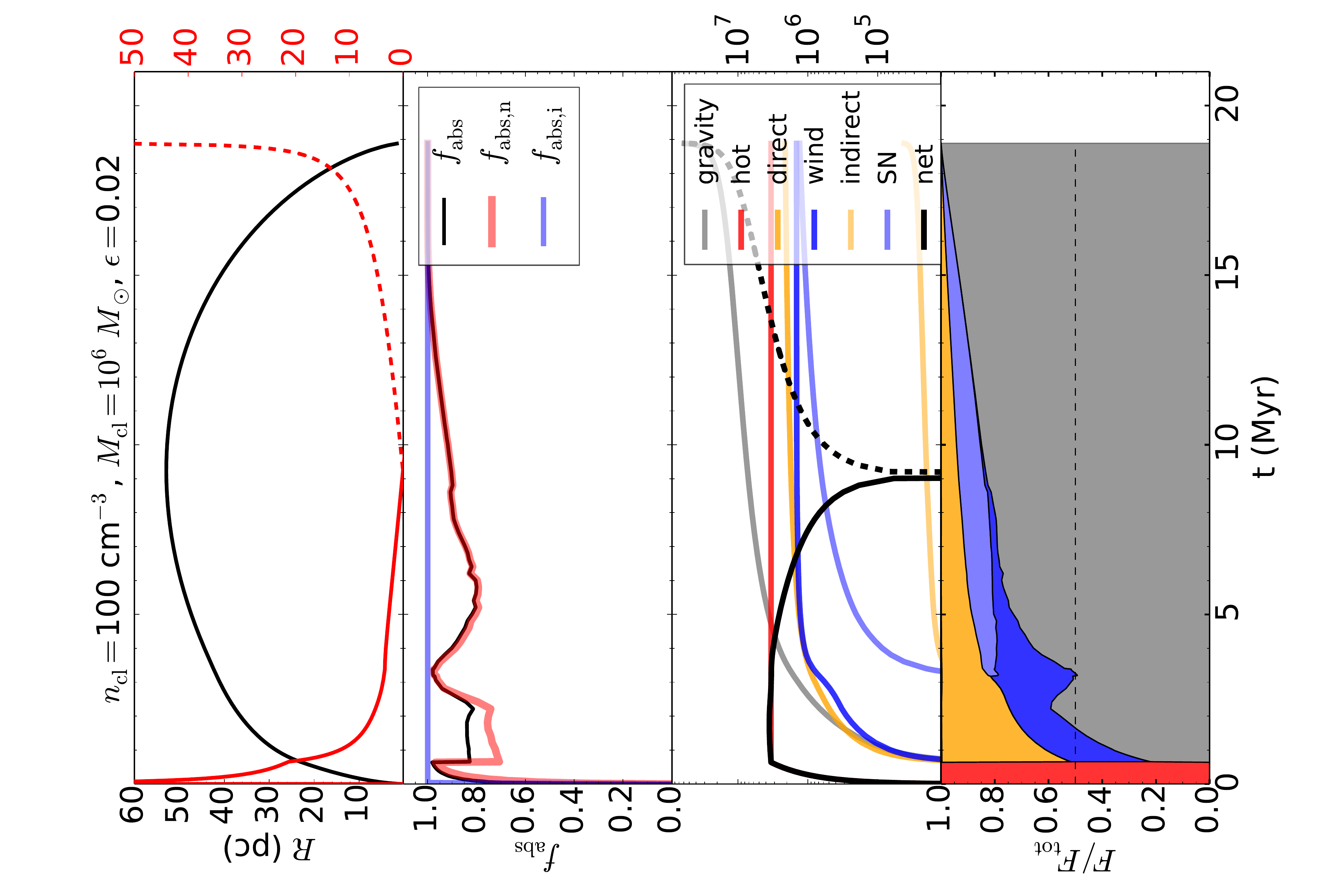}
	\vspace{-0.6mm}
		\caption{Models for clouds with $M_{\rm{cl}} = 10^6\,M_{\odot}$ and $\epsilon = 0.02, 0.05, 0.1,$ and 0.25.}
		\label{fig:Overview6}
\end{figure*}

\begin{figure*}
	\includegraphics[width=0.33\textwidth, angle = 90]{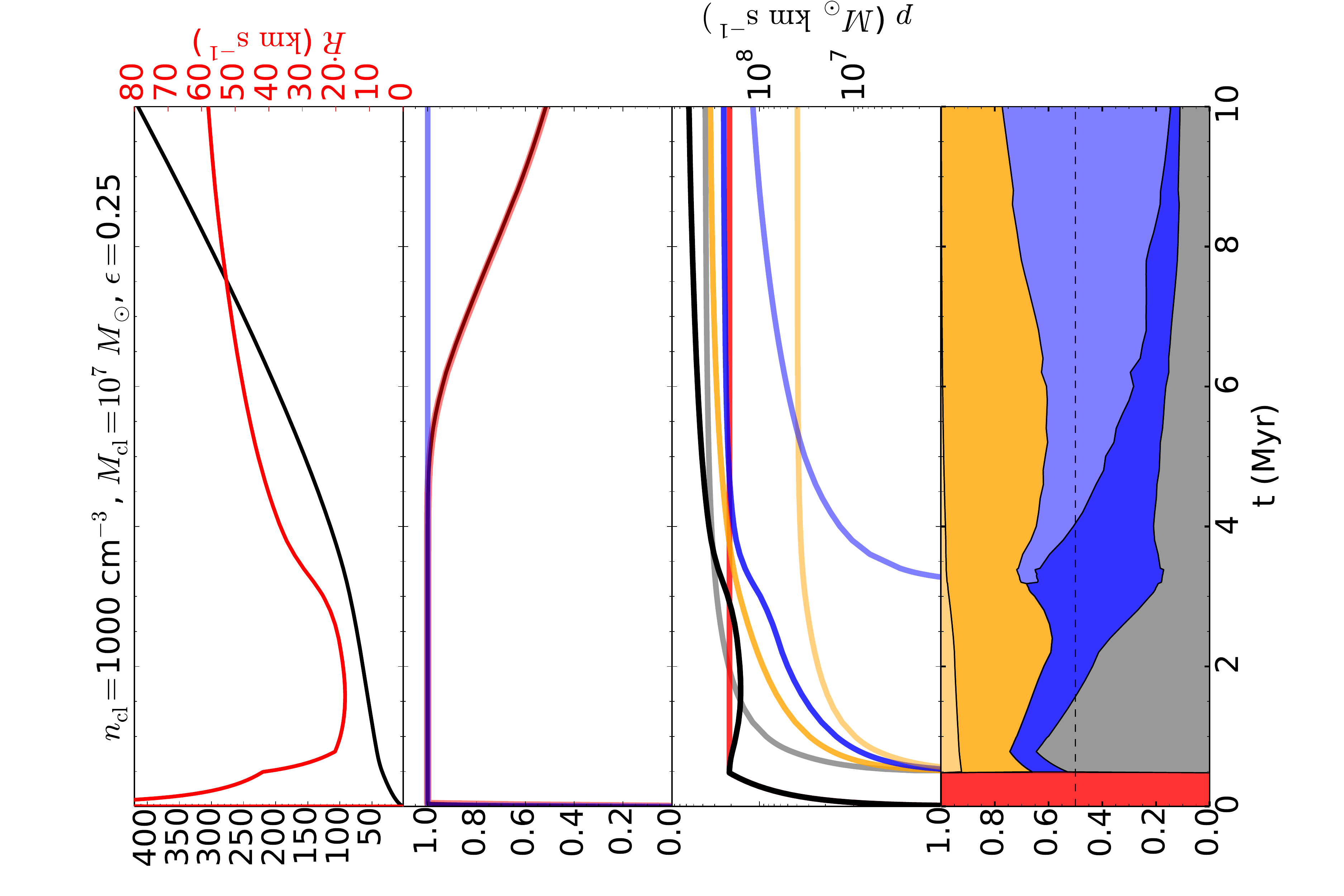}
	\hspace{-4mm}
	\includegraphics[width=0.33\textwidth, angle = 90]{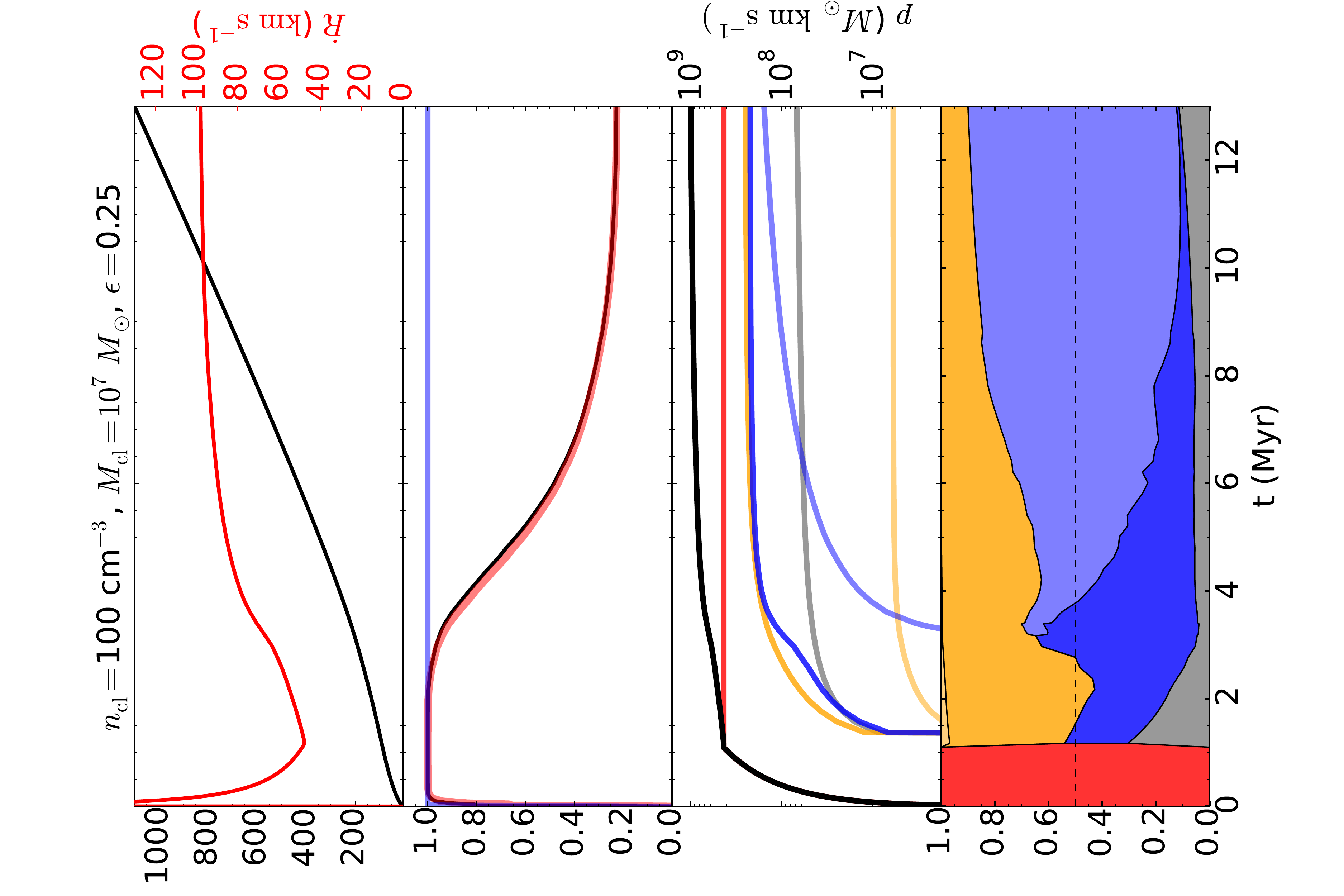}
	\vspace{-0.6mm}\\
	\includegraphics[width=0.33\textwidth, angle = 90]{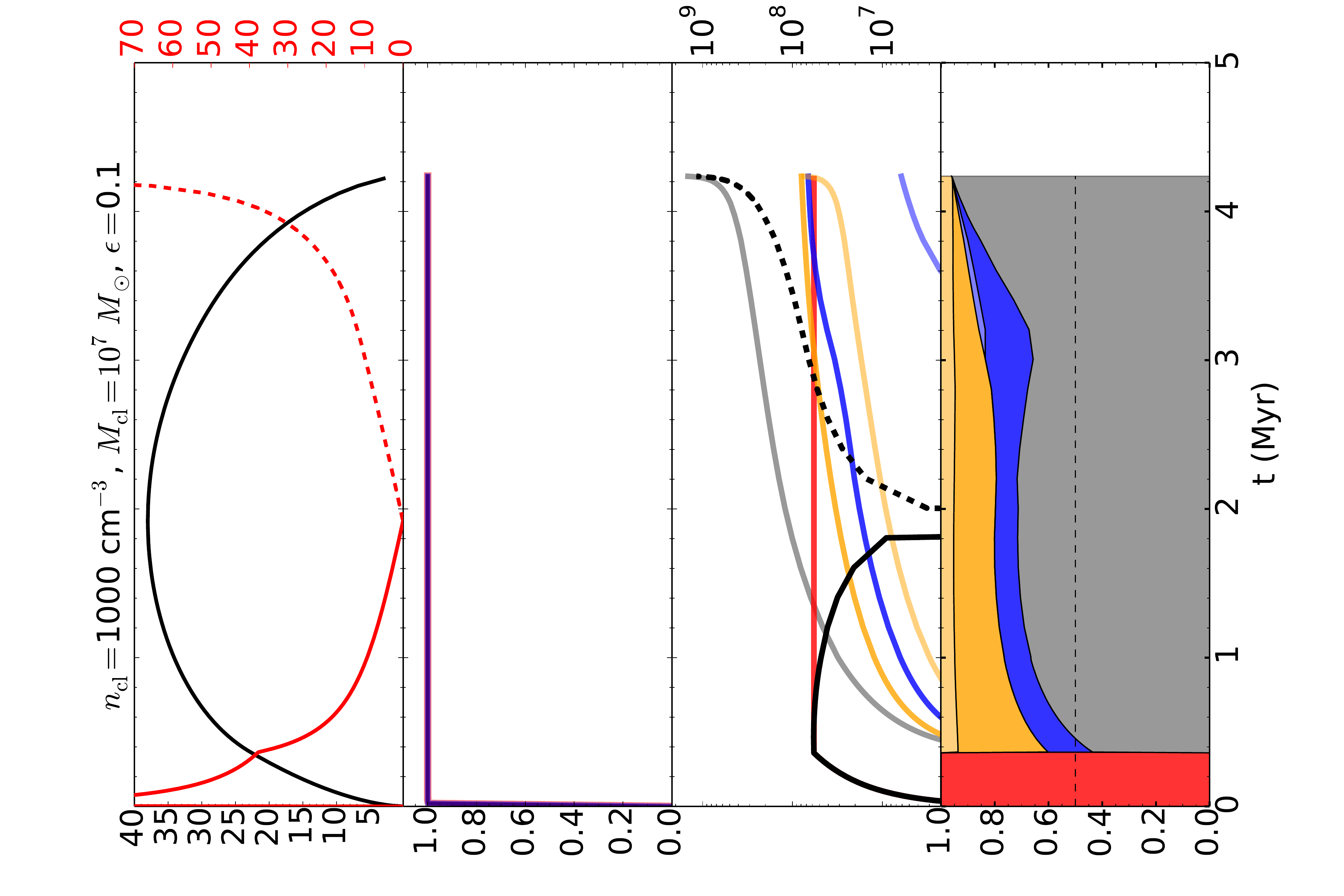}
	\hspace{-4mm}
	\includegraphics[width=0.33\textwidth, angle = 90]{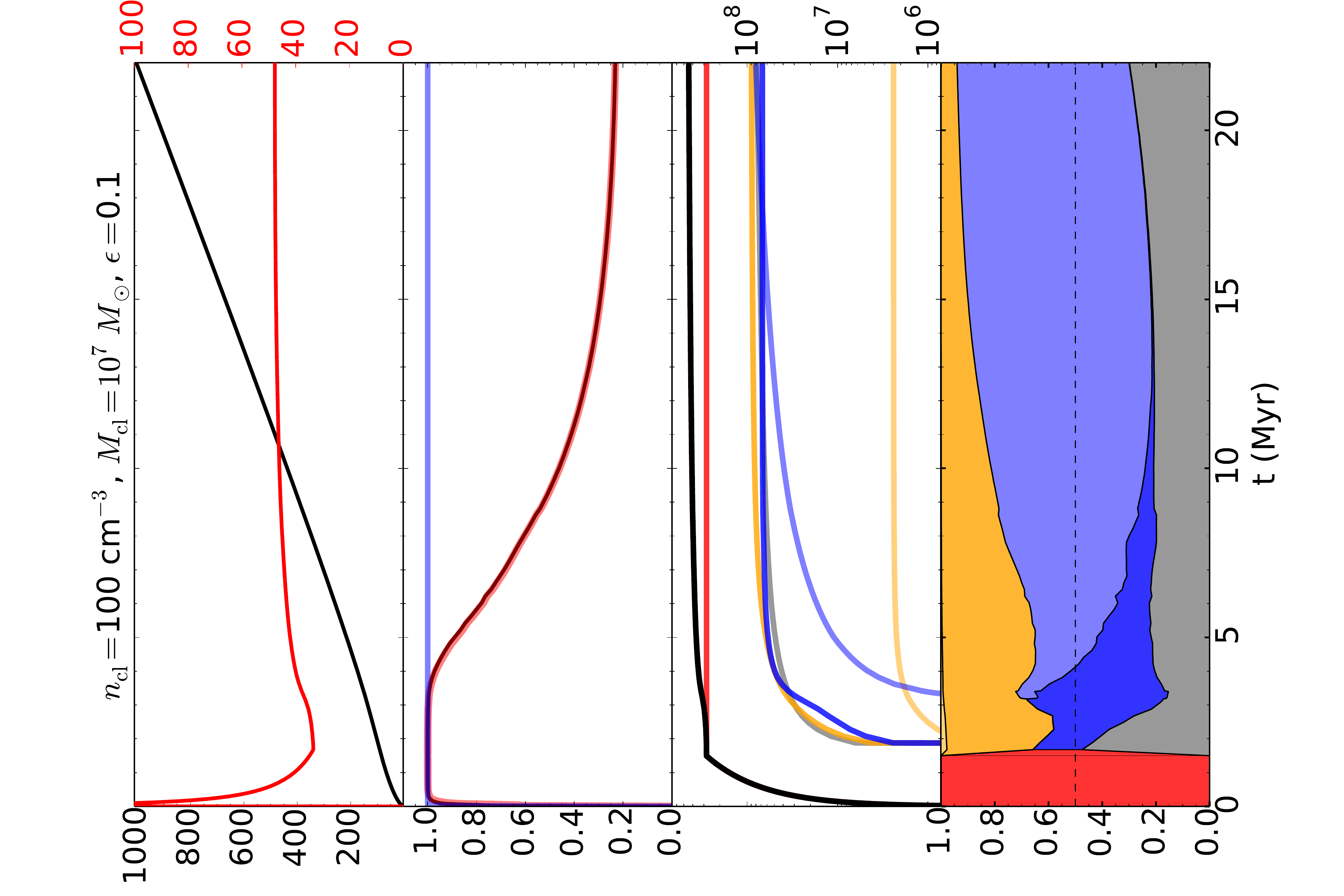}
	\vspace{-0.6mm}
	\includegraphics[width=0.33\textwidth, angle = 90]{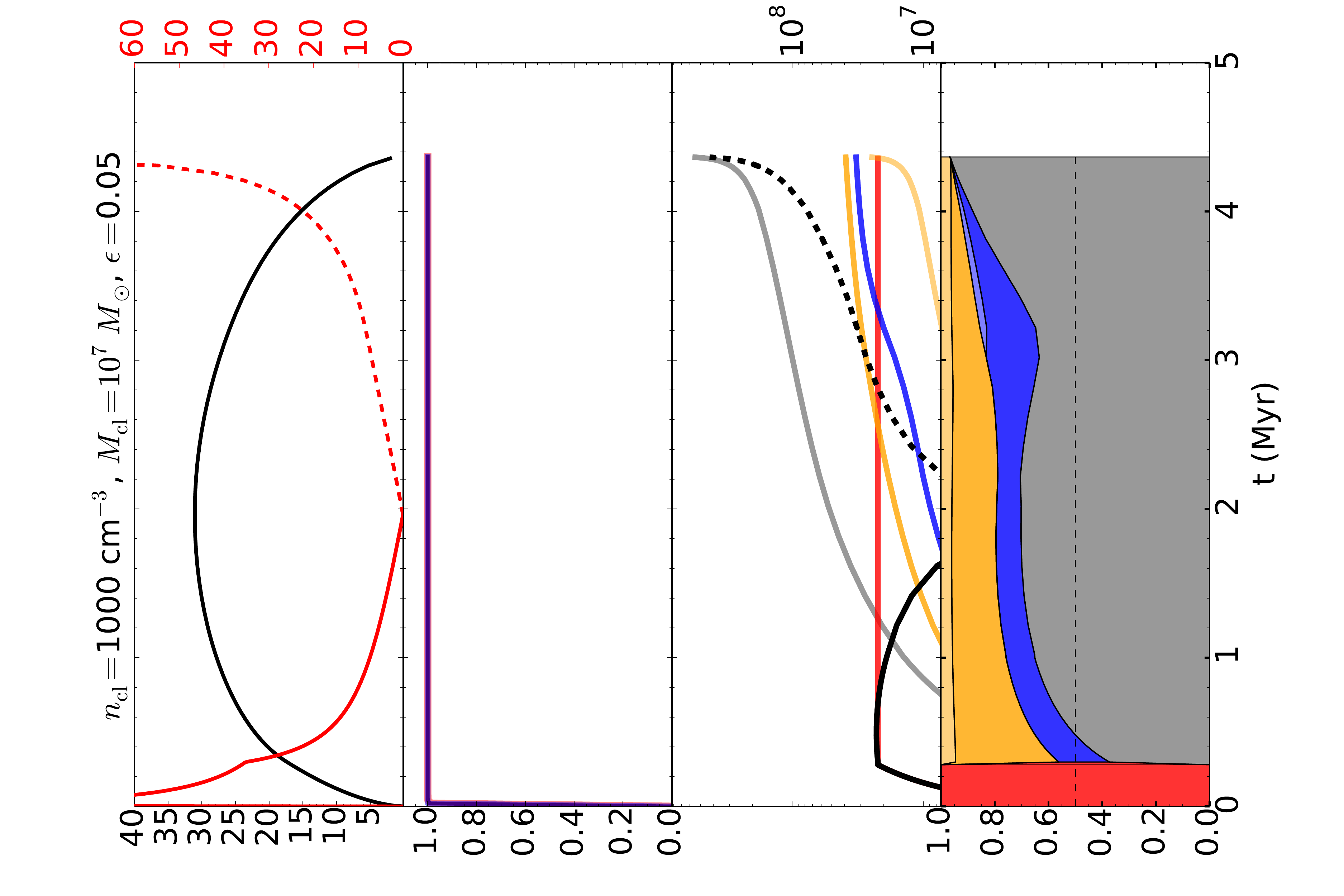}
	\hspace{-4mm}
	\includegraphics[width=0.33\textwidth, angle = 90]{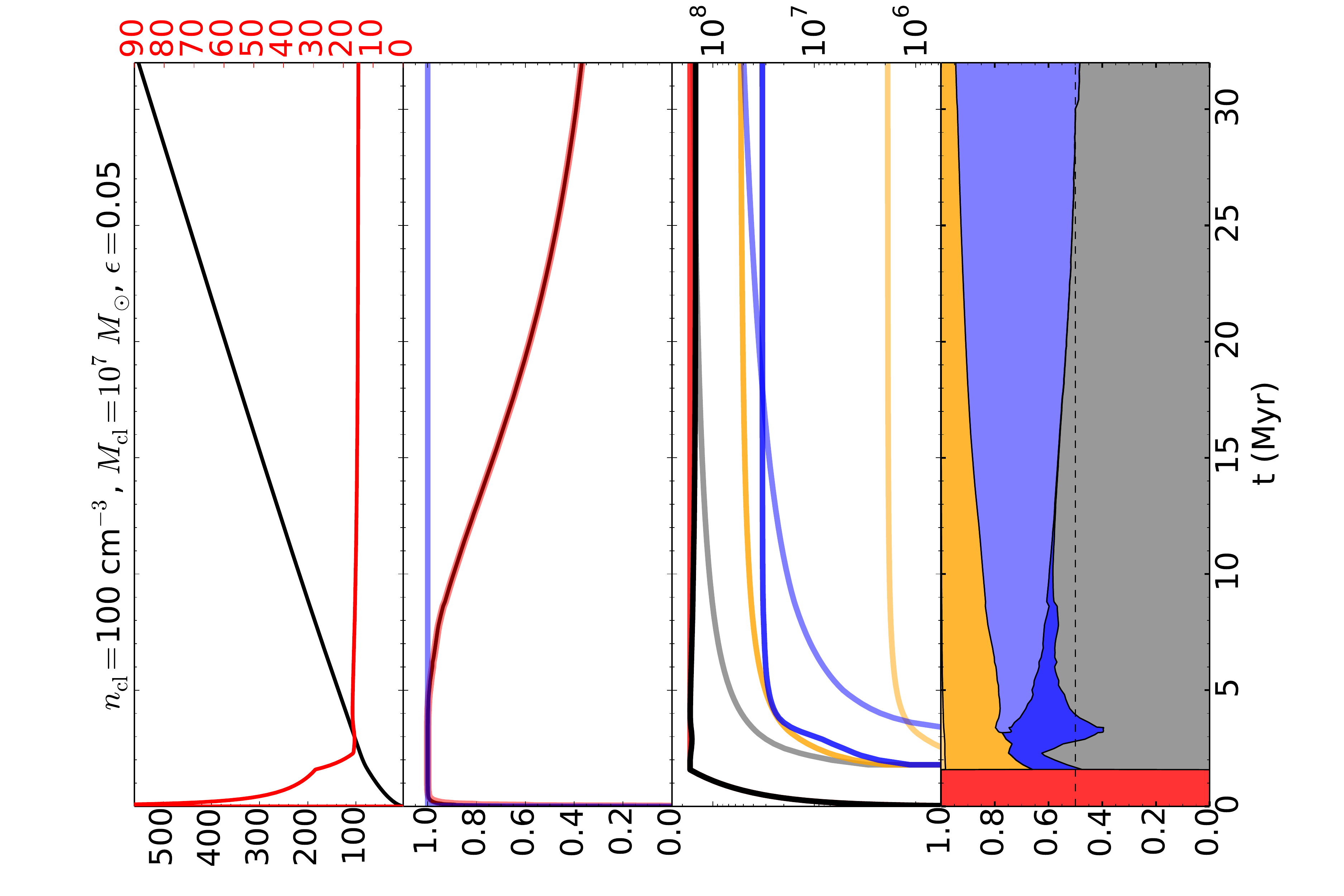}
	\vspace{-0.6mm}\\
	\includegraphics[width=0.33\textwidth, angle = 90]{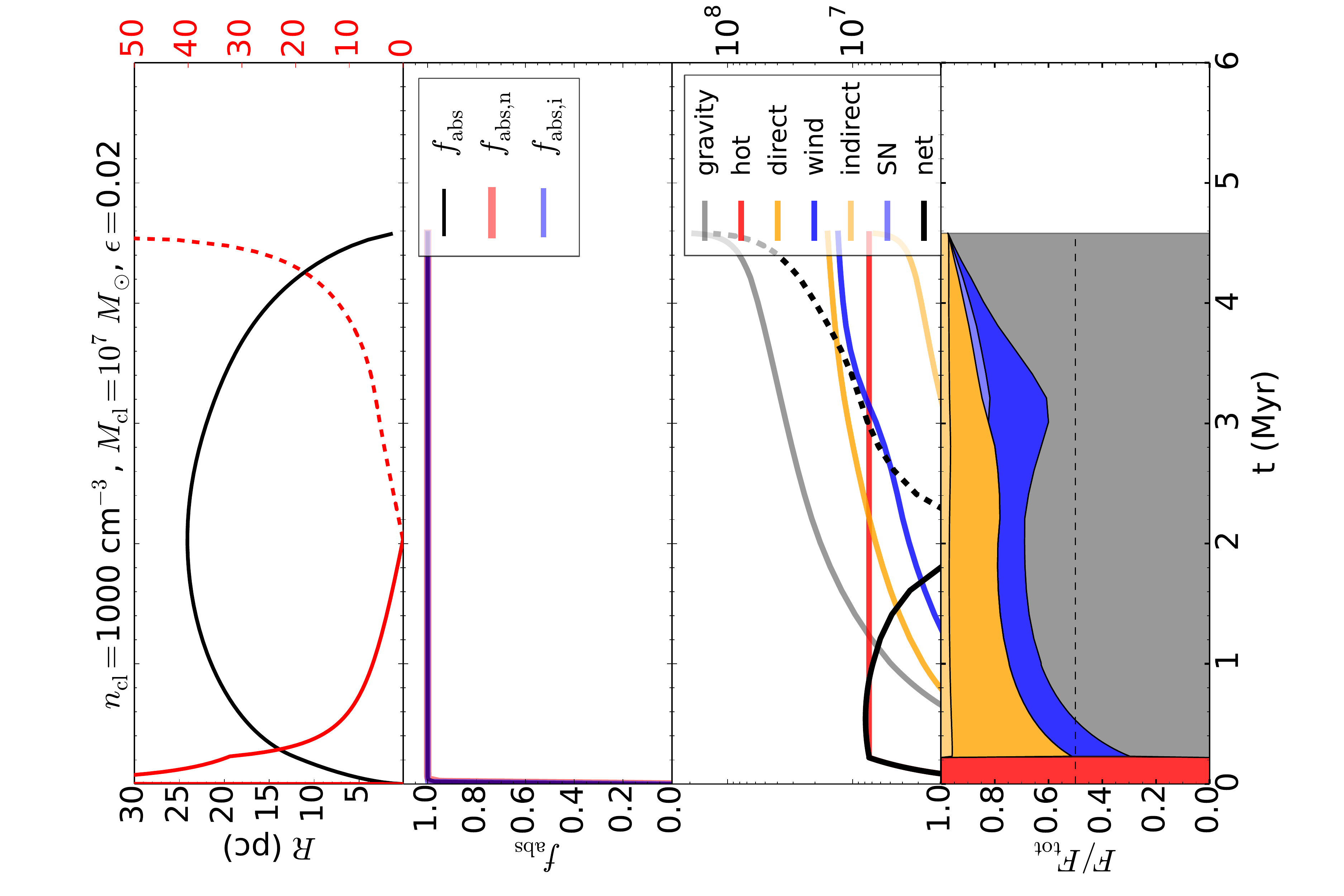}
	\hspace{-4mm}
	\includegraphics[width=0.33\textwidth, angle = 90]{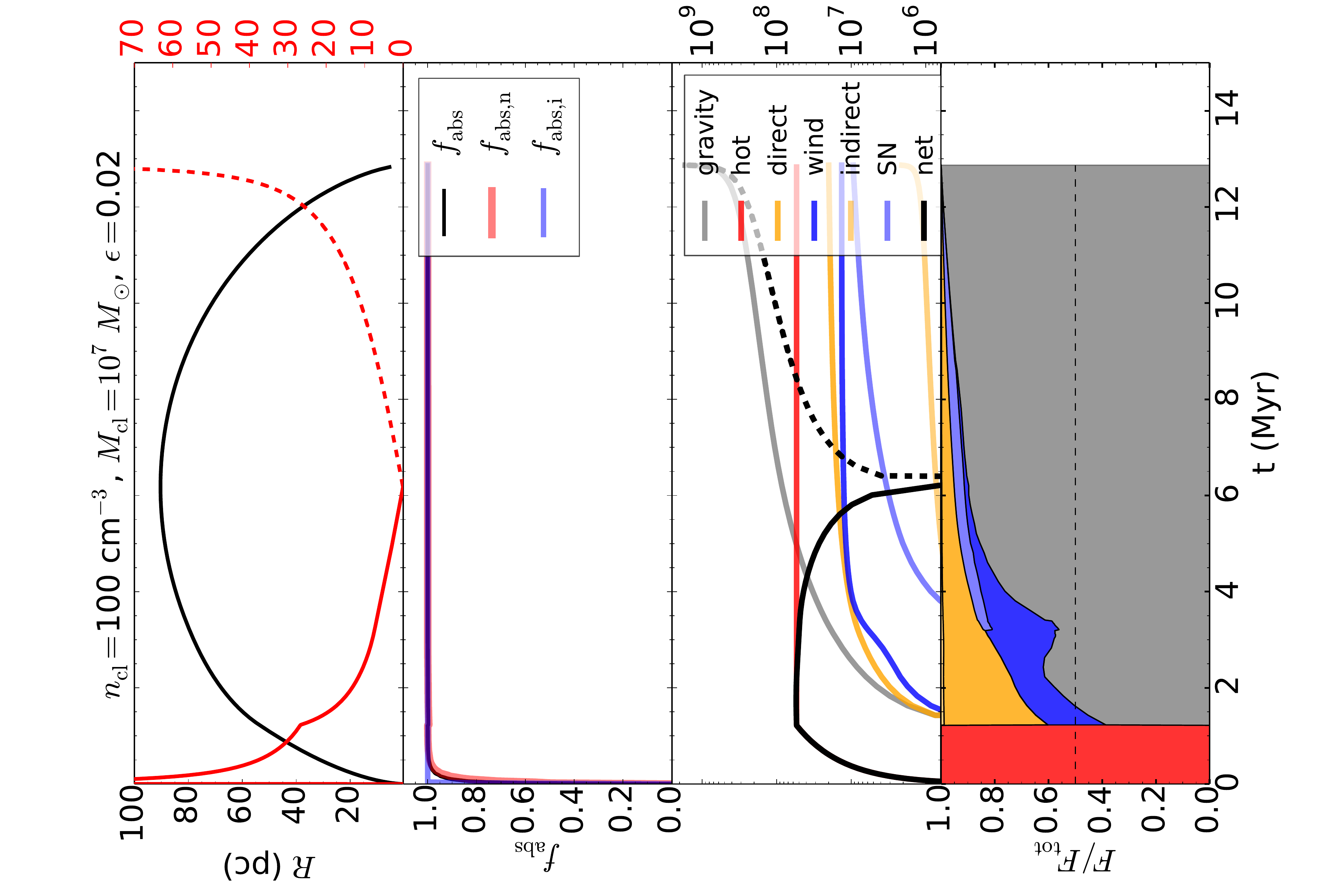}
	\vspace{-0.6mm}
		\caption{Models for clouds with $M_{\rm{cl}} = 10^7\,M_{\odot}$ and $\epsilon = 0.02, 0.05, 0.1,$ and 0.25.}
		\label{fig:Overview7}
\end{figure*}

\begin{figure*}
	\includegraphics[width=0.33\textwidth, angle = 90]{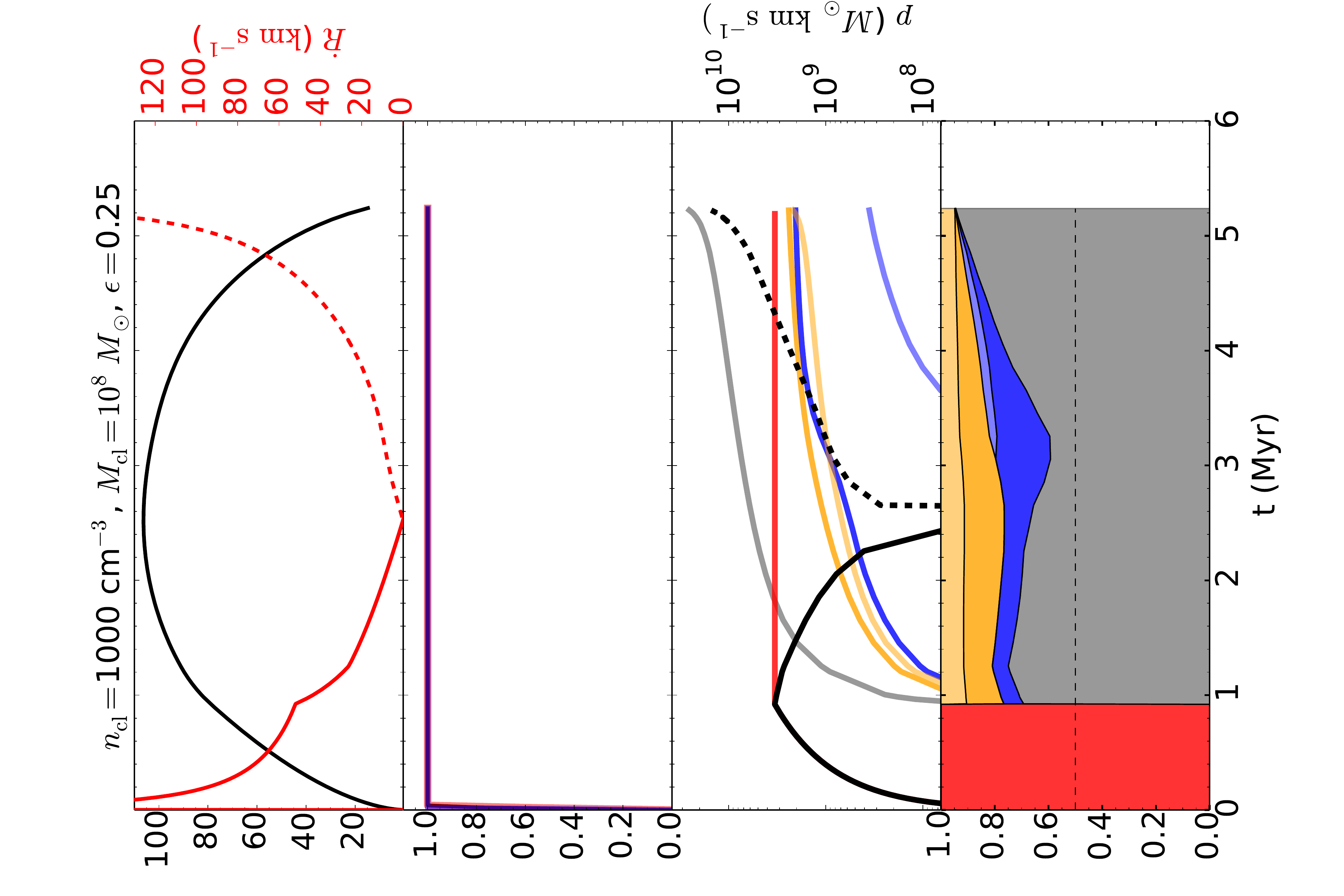}
	\hspace{-4mm}
	\includegraphics[width=0.33\textwidth, angle = 90]{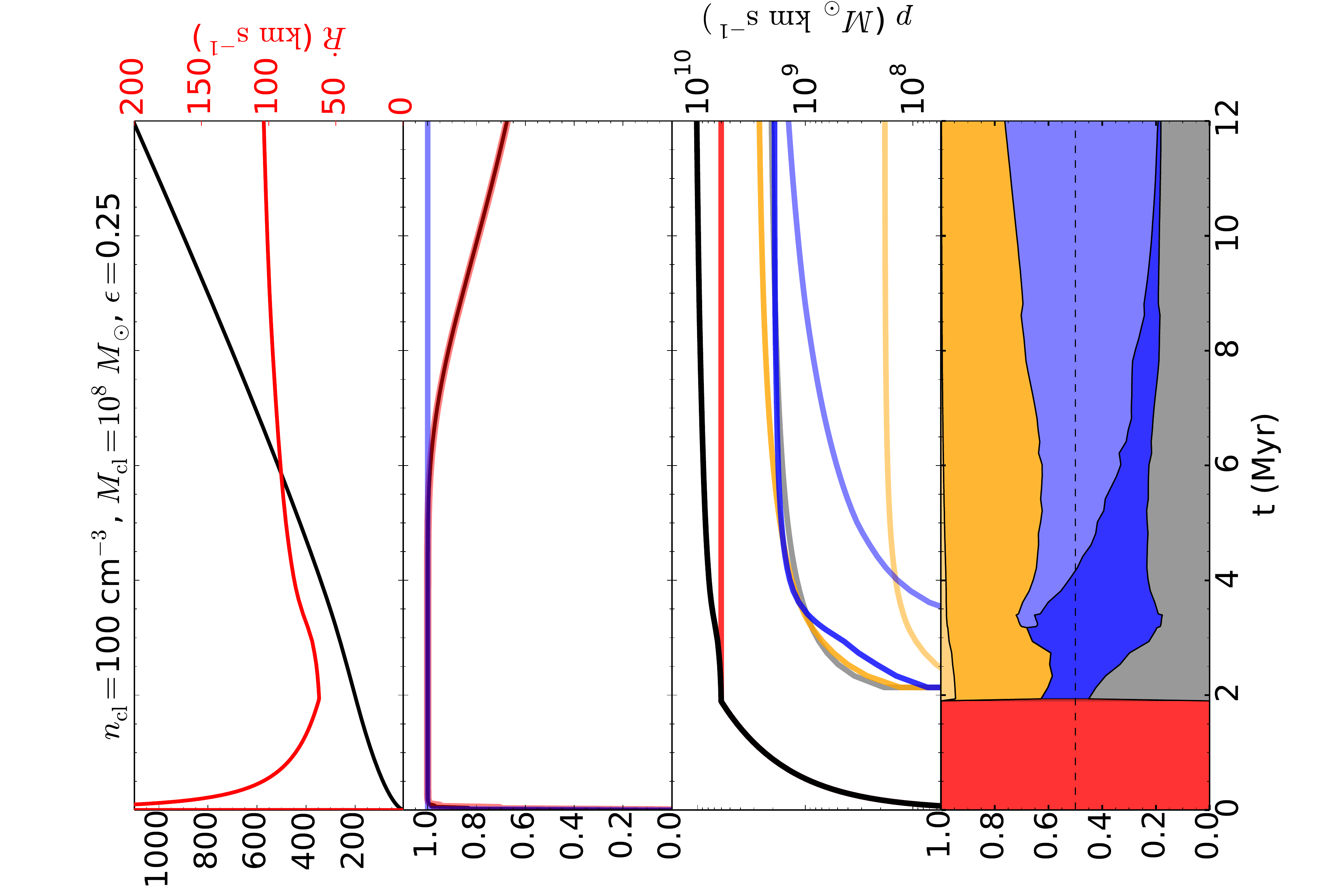}
	\vspace{-0.6mm}\\
	\includegraphics[width=0.33\textwidth, angle = 90]{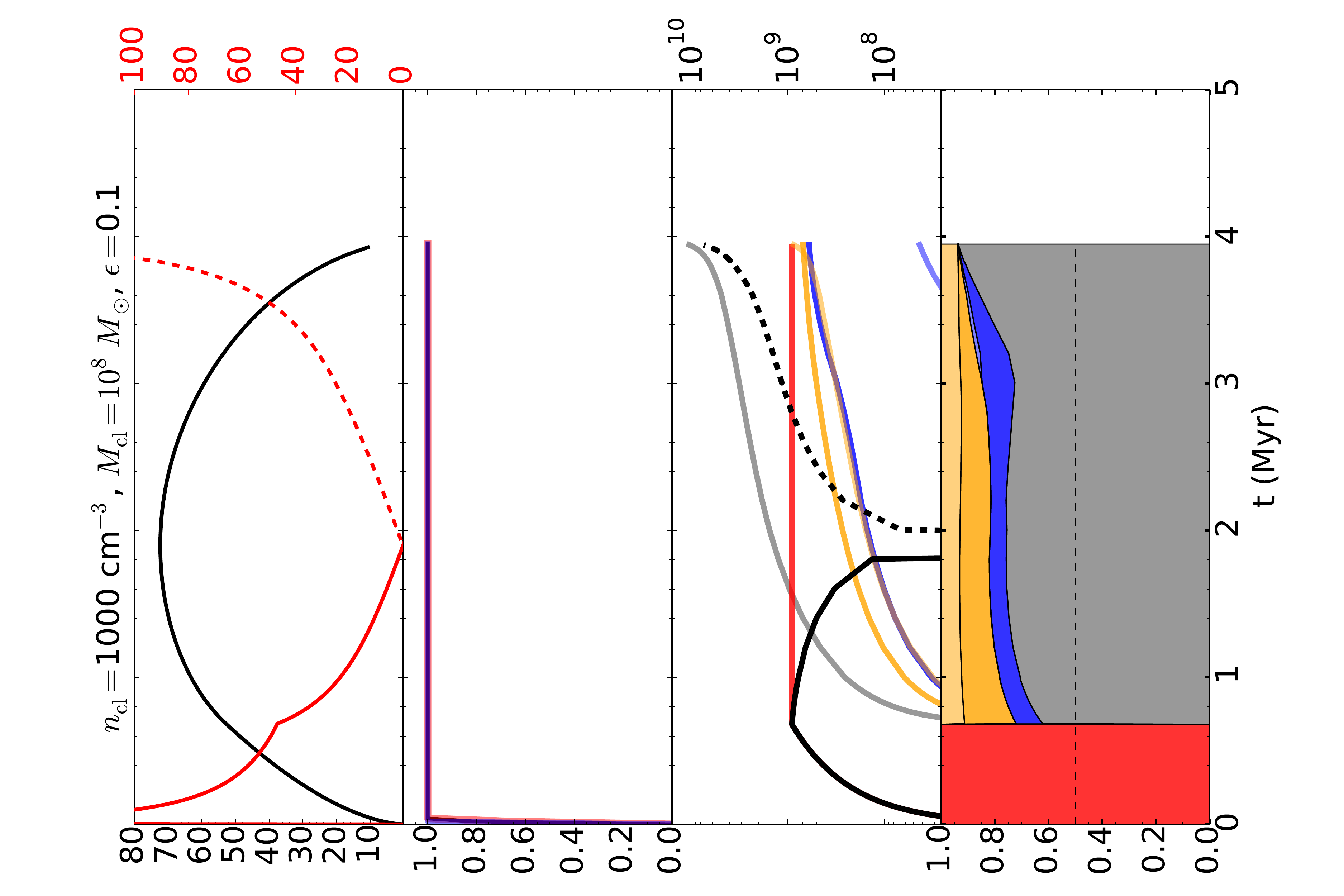}
	\hspace{-4mm}
	\includegraphics[width=0.33\textwidth, angle = 90]{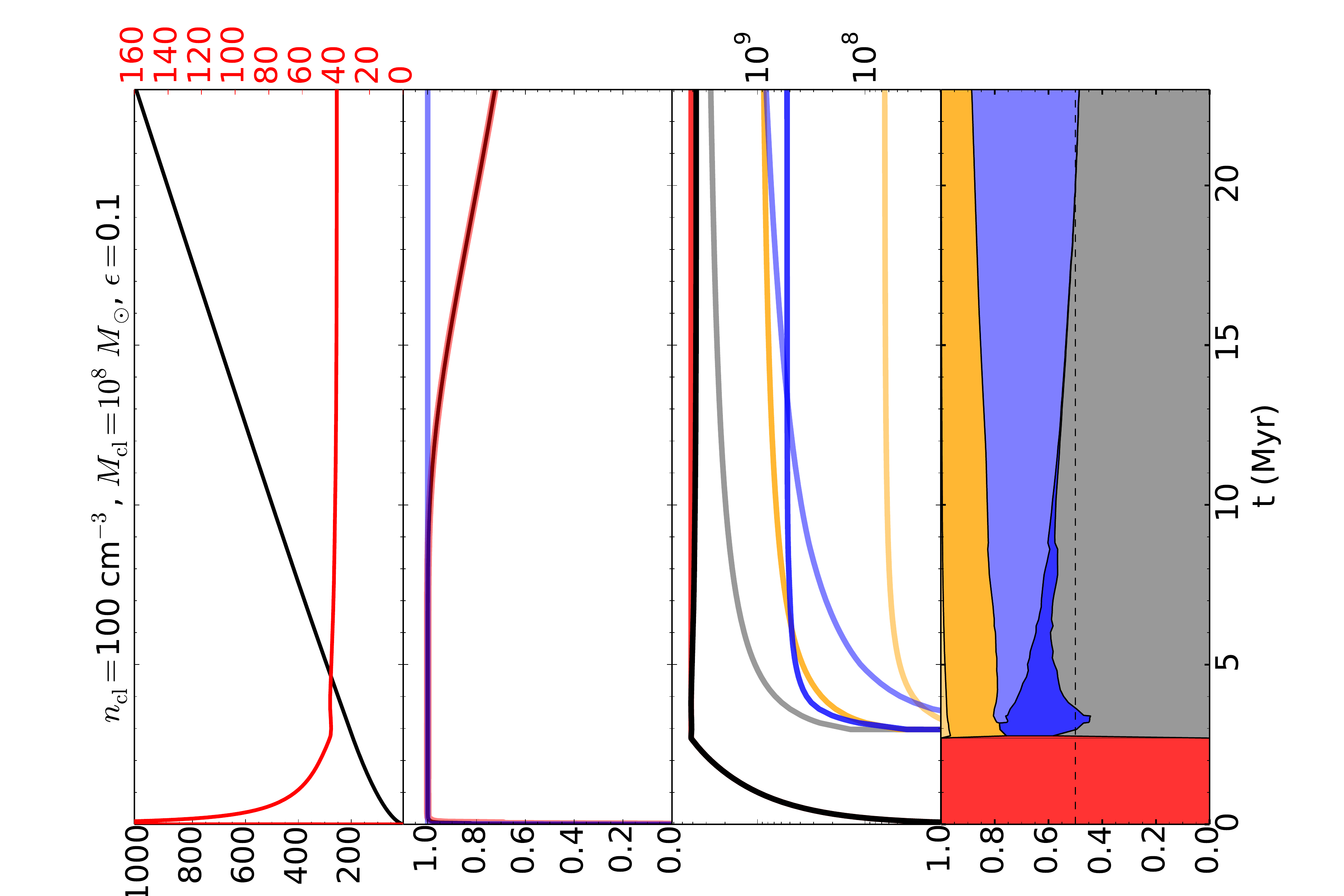}
	\vspace{-0.6mm}
	\includegraphics[width=0.33\textwidth, angle = 90]{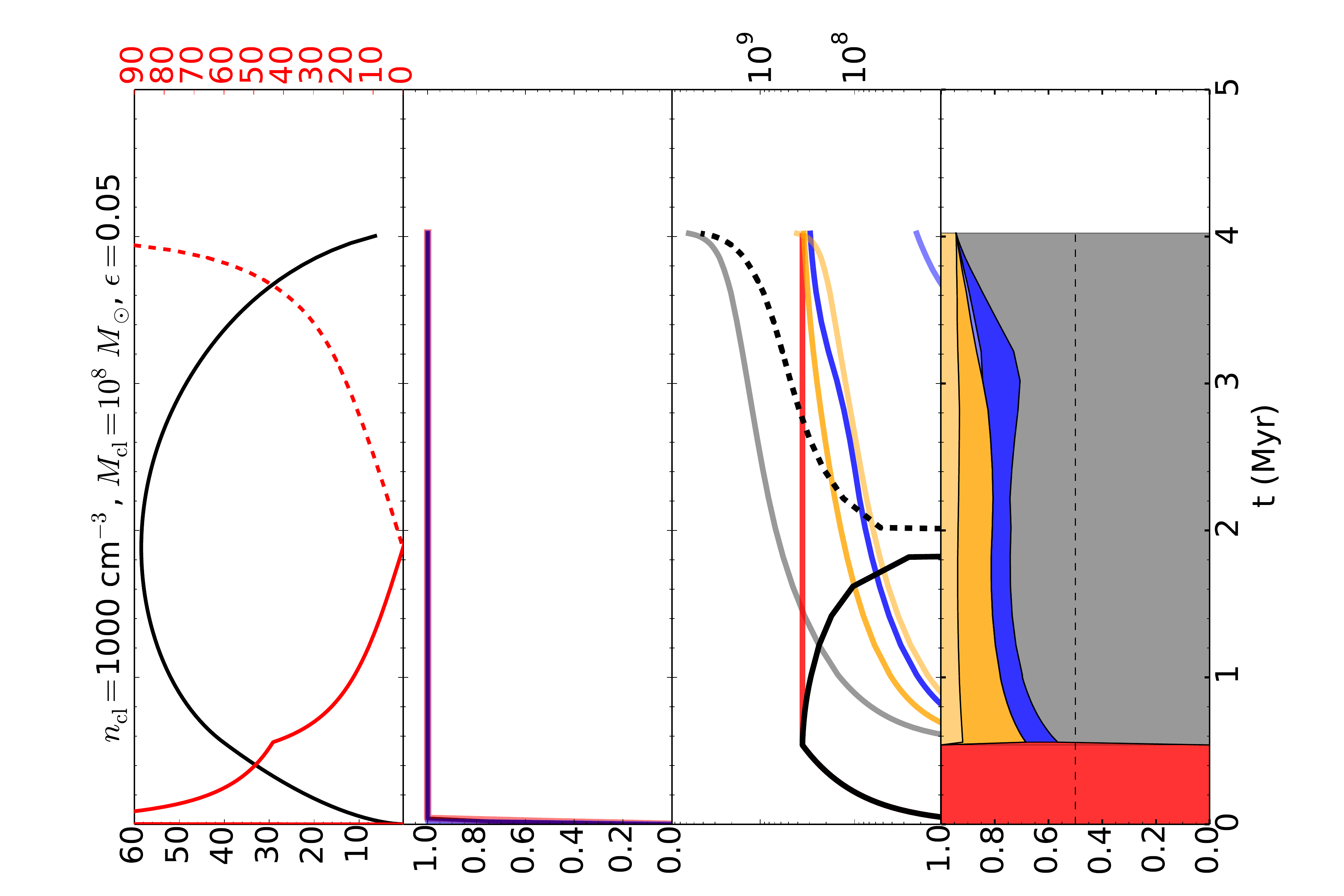}
	\hspace{-4mm}
	\includegraphics[width=0.33\textwidth, angle = 90]{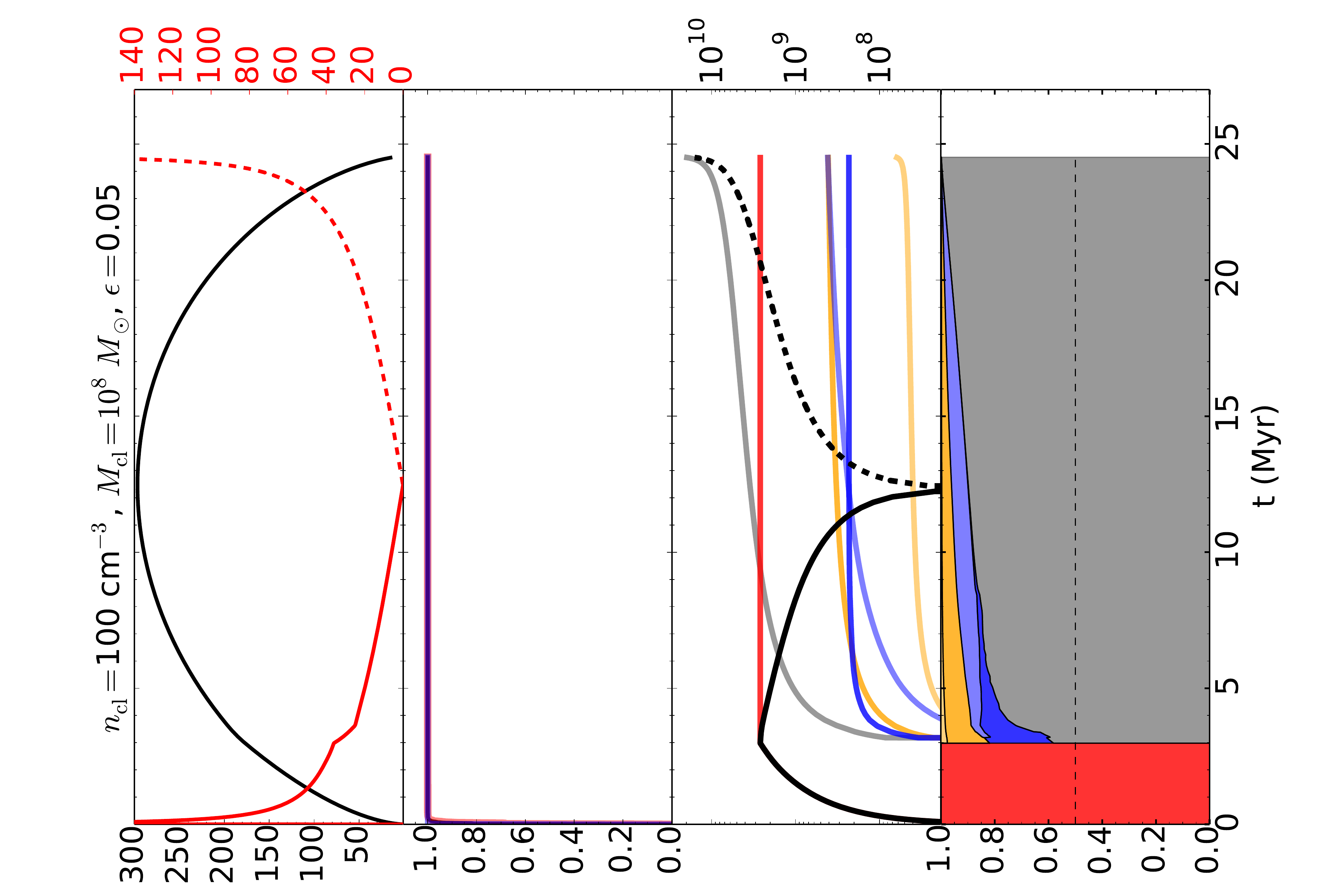}
	\vspace{-0.6mm}\\
	\includegraphics[width=0.33\textwidth, angle = 90]{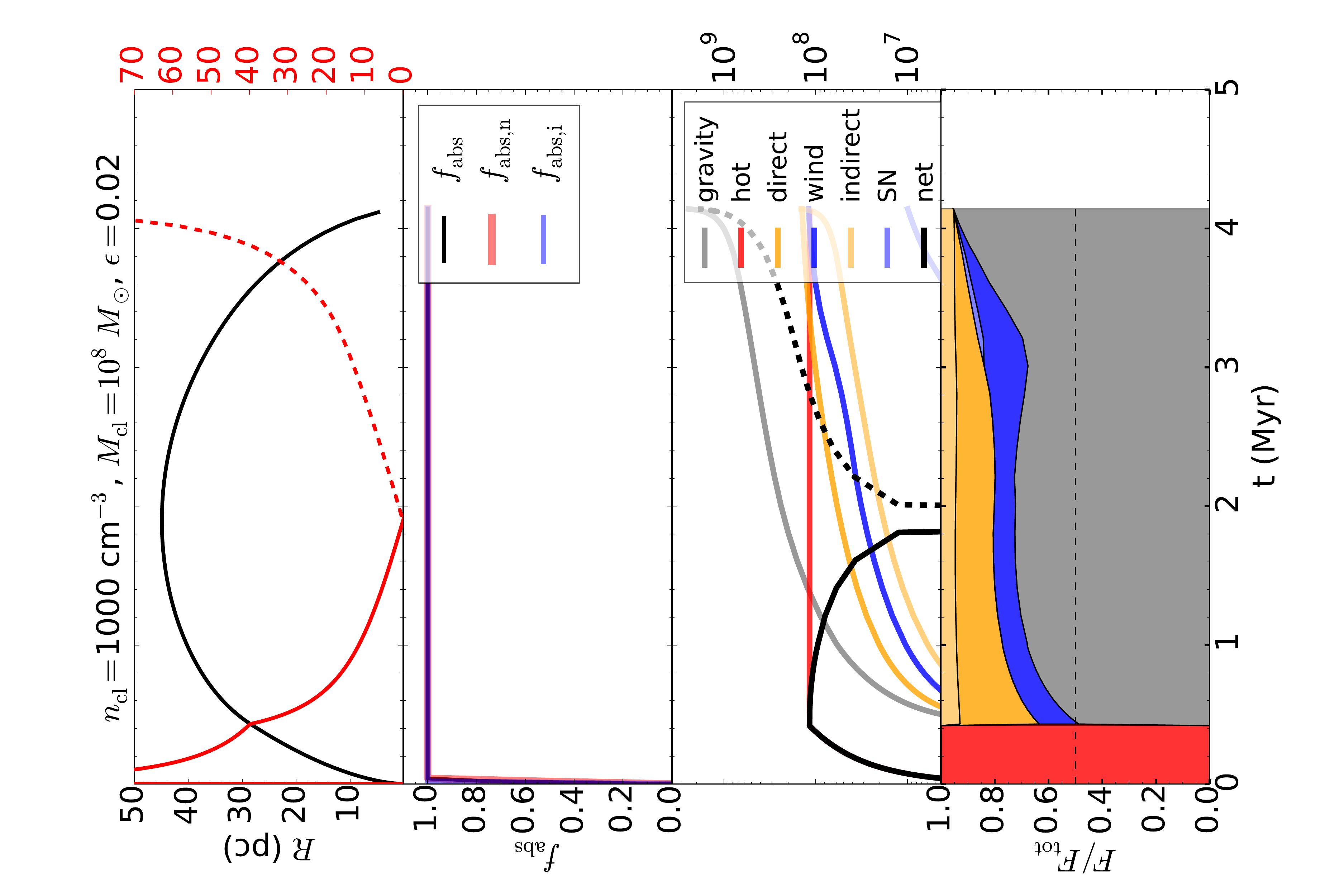}
	\hspace{-4mm}
	\includegraphics[width=0.33\textwidth, angle = 90]{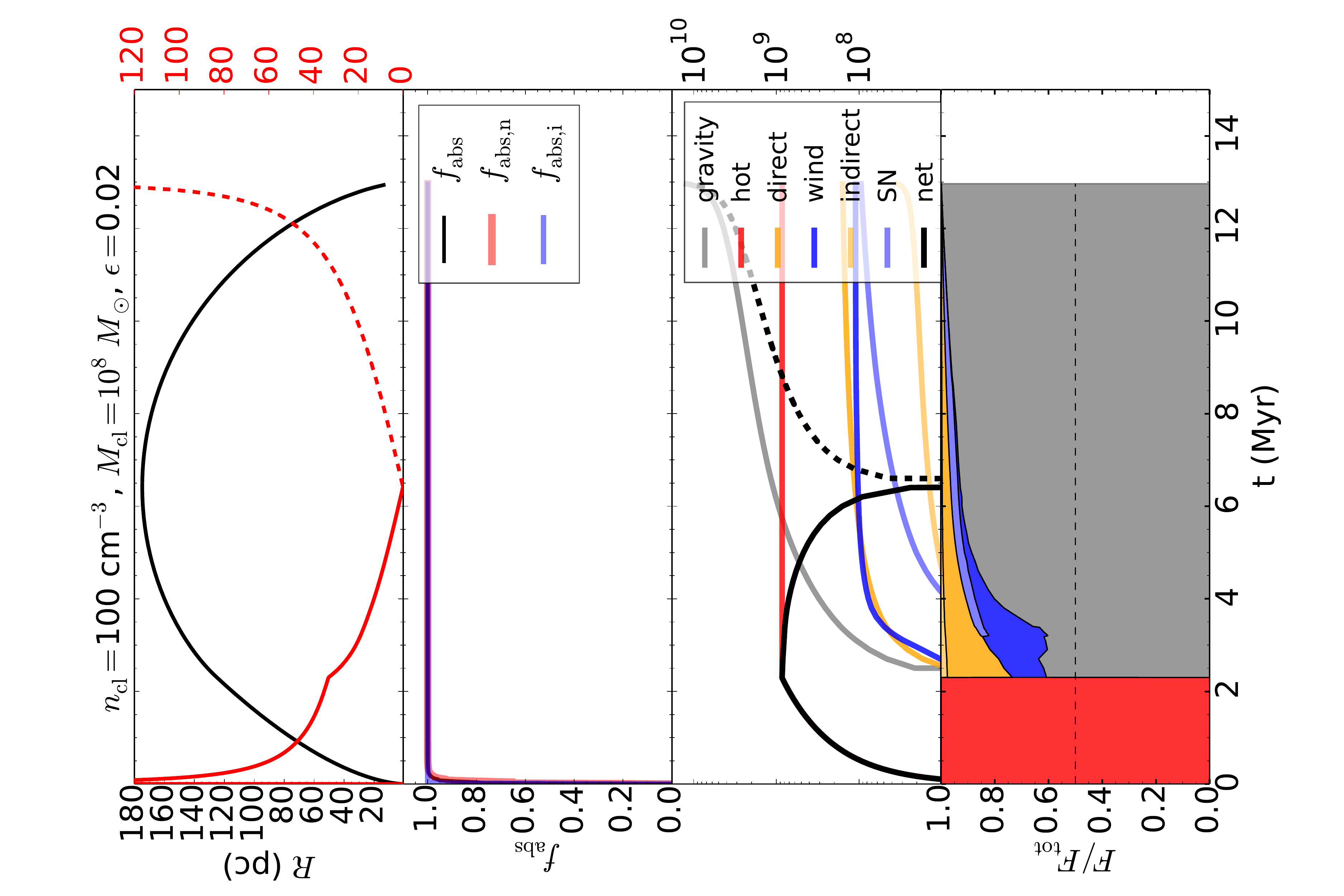}
	\vspace{-0.6mm}
		\caption{Models for clouds with $M_{\rm{cl}} = 10^8\,M_{\odot}$ and $\epsilon = 0.02, 0.05, 0.1,$ and 0.25.}
		\label{fig:Overview8}
\end{figure*}

On the following pages we provide figures showing the shell radius and velocity, the absorption fraction of ionizing and non-ionizing radiation as well as momentum and force comparisons for models with a cloud mass of $10^5M_{\odot}$ and star formation efficiencies of 0.1, 0.15, 0.2, and 0.25, \blue{(Figure~\ref{fig:Overview5})} and models with cloud masses $M_{\rm{cl}}= 10^6, 10^7, 10^8M_{\odot}$ and star formation efficiencies $\epsilon = 0.02, 0.05, 0.1$, and 0.25 \blue{(Figures~\ref{fig:Overview6}, \ref{fig:Overview7}, and \ref{fig:Overview8})}. Densities of $n_{\rm{cl}} = 1000, 100$\,cm$^{-3}$ are shown; the metallicity is solar. Dashed lines in the expansion velocity and momentum plots show negative values.


\label{lastpage}
\end{document}